\documentclass[%
 superscriptaddress,
 amsmath,amssymb,
 prx,
 twocolumn,
]{revtex4-2}

\usepackage{graphicx}
\usepackage{dcolumn}
\usepackage{bm}

\usepackage[table,svgnames]{xcolor} 
\usepackage{color}
\usepackage[normalem]{ulem}
\usepackage{hyperref}

\definecolor{eggplant}{RGB}{180,33,147}
 
\definecolor{grey}{rgb}{.6,.6,.6}

\newcommand{\ket}[1]{|#1\rangle}
\newcommand{\bra}[1]{\langle #1|}
\newcommand{\bracket}[2]{\langle #1| #2 \rangle}

\newcommand{\QIGA}{QiGA }
\newcommand{\QIGAbis}{QiGA-2 }

\newcommand{\circled}[1]{\raisebox{.5pt}{\textcircled{\raisebox{-.9pt} {#1}}}}

\begin{document}

\title{Opening the Black Box Inside Grover's Algorithm}

\author{E.M.\ Stoudenmire}
\affiliation{Center for Computational Quantum Physics, Flatiron Institute, 162 5th Avenue, New York, NY 10010, USA}
\author{Xavier Waintal}
\affiliation{Univ. Grenoble Alpes, CEA, Grenoble INP, IRIG, Pheliqs, F-38000 Grenoble, France}

\date{\today} 

\begin{abstract}
Grover's algorithm is one of the primary algorithms offered as evidence that quantum computers can provide an advantage over classical computers. 
It involves an ``oracle'' (external quantum subroutine) which must be specified for a given application
and whose internal structure is not part of the formal scaling of the quadratic quantum speedup guaranteed by the algorithm. 
Grover's algorithm also requires exponentially
many calls to the quantum oracle to succeed ($\sim \sqrt{2^n}$ calls where $n$ is the number of qubits), raising the question of its implementation on both noisy and error-corrected quantum computers.  In this work, we construct a quantum inspired algorithm, executable on a classical computer, that performs Grover's task in a {\it linear} number of calls to (simulations of) the oracle---an exponentially smaller number than Grover's algorithm---and demonstrate this algorithm explicitly for Boolean satisfiability problems. 
The complexity of our algorithm depends on the cost to simulate the oracle once which may or may not be exponential, depending on its internal structure.
Indeed, Grover's algorithm does not have  an
{\it a priori} quantum speed-up as soon as one is given access to the ``source code'' of the oracle which may reveal an internal structure of the problem. Our findings illustrate this point explicitly as our algorithm exploits the structure of the quantum circuit used to program the quantum computer to speed up the search.
There are still problems where Grover's algorithm would provide an asymptotic speedup if it could be run accurately for large enough sizes. Our quantum inspired algorithm provides lower bounds, in terms of the quantum circuit complexity, for the quantum hardware to beat classical approaches for these problems.
These estimates, combined with the unfavorable scaling of the success probability of Grover’s algorithm -- which in the presence of noise decays as the exponential of the exponential of the number of qubits -- makes a practical speed-up unrealistic even under extremely optimistic assumptions of the evolution of both hardware quality and availability. 
\end{abstract}

\maketitle

\section{\label{sec:intro} Introduction}

Two classes of algorithms dominate the landscape of possible applications for quantum computing. 
The first class computes a non-trivial result then extracts this result using the quantum Fourier transform. 
This class includes the seminal Shor's algorithm for integer factorization \cite{Shor:1994,Beauregard,Kitaev} 
as well as the quantum phase estimation algorithm \cite{Kitaev} 
proposed for solving quantum chemistry problems and several other algorithms \cite{HHL}.  
Some of these algorithms, in particular Shor's, offer an exponential speedup over any known classical methods,
though only for a handful of rather specific applications.

The second class includes Grover's algorithm (GA) and its generalizations, such as amplitude amplification \cite{Grover,GroverReview,Brassard}. 
Grover's algorithm promises a less spectacular quadratic speedup, but in return enjoys 
wide popularity due to its many possible use cases.
Theoretically, quite a large number of problems could be accelerated by simply replacing the critical part of the classical
algorithm by a call to a Grover's routine implemented on a quantum computer. 
It is also appealing that the quadratic speedup
of Grover's algorithm can be put on firm mathematical grounds and is provably optimal under certain assumptions \cite{Bennett}, in contrast to Shor's where the speedup is only conjectured.
This quadratic speedup is very convenient theoretically as it does not require any knowledge of the oracle which encodes the problem
into the quantum algorithm. 
The class of problems for which Grover's algorithm can be applied include instances of 
NP-complete \cite{NP_footnote} problems which are extremely challenging computationally. 
Applications where Grover's algorithm is the main subroutine range from optimizing functions \cite{Baritompa} for e.g.\ analyzing high energy physics data \cite{Wei} to solving various graph 
problems \cite{Durr} to option pricing \cite{Stamatopoulos2020} to pattern recognition (finding a string in a text) \cite{Ramesh} to
various form of machine learning (including supervised learning \cite{Aimeur}, perceptrons \cite{Kapoor}, active learning agents \cite{Kapoor} and reinforcement learning \cite{Dong}).   Specific versions have been implemented on small quantum processors with up to $n \leq 5$ qubits \cite{Dewes,Figgatt,Mandviwalla,Pokharel} but the success probability is still low for the largest systems.

The problem that Grover's algorithm solves (hereafter the ``Grover's problem'') is inverting an unknown function.
Given a function \mbox{$y=f(b)$} where $b=(b_1,b_2\dots b_n)$ is a configuration of $n$ bits $b_i\in\{0,1\}$, the problem is to find a value $b^*$ for a given $y^*$ such that 
$y^*=f(b^*)$. A naive exhaustive search would require an exponentially large number $N=2^n$ of calls to the corresponding classical function. Grover's algorithm solves this problem
in only $\sqrt{N}$ calls to a function which implements $\ket{b}\rightarrow \ket{f(b)}$
on a quantum computer, the ``quantum oracle''.

\begin{figure*}[t]
    \centering
    \includegraphics[width=1.5\columnwidth]{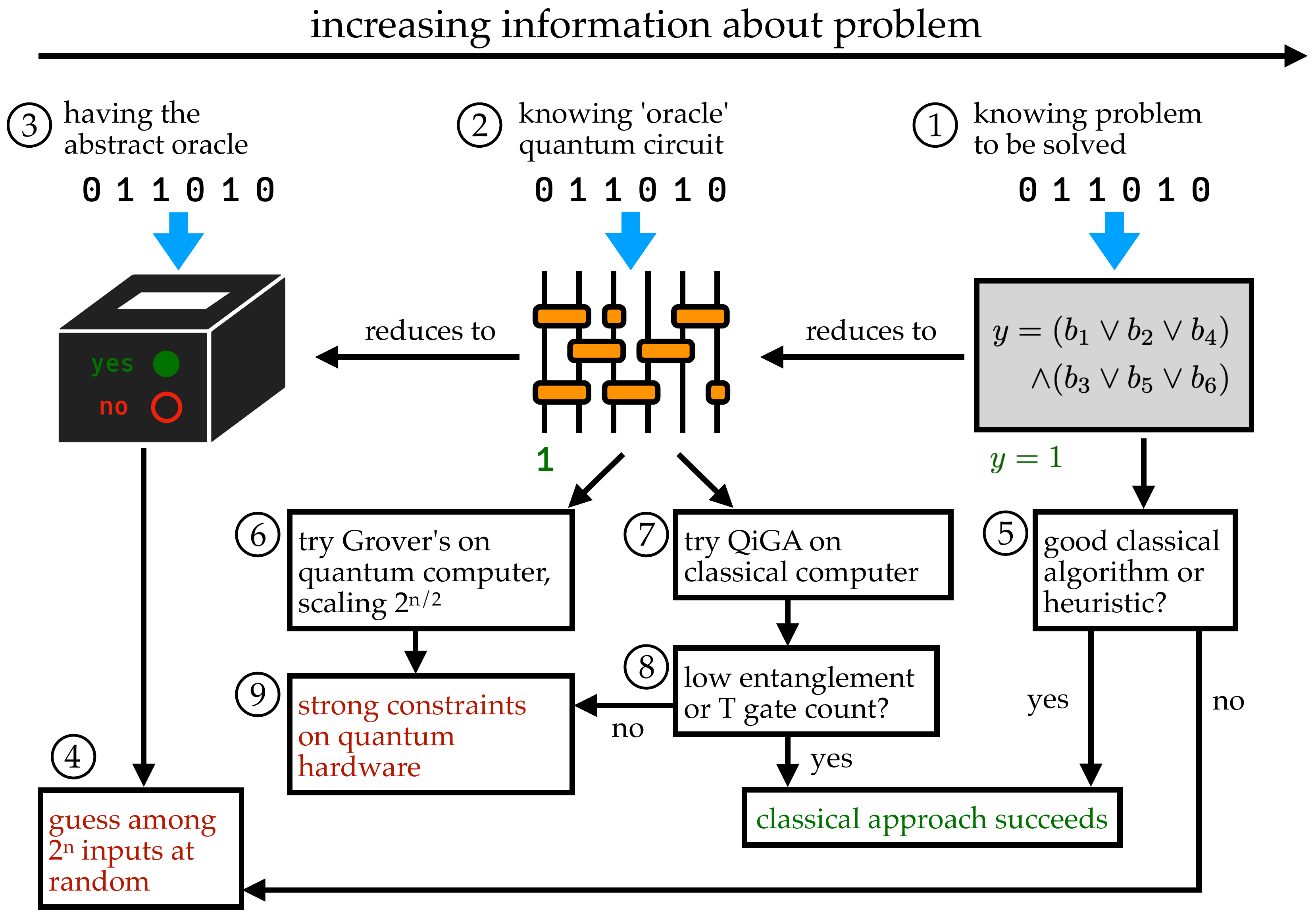}
    \caption{ 
    \label{fig:QigaContext} Diagram showing the context of this work. See text for explanations.
    }
\end{figure*}

The goal of this article is three-fold. First, we build a quantum inspired algorithm (\QIGA\!\!)
that takes the same input as a quantum computer (the quantum circuit that implements the oracle)
and is able to solve the problem in a single call (simulation) of this circuit. Second, we use
\QIGA to obtain necessary conditions for a Grover problem to be hard classically. Third, we analyze the corresponding constraints on the quantum hardware and discuss whether there is room for a possible quantum
advantage, for some of the Grover problems, in the future.

Our conclusions are also three-fold. First, we prove a link between the theoretical complexity of a
Grover problem and the level of quantum entanglement that the quantum computer will encounter when looking for its solution. 
Second, we argue that the quantum advantage of Grover's algorithm is not generic: for many Grover problems, Grover's algorithm does not provide an advantage, even theoretically (this statement does not contradict previous knowledge, as we shall see, but is not widely acknowledged). Third, for the Grover's problems for which \QIGA would be
too computationally expensive on classical hardware, we estimate the resources needed on quantum hardware. 
We arrive at time-to-solutions of thousand of years under highly optimistic assumptions about the evolution of quantum hardware.

\section{Point of view taken in this work and main results}
\label{sec:point_of_view}

The theoretical speedup of Grover's algorithm is of a peculiar nature: it is an abstract speedup that considers the oracle as a black box function and counts the computational cost solely in terms of the number of calls to the oracle (the cost of one call to the $f(b)$ oracle almost always depend on $N$ in some way).
Because this peculiarity implies some rather subtle points, we first carefully define our working hypothesis and the definition of quantum advantage that will be used throughout this work. 
The discussion below and the numbered items (\circled{1}, \circled{2}, etc.) correspond to the diagram of Fig.~\ref{fig:QigaContext}.

Quantum advantage compares the performance of a classical computer with those of a quantum one.
Any proper definition of this concept must define what information about the problem is given to the computers. We define three different levels of information.  One can be given the actual problem \circled{1} to be solved. An example of such a problem is the Hamiltonian path problem: given a set of cities, and the roads that connect them, find a path that goes through all the cities once and just once.
Second, it is easy to write an algorithm that takes a path as its input (encoded through the binary variables $b_i$), follows the path, checks how many times each city is visited,
returns one if they are visited exactly once, and returns zero otherwise. 
This algorithm implements the function $f(b)$. From this algorithm, using reversible logic \cite{Nielsen}, one can define the quantum circuit \circled{2} (the sequence of one and two qubits gates) that implements the quantum version
$\ket{b}\rightarrow \ket{f(b)}$ of the oracle. Third, we can imagine that we can call $f(b)$ for any given
input $b$ but do not know the problem it comes from nor the program or ``source code'' of the classical oracle or the quantum circuit. This is the abstract oracle \circled{3}.

If we only possess the abstract oracle \circled{3}, there is not much we can do classically: we can only  \circled{4} loop over the $2^n$ different inputs $b$, and call $f(b)$ until it returns the value $y^*$.
It is in this context that the quantum speed up of Grover's algorithm has been proven: 
Grover's algorithm needs to call the quantum oracle only $\sqrt{2^n}$ times, which is parametrically faster than $2^n$ calls to the abstract oracle.
On the other hand, if we know the problem \circled{1}, there might be a good classical algorithm \circled{5} to solve this problem. For many problems, this specialized algorithm might be faster than even an ideal application of Grover's algorithm. For instance, if the problem is to find an integer $b$ (written in binary form) such that
$b^2 - 26b + 169 =0$, then we obviously have better methods to find $b^*$ than to loop over the different values of $b$. There is a subclass of Grover's problems, however, where the best classical algorithm scales as $2^n$, for instance breaking certain cryptographic functions or solving circuit-SAT \cite{SAT_footnote}. 
For these problems knowing \circled{1} does not provide any benefit over knowing \circled{3} in the context of classical computing. 
Note that it is not sufficient for a problem to be NP-complete to fall into this category. For instance the 3-SAT problem, that we discuss below, is NP-complete, 
yet classical algorithms to solve it scale better than $O(\sqrt{N})$ (but are still exponential).
We will also later discuss how the generalization of Grover's algorithm known as ``amplitude amplification'' relates to the above statements (in Section~\ref{sec:advantage_theo}).

We are interested in the situation where we are given the quantum circuit \circled{2} as the input. For these problems, we could use the quantum circuit as the input to run Grover's algorithm on a quantum computer \circled{6}, in a future where a quantum computer would be available with the correct specifications
(necessary of number of qubits, amount of error per gate, and speed). 
In the first part of this paper, we construct a quantum inspired algorithm
\QIGA \circled{7} that runs on a classical computer and takes the same input \circled{2} as the quantum computer. Hence the comparison between the two is fair: both computers are given the same input.
\QIGA solves the problem in at most $O(\log N)$ calls and often a single simulation of the quantum circuit. 
To obtain the same result, a quantum computer would need to make exponentially many calls to the oracle.
So the simulation of a single run of the oracle quantum circuit must be exponentially hard for \QIGA in order for the quantum computer to provide an advantage. 
Hence, \QIGA provides necessary conditions on the quantum circuit for the problem to be hard classically \circled{8}: the quantum entanglement must be high 
which translates into a minimum depth of the quantum circuit, among other aspects. Also, the number of non-Clifford gates (such as the T gate count) must be high enough.
These necessary conditions on the quantum circuit translate in turns \circled{9} into constraints on the hardware in terms of noise, speed, number of qubits, and parallelism. 
All the conclusions drawn in this article assume that the classical computer is given the quantum circuit that will be used to program the quantum oracle on the quantum computer. 
As far as the authors know, this is not a controversial position \cite{Aaronson_QCLec18} and the common use of the
``abstract oracle'' \circled{3} as the input for the classical computer should be understood as 
``using the knowledge \circled{1} of a problem that is classically so hard that it would take $2^n$ operations to solve anyway.'' 
This, of course, considerably reduces the number of problems for which Grover's algorithm could be expected to provide a speed-up.

\begin{figure}[t]
    \centering
    \includegraphics[width=\columnwidth]{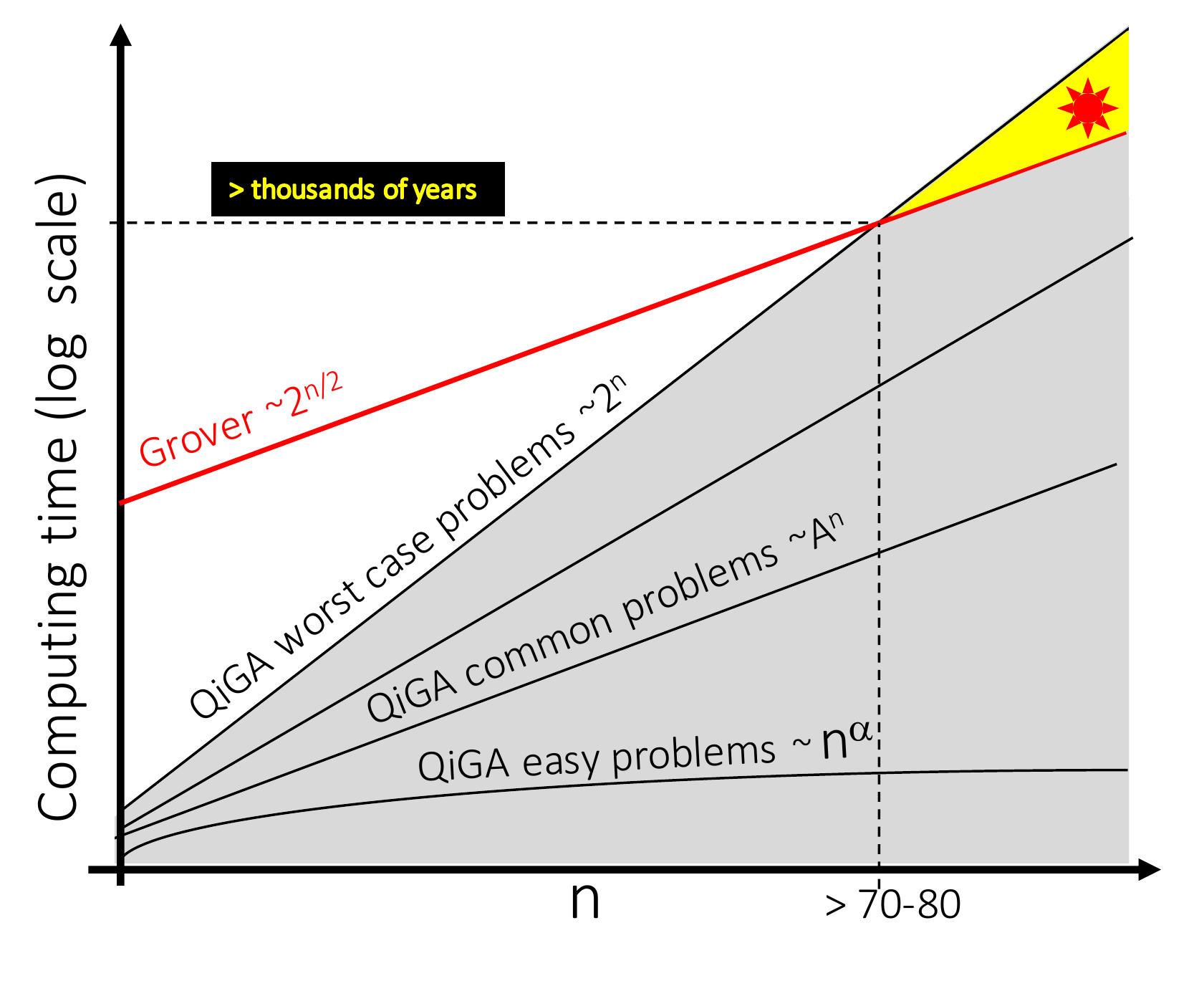}
    \caption{ 
    \label{fig:QiGA_2} Schematic of the computing time versus number of qubits $n$ for a quantum computer using Grover (red) and our \QIGA algorithm (black). A quantum advantage is only possible for very specific problems and astronomically large computing times (yellow region with the red star). Different problems have different classical difficulties whose scaling ranges from polynomial $n^\alpha$ (where $\alpha\ge 0$ is a problem dependent exponent) to exponential $A^n$ with $1<A<2$ up to the worst case $A=2$ (that is believed to be reached in some cryptographic problems).
    }
\end{figure}

Figure~\ref{fig:QiGA_2} summarizes the main practical implications of our work. It sketches the computational time versus problem size for a quantum computer using Grover (red) and various types of problem instances using \QIGA (black, see the rest of the article for examples).
For a large class of problems, \QIGA scales better than Grover for any problem size. A fraction of the problems will be hard for QiGA.  However, even for these hard instances, the crossing point where one enters the asymptotic regime and a quantum computer would provide a quantum speed-up corresponds to $n > 80$ which already is in a regime which is impossible to reach in practice (thousands or more years of uninterrupted computing time while holding the error rate per step below $10^{-15}$).

This article starts with Section \ref{sec:review} where we introduce our notations and some standard material on Grover algorithm. The rest of the article is split into two parts. In the first (Sections  \ref{sec:classical}, \ref{sec:3SAT} , and \ref{sec:quiga2}), we construct the quantum inspired algorithm that mimics Grover's algorithm using a classical computer but takes advantage of its internal structure to require exponentially fewer calls to the oracle. This algorithm is based on tensor networks, and more precisely on matrix product states (MPS). We explicitly demonstrate and test this algorithm using both random and quasi-one-dimensional instances of NP-complete random boolean satisfiability problems (3-SAT) in Section \ref{sec:3SAT}. Section \ref{sec:quiga2} discusses an alternative quantum inspired algorithm requiring only amplitude calculations. (Note that this second algorithm does not necessarily require the use of tensor networks, so may be more accessible for readers unfamiliar with that subject.)
In the second part of the article, we examine the implication of our findings on the possibility of a quantum speed-up (Sections \ref{sec:scaling}
and \ref{sec:advantage}). 
In particular, Section~\ref{sec:scaling} establishes the lack of resilience of Grover's algorithm to noise, an important aspect needed for the discussion of a possible quantum advantage.

\section{\label{sec:review} Problem formulation}

Grover's algorithm \cite{Grover,GroverReview} aims to harness the potential of quantum parallelism by gradually extracting the result of a parallel function evaluation through a series of \emph{Grover iterations}. For a problem involving
classical states of length $N$, for which $n$ qubits are needed such that $N=2^n$, the number of 
Grover iterations needed to extract the result of the problem scales as $\sqrt{N}$, which is to be compared to worst-case classical
strategies such as guessing solutions at random or performing an exhaustive search which have a cost scaling as $N$.
 
\subsection{\label{sec:notations} Notation and Problem Setting}
Let $b\in\{0,1,\cdots,N-1\}$ be an integer whose binary representation is
$b_{n-1}\cdots b_1 b_0$, that is $b = \sum_{i=0}^{n-1} b_i 2^i$ with $b_i\in\{0,1\}$. We denote by $\ket{b} = 
\ket{b_{n-1} \cdots b_1 b_0}$ the corresponding $n$-qubit state in the computational basis.

Let $f(b)$, be a function that takes a bitstring as an input $b$
and returns 
\begin{align}
    f(b) = \begin{cases} 1, & \text{if}\ b = w \\ 0, & \text{if}\ b \neq w \end{cases} \ \ .
\end{align}
Here $w$ is a (unknown) specific bitstring. 
GA aims to solve the problem of finding the value of $w$ in as few calls to the function $f$ as possible.
This problem can be viewed as  inverting $f$, that is, computing $w = f^{-1}(1)$.

GA also assumes one can implement an \emph{oracle operator} $U_w$ such that for states $\ket{b}$
in the computational basis
\begin{align}
    U_w \ket{b} = (-1)^{f(b)} \ket{b}.
    \label{eq:oracle_action_basic}
\end{align}
Since a quantum computer can perform classical logic (at the price of adding ancilla qubits to ensure reversibility), a classical algorithm that computes $f(b)$ can be turned into a quantum circuit
that implements Eq.~\eqref{eq:oracle_action_basic}. The explicit form of $U_w$ reads,
\begin{align}
    U_w \ket{b} = \begin{cases} -\ket{b}, & \text{if}\ b = w \\ +\ket{b}, & \text{if}\ b \neq w \end{cases}
    \label{eq:oracle_action}
\end{align}
therefore  $U_w$ is equivalent to the operator
\begin{align}
    U_w = 1 - 2 \ket{w} \bra{w} \ . \label{eq:oracle_operator}
\end{align}
However, for any real application of practical interest, one does not know the value of $w$ ahead of time and only knows how to implement 
$U_w$ as a quantum circuit based on the function $f$.

GA straightforwardly generalizes to the case of multiple solutions $\{w^\alpha\}_{\alpha=1}^{S}$ such that $f(w^\alpha)\equiv 1$. 
One defines the oracle as 
\begin{align}
    U_w = 1 - 2 \sum_{\alpha=1}^S \ket{w^\alpha} \bra{w^\alpha}.
    \label{eq:multi_oracle}
\end{align}
Each solution $w^\alpha$ has a binary representation $w^\alpha_{n-1} \cdot w^\alpha_2 w^\alpha_1 w^\alpha_0$.
In this article, we focus on the case where the problem has a fixed number of solutions $S$ (or more generically where $S$ grows at most polynomially with $n$). For problems that have an exponential number of solutions, our algorithm would have to be revisited, but we conjecture that easy classical solutions exist in that case. For each qubit, we define the two states $\ket{+}$ and $\ket{-}$ as,
\begin{align}
\ket{\pm} = \frac{\ket{0}\pm\ket{1}}{\sqrt{2}}
\end{align}
and the equal weight superposition state $\ket{s}$ as,
\begin{align}
    \ket{s} &= \ket{+++\cdots+} \\
            &=\frac{1}{\sqrt{2^n}} \sum_{x_{n-1} \cdots x_0 \in \{0,1\}^n} \ket{x_{n-1}\cdots x_0} \ .
\end{align}
Last, GA requires a second operator, the diffusion operator $U_s$ that has a structure
similar to the oracle but with respect to the known state $\ket{s}$:
\begin{align}
    U_s= 1 - 2 \ket{s} \bra{s} \ . \label{eq:diffusion_operator}
\end{align}

\subsection{\label{sec:grover_def} Definition of the Grover Algorithm}
Given an oracle $U_w$, GA proceeds as follows
\begin{enumerate}
    \item initiate the qubits in state $\ket{000\cdots0}$
    \item apply a Hadamard gate on each qubit to obtain $\ket{s}$
    \item apply the oracle operator $U_w$
    \item apply the diffusion operator $U_s$
    \item repeat steps 3 and 4 each $q$ times
    \item measure the qubits in the computational basis and find $\ket{w}$ with a probability very close to one
\end{enumerate}
The optimal number of steps of GA can be estimated to be about $q\approx r$ with 
$r \equiv \frac{\pi}{4} \sqrt{N} = \frac{\pi}{4} 2^{n/2}$. In the case where there are multiple solutions, the measurement at the end produces one of the $w^\alpha$ with uniform probability. GA has an appealing geometrical interpretation \cite{Nielsen}: $U_w$ and $U_s$ are mirror operators with respect to the hyper-planes perpendicular to $w$ and $s$. It follows that the product $U_sU_w$ is a rotation inside the ($\ket{s},\ket{w}$) plane that gradually rotates the state from $\ket{s}$ to $\ket{w}$.

\subsection{\label{sec:mps} On the level of entanglement inside Grover's algorithm}

At the root of this work is the observation that, {\it in between} the calls to the oracle, the level
of entanglement present in the internal state of a quantum computer running GA is extremely low \cite{Ma}. 
In between each iteration of GA, the entanglement entropy between any two subgroups of qubits is at most $\log(2)$.
Indeed, after any application of
$U_w$ or $U_s$, the internal state $\ket{\Psi}$ of the quantum computer lies in a superposition of $\ket{s}$ and $\ket{w}$ \cite{Nielsen},
\begin{align}
\ket{\Psi} = \alpha \ket{s} + \beta \ket{w}
\end{align}
with $|\alpha|^2+|\beta|^2=1$, i.e.\ in the superposition of two (unentangled) product states ($1+S$ states in the general case).   
It follows that $\ket{\Psi}$ can be cast into the form,
\begin{align}
\label{eq:mps}
\ket{\Psi} &=& 
\begin{bmatrix} \alpha  & \beta  \end{bmatrix}
\begin{bmatrix} \ket{w_1} & 0 \\ 0 & \ket{+} \end{bmatrix}
\begin{bmatrix} \ket{w_2} & 0 \\ 0 & \ket{+} \end{bmatrix} \ \ \ \nonumber\\
\ \ \ \ & ... & \begin{bmatrix} \ket{w_{n-1}} & 0 \\ 0 & \ket{+} \end{bmatrix}
\begin{bmatrix} \ket{w_n} & 0 \\ 0 & \ket{+} \end{bmatrix}
\begin{bmatrix} 1 \\ 1 \end{bmatrix}
\end{align} 
As we shall see, such a state is manifestly a rank $\chi=2$  \emph{matrix product state} \cite{Ostlund,Vidal,Perez-Garcia}(MPS) or $\chi=1+S$ in the general case. In fact, any state which is a sum of $P$ product states can be explicitly written as an MPS of rank $\chi=P$ \cite{McCulloch}. Such a state has low entanglement and can easily be kept inside the memory of a classical computer at a linear cost in the number of qubits. In other words, while Grover's algorithm takes advantage of quantum parallelism (i.e.\ superposition), it uses very little entanglement for most of the algorithm. The only possible exception is while the oracle circuit has only been partially applied (see the illustration below). Similarly, the oracle itself is a rank-2 \emph{matrix product operator} \cite{McCulloch,Verstraete:2004} (MPO), another standard object of tensor network theory (that, again, may or may not be computationally hard to obtain, as we shall explain.)

\begin{figure}[t]
    \centering
    \includegraphics[width=\columnwidth]{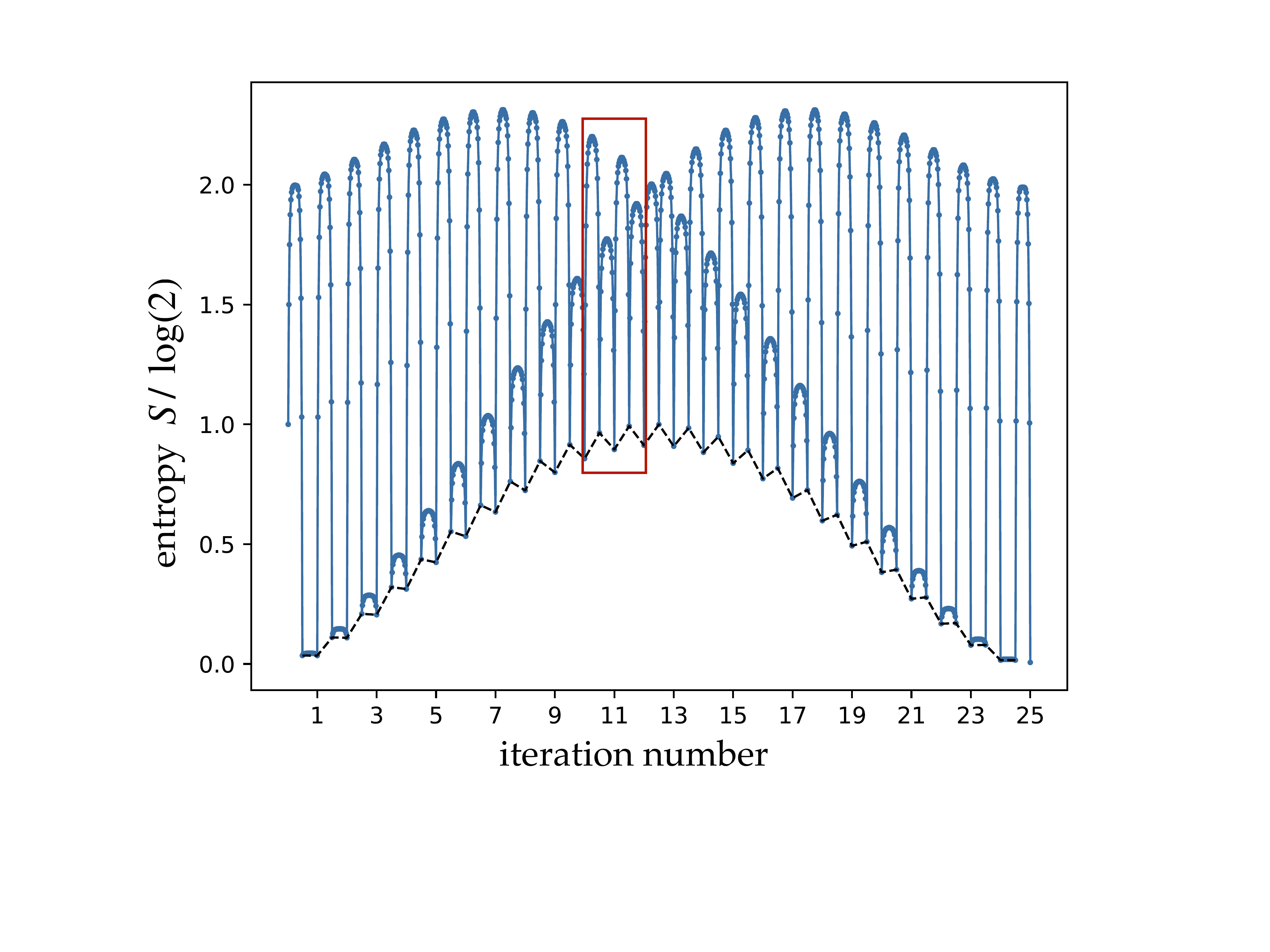}
    \includegraphics[width=\columnwidth]{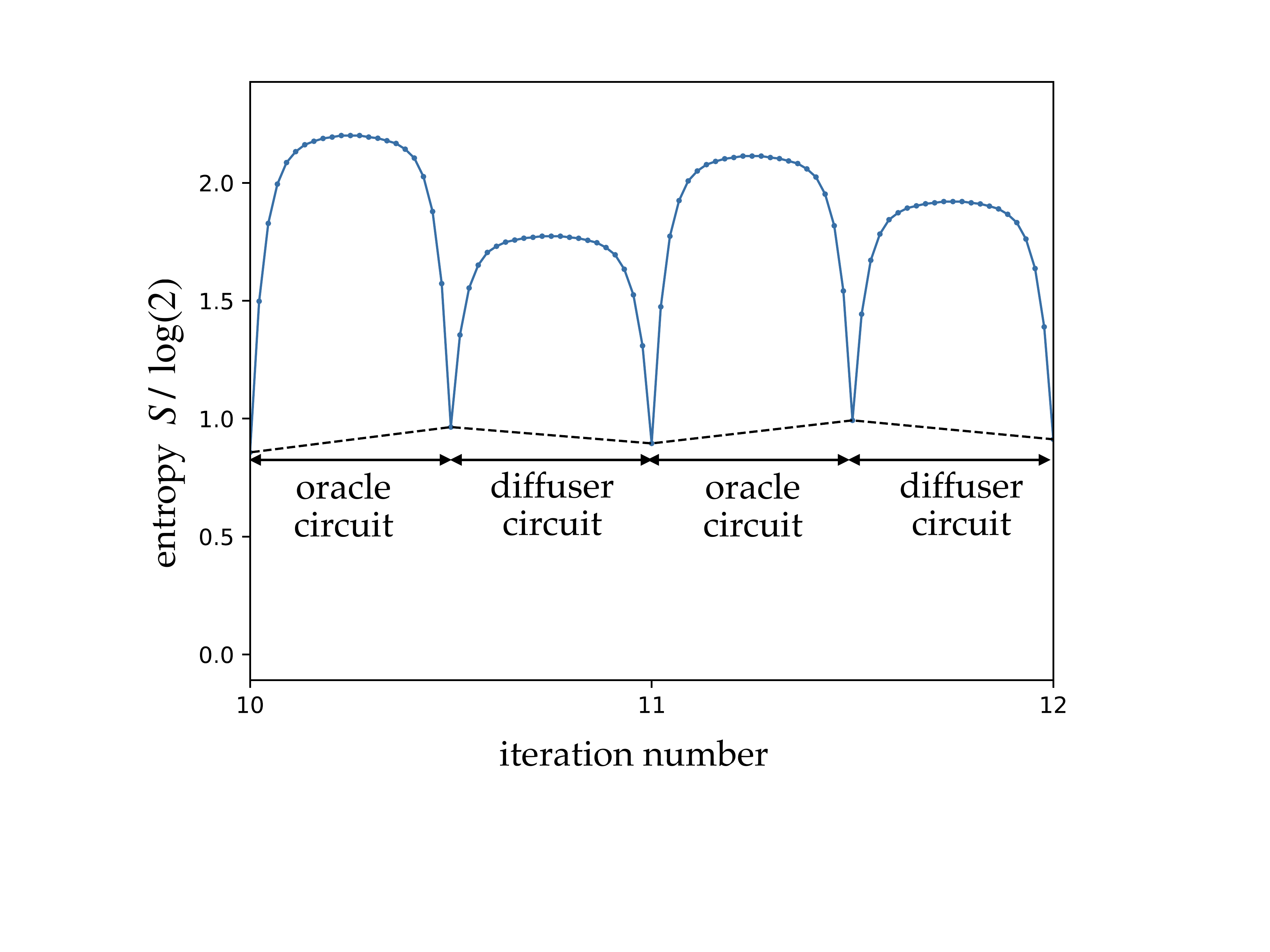}
    \caption{Entanglement entropy, in units of $\log(2)$, of the quantum state during a simulation of Grover's algorithm for $n=10$ qubits for the optimal number $r=25$ of iterations.
    Substeps between each complete iteration show the properties of the state after each layer of the quantum circuits described in Appendix~\ref{appendix:circuit} which implement the oracle and diffusion operators. The entropy is dominated by the oracle step for the first half of the algorithm, then becomes dominated by the diffusion step for the second half of the algorithm.
    During the simulation the matrix product state (MPS) rank (not shown) goes up and down in a sawtooth shape with maxima of $\chi=11$ during an iteration and minima of $\chi=2$ between each iteration. The dashed line along the lower envelope is the theoretical prediction for the entanglement between iterations of the algorithm (the oscillations are a small $1/N$ effect). The lower panel shows a zoom of the region in the box in the upper panel, labeling substeps where the oracle and diffuser circuits act.
    \label{fig:entropy}
    }
\end{figure}

Before describing our ``quantum inspired'' GA (where we use the resources available in 
a classical computer in the most efficient way), we start with a ``simulation'' of GA (where we only use operations implementable on a quantum computer).
Figure~\ref{fig:entropy} shows the entanglement entropy of the quantum state during a simulation of Grover's algorithm for $n=10$ qubits, using the quantum circuit described in Appendix~\ref{appendix:circuit} to implement a ``toy'' oracle and the diffusion operator. As claimed, the entanglement in between iterations never exceeds $\log(2)$, a value which is reached when the algorithm is halfway done. The entanglement entropy does become higher during the partial application of the oracle and diffusion operator circuits. The value of this intra-oracle entanglement barrier is strongly problem dependent and will determine the performance of the quantum inspired GA.

\section{\label{sec:classical} A quantum inspired algorithm for simulating Grover's algorithm in a single call to the oracle}

We now detail the different steps of our quantum inspired Grover's algorithm (\QIGA\!\!). Although we use MPS and MPO tensor network technology for both \QIGA as well as mere simulations of GA, we emphasize that the goals are very different. In the first, we aim at solving the Grover problem with as few computations as possible while in the latter we want to mimic what would happen in a (possibly noisy) actual quantum computer.

\subsection{\label{sec:tensor} Review: Tensor Networks, Matrix Product States, and Matrix Product Operators}

\begin{figure}[t]
    \centering
    \includegraphics[width=\columnwidth]{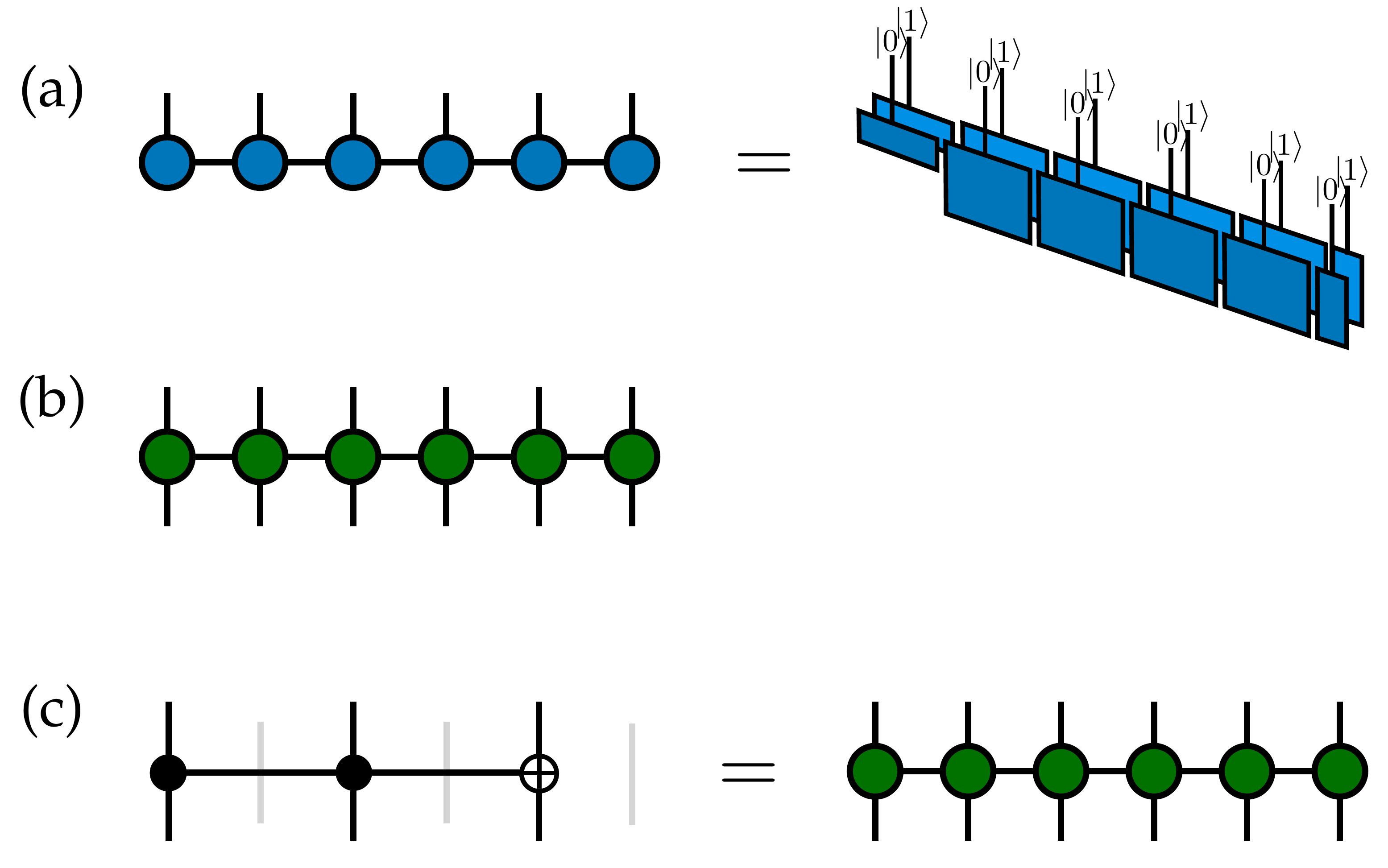}
    \caption{Tensor diagrams for (a) matrix product state (MPS) tensor network for representing pure quantum states and (b) matrix product operator (MPO) for representing operators. Panel (c) illustrates the idea that common quantum operations such as Toffoli gates can be exactly and efficiently represented by MPOs. The right-hand side of panel (a) shows that an MPS can be viewed as a collection of $2n$ matrices.
     \label{fig:mps_and_mpo}
    }
\end{figure}

Since the time that Grover's algorithm was formulated, there have been profound developments in classical algorithms 
 which can not only simulate some quantum algorithms, but strongly resemble non-error-corrected quantum computers in many important respects \cite{VidalEfficient,Zhou}.
These algorithms, which scale linearly in the number of qubits, allow
one to apply arbitrary quantum gates and perform unbiased measurements of the state, 
plus other operations quantum computers cannot perform 
without exponential effort, such as efficient post-selection. Their main limitation is that only moderately entangled states can be handled efficiently \cite{Orus_2014, Bridgeman_2017}.

The basis of these algorithms are tensor networks, the most notable examples being
matrix product states (MPS) for representing pure states and  matrix product operators (MPO) for 
 operators and mixed states. The tensor diagrams for MPS and MPO are shown in Fig.~\ref{fig:mps_and_mpo}(a--b).  Figure~\ref{fig:mps_and_mpo}(c) emphasizes that common quantum gates can be written exactly as MPO tensor networks.
References \onlinecite{Orus_2014, Bridgeman_2017} provide accessible reviews of the subject. An MPS compresses a quantum state by factoring it into a network of smaller
tensors contracted in a one-dimensional chain-like structure. In Fig.~\ref{fig:mps_and_mpo}(a), the open lines are the qubit indices (or physical indices) while the lines connecting two tensors (the circles) correspond to ``virtual'' or ``bond'' indices. The explicit form of an MPS state as sketched in 
Fig.~\ref{fig:mps_and_mpo}(a) is
\begin{align}
    \ket{\Psi}  
            &= \sum_{x_{n-1}\cdots x_0 \in \{0,1\}^n} M_{n-1} (x_{n-1})\cdots M_{0} (x_{0}) \ket{x_{n-1}\cdots x_0} 
\end{align}
where the $M_i(x)$ are $\chi\times\chi$ matrices  ($1\times\chi$ and $\chi\times 1$ respectively for $i=n-1$ and $i=0$).
MPS and MPO networks are a type of low-rank factorization of a high-order (many-index)
tensor, with the ranks $\chi$ referring to the dimensions of the indices linking the tensors together. Practically this means one can approximate
tensors naively having $2^n$ parameters by  storing just $2n$ matrices of size $\chi \times \chi$ [right side of Fig.~\ref{fig:mps_and_mpo}(a)] or  
equivalently $n$ tensors of dimensions $\chi \times 2 \times \chi$ [left side of Fig.~\ref{fig:mps_and_mpo}(a)].
The full high-order tensor would be recovered if all the smaller tensors were contracted together, but doing so would incur an exponential cost. 
The main trick of using tensor networks is to implicitly manipulate the high-order tensor by operating on a few of the small tensors at a
time, which keeps the costs polynomial in the rank $\chi$ (usually $\chi^3$ or $\chi^2$) and linear in the number of indices $n$. Quantum states such as GHZ or W states are exactly MPS of rank $\chi=2$ and product states such as the initial state $\ket{s}$ of Grover's algorithm are $\chi=1$ MPS. As evident from Eq.~(\ref{eq:mps}),  the internal state of GA after a call to the oracle or diffusion operator is a rank $\chi=2$ MPS (in the case of a single solution, and $\chi$ linear in the number of solutions more generally).

\begin{figure}[t]
    \centering
    \includegraphics[width=\columnwidth]{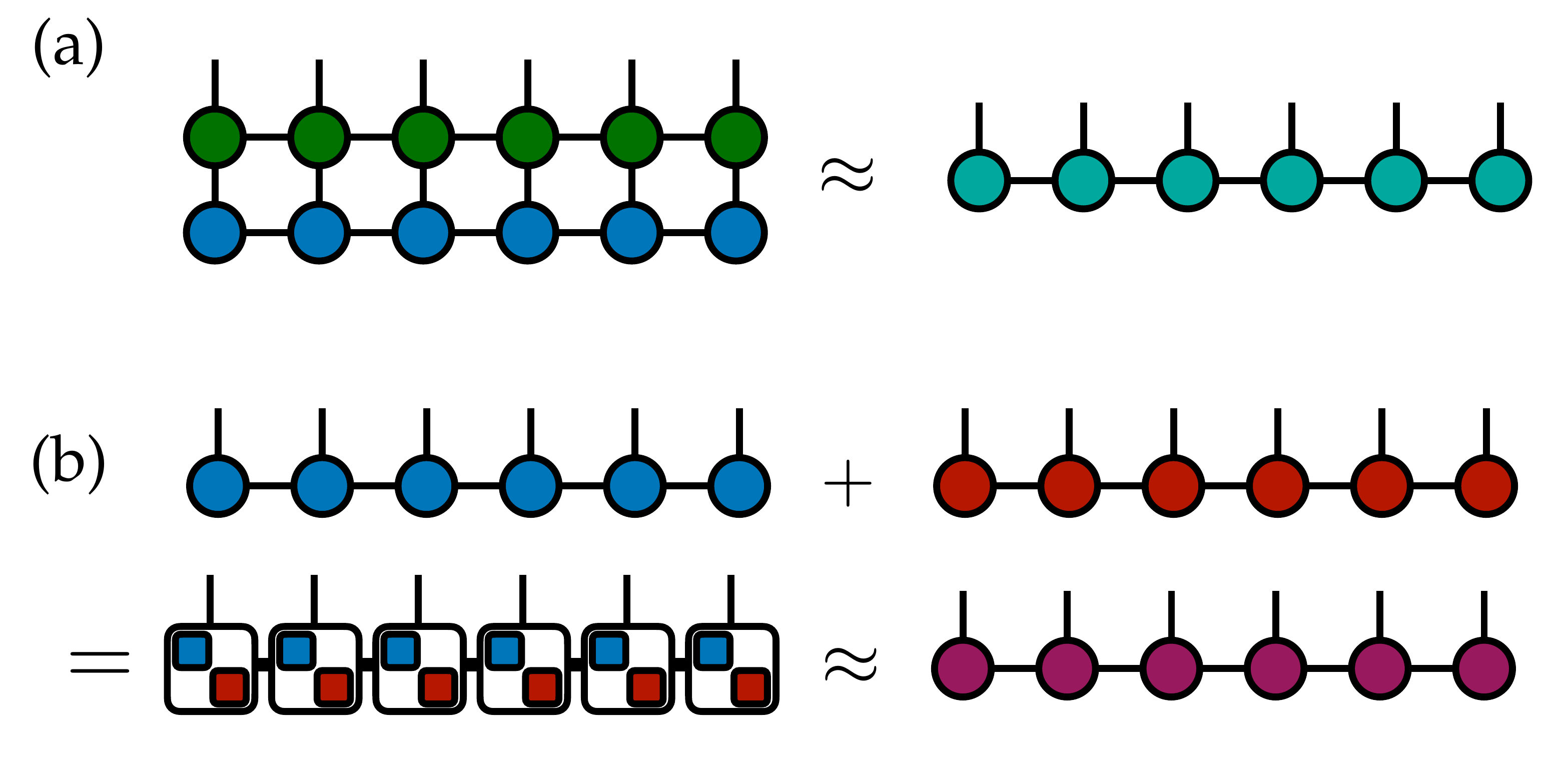}
    \includegraphics[width=\columnwidth]{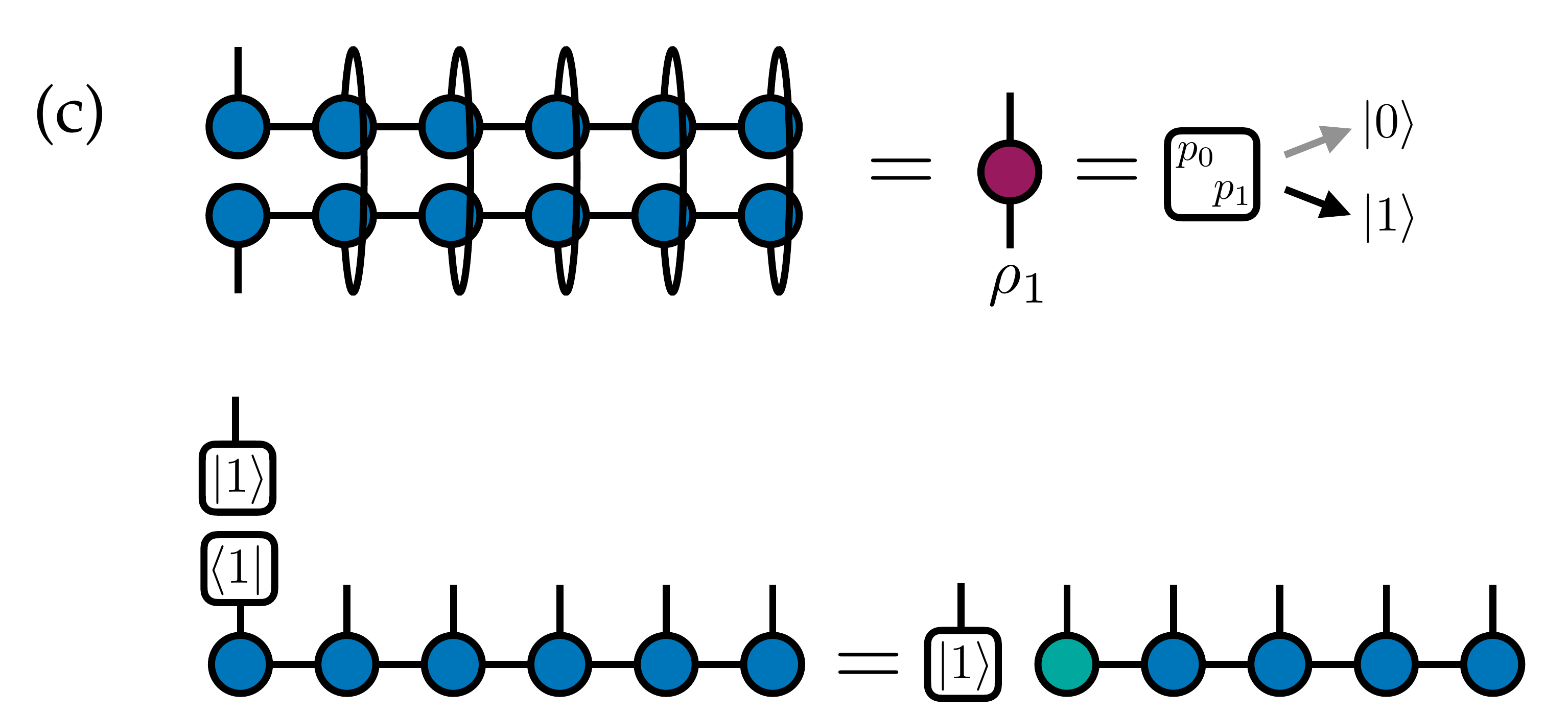}
    \caption{Illustration of three essential MPS and MPO algorithms. (a) An MPS can be multiplied by an MPO to obtain another MPS with the new rank $\chi'$ determined adaptively. (b) Two MPS can be summed by forming a new MPS representing the sum with a block-diagonal structure then adaptively compressing it. (c) An MPS can be ``perfectly'' sampled by computing the reduced density matrix $\rho_1$ for the first qubit then randomly selecting $\ket{0}$ or $\ket{1}$ with probabilities given by diagonal entries of $\rho_1$. After projecting the first qubit into the selected state, one can recursively sample the second qubit and so on.  \label{fig:mps_operations}
    }
\end{figure}

Figure~\ref{fig:mps_operations} depicts three of the main tensor network algorithms used in this work, which are (a) applying
MPOs to MPS \cite{Vandamme2023efficient}, (b) summing MPS \cite{McCulloch}, and (c) sampling from MPS \cite{Ferris:2012,METTS}. 
MPO tensor networks can be used to represent any two-qubit operator
by using a rank of at most $\chi=4$. Multi-qubit operators such as Toffoli gates can also be represented exactly as an MPO (rank $\chi=2$). 
Applying an MPO of 
rank $k$ to an MPS of rank $\chi$ can be performed to high accuracy with a cost scaling as $n \chi^3 k^2$ \cite{Vandamme2023efficient}. 
Two MPS of ranks $\chi_1$
and $\chi_2$ can be summed in a controlled way, by forming a larger MPS exactly representing the sum with a rank $\chi_1+\chi_2$
then recompressing this MPS to some final lower rank, which can be determined adaptively for a fixed error goal \cite{McCulloch}. 
An MPS can be sampled in any local basis using a recursive algorithm scaling as $n \chi^2$ \cite{Ferris:2012,METTS} \footnote{the $n \chi^2$ scaling for MPS sampling assumes the MPS has been brought into an ``orthogonal form'' which is typically the case for MPS computed by circuit evolution algorithms. Otherwise an MPS can be brought into this form at a cost of $n \chi^3$}. 
This sampling procedure does not use a Markov chain, so it produces ``perfect'', independent samples of the state.

The above algorithms reflect only some of the capabilities of tensor networks, which are an actively developing field \cite{Orus_2019}. 
Other approaches not used in this work include two-dimensional tensor networks capable of capturing much more entanglement than MPS \cite{Evenbly2011tensor}, 
algorithms for applying multiple operators simultaneously with higher fidelity than sequential application \cite{Ayral}, 
and methods for removing inessential entanglement (thus reducing computational costs) to allow physical simulations
to reach very long times \cite{White_DMT, Rakovszky}.

\subsection{Solving the problem in a single call to the oracle}

\begin{figure}[t]
    \centering
    \includegraphics[width=\columnwidth]{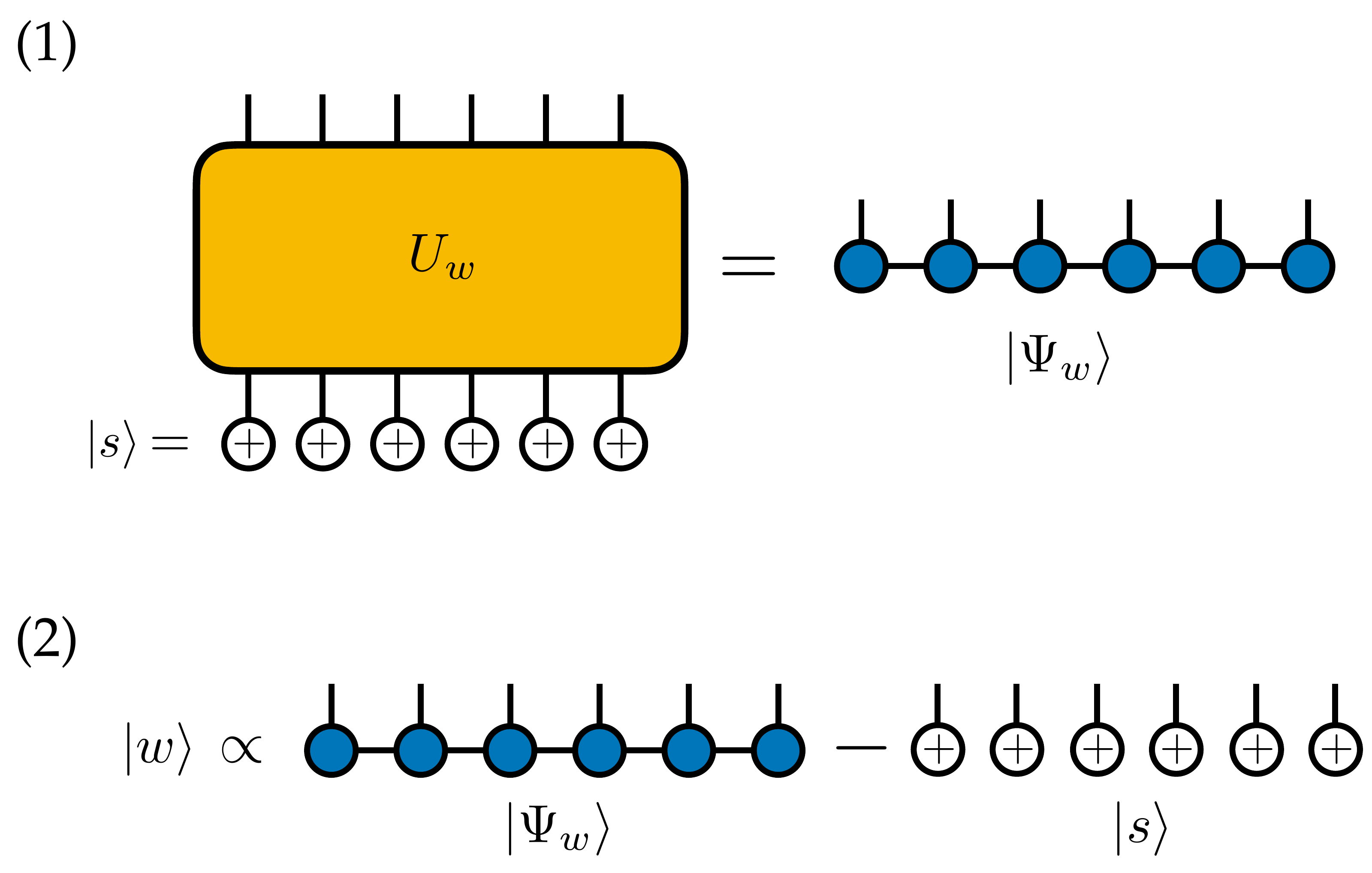}
    \caption{Steps of the quantum-inspired Grover's algorithm (QiGA) for the case of an \emph{open} simulation 
      of the oracle $U_w$ and the case of a single solution $\ket{w}$. 
     In step (1), the post-oracle state $\ket{\Psi_w}$ is computed by acting the oracle operator $U_w$ onto the superposition state $\ket{s}$,
     using tensor network techniques.  
     In step (2), the state $\ket{s}$ is subtracted from $\ket{\Psi_w}$. The result is proportional to the target state. \label{fig:qiga_open}
    }
\end{figure}

The first step of our \QIGA algorithm involves constructing the MPS representation
of the state \mbox{$\ket{\Psi_w} = U_w \ket{s}$} by applying the oracle (i.e.\ simulating the oracle) exactly once.
There are systematic techniques to perform this application using MPO-MPS products since all gates present in quantum circuits have a known low rank MPO representation. The classical cost for doing so can be anywhere from $O(n)$ to $O(2^n)$ depending on the entanglement generated by the particular oracle.

Then from the definition  of the oracle Eq.~(\ref{eq:multi_oracle}) and the fact that
$\bracket{w^\alpha}{s}=\frac{1}{\sqrt{2^n}}$, we can express the post-oracle state $\ket{\Psi_w}$ as
\begin{align}
\ket{\Psi_w} = \ket{s} - \frac{2}{\sqrt{2^n}} \sum_{\alpha=1}^{S} \ket{w^\alpha} \label{eq:sum_of_products} \ .
\end{align}
This expression is explicitly the sum of $1+S$ product states, thus $\ket{\Psi_w} = U_w \ket{s}$ is exactly an MPS of bond dimension $\chi=1+S$.

The form of $\ket{\Psi_w}$ as a sum of product states in Eq.~(\ref{eq:sum_of_products}) immediately presents a classical strategy to extract the solution states $\{\ket{w^\alpha}\}$ in a single step: one simply subtracts the state $\ket{s}$. Subtracting a product state such as $\ket{s}$ from an MPS is a straightforward operation with a cost scaling as $n \chi^3 \propto \log{N}$.
For example, in the case of $n=100$ qubits and $\chi=50$ the subtraction takes about 0.2 seconds on a personal computer.

It is important to note that this subtraction operation has no quantum equivalent. This can be seen easily with an
argument analogous to the no-cloning theorem: if there existed a unitary matrix that maps $\ket{\Psi}\otimes \ket{s}$
to $(\ket{\Psi}-\ket{s}) \otimes \ket{\Phi}$ for all $\ket{\Psi}$, then this matrix would map  $\ket{s}\otimes \ket{s}$
to the null vector which contradicts the assumption that the matrix is unitary. It follows that our algorithm
cannot be used as a ``classical preconditioner'' for amplitude amplification \cite{Brassard}. See the associated discussion in
section~\ref{sec:advantage_theo}.

In summary the different steps of \QIGA are:
\begin{enumerate}
    \item Use classical simulation techniques to compute $\ket{\Psi_w} = U_w \ket{s}$ as an MPS of rank $\chi=1+S$
    \item Compute $\ket{\tilde{W}} = \ket{\Psi_w}-\ket{s}$ and normalize it to obtain $\ket{W}$, an MPS of rank $\chi=S$
    \item Sample from $\ket{W}$ to obtain the states $\ket{w^\alpha}$ with uniform probability, using the fact that perfect sampling of MPS can be performed with a cost $n \chi^2$ \cite{Ferris:2012,METTS}
\end{enumerate}
If $S$ is small enough, the states $\ket{w^\alpha}$ can also be fully enumerated.
The above steps of \QIGA are depicted in Fig.~\ref{fig:qiga_open} for the case of a single solution $\ket{w}$.

One can modify the classical approach described above not only to sample individual solutions $w^\alpha$ but even to
count the number of solutions. To do so, one acts with $U_w$ on the \emph{unnormalized} state $\sum_b \ket{b}$. Then
the squared norm of the resulting state gives the number of solutions.

\section{Illustration with an explicit construction for the 3-SAT problem \label{sec:3SAT}}

To make our our quantum inspired GA (QiGA) more concrete, we will illustrate it with a practical application, which is a simulation of the
oracle for the 3-SAT boolean satisfiability problem. 3-SAT is an NP-complete problem and finding fast, possibly heuristic, algorithms for solving it is the subject of active research, with applications including cryptanalysis \cite{Massacci:2000,Mironov:2006}, industrial operations research  \cite{ortools},  and computational biology \cite{corblin2007sat}. 

In a SAT problem, the function $f(b)$ is given by a set of $p$ clauses that must all be satisfied for $b$
to be a solution. In 3-SAT, each clause $\delta$ is constructed out of 3 variables 
$b_{i_\delta}$, $b_{j_\delta}$ and $b_{k_\delta}$ from the binary representation of $b$. $f(b)$ takes the form,
\begin{align}
f(b) = (\tilde b_{i_1} \lor \tilde b_{j_1} \lor \tilde  b_{k_1}) \land (\tilde b_{i_2} \lor \tilde b_{j_2} \lor \tilde b_{k_2})
\land \ldots \nonumber \\
 \ldots \land (\tilde b_{i_p} \lor \tilde b_{j_p} \lor  \tilde b_{k_p})
\end{align}
where  $\lor$ means logical ``or'',  $\land$ logical ``and'', and $\tilde b_a = b_a$ or $\tilde b_a = \lnot b_a$ (not $b_a$) depending on the clause. 

The problems we consider are defined by choosing the ratio $\alpha=p/n$ of clauses to variables (or qubits) to be fixed, usually between $4 < \alpha < 5$ since in this range the number of satisfying solutions $S$ becomes small. 
Otherwise the choice of which variables 
enter into each clause and whether a variable is negated is made with uniform probability.  
Below we will consider totally random SAT clauses in section \ref{sec:randomSAT} then clauses with quasi-one-dimensional 
structure in section \ref{sec:1dSAT}.

\subsection{Tensor Network SAT Approach \label{sec:sat_approach}}

To explicitly implement the Grover's oracle operator for 3-SAT and construct the post-oracle state $\ket{\Psi_w}$, first prepare the state of
the system to be
\begin{align}
\ket{+}_1 \ket{+}_2 \ket{+}_3 \cdots \ket{+}_n \ket{1}_A  = \ket{s} \ket{1}_A
\end{align}
where the extra qubit in the $\ket{1}_A$ state acts as an ancilla whose role is to record which states of the previous $n$ qubits
either satisfy ($\ket{1}_A$) or do not satisfy ($\ket{0}_A$) all of the SAT clauses applied so far.

Next, each SAT clause $C$ such as $C=(b_3 \lor \lnot b_{7} \lor b_{8})$ is mapped to an operator by noting
that there is only one bitstring which \emph{fails to satisfy} the clause. In the example above, this bitstring is $0_3, 1_7, 0_8$. 
Using this bitstring, one defines an operator
\begin{align}
O_C & = P^0_{3} \otimes P^1_{7} \otimes P^0_{8} \otimes P^0_{A} + (1 - P^0_{3} \otimes P^1_{7} \otimes P^0_{8}) \otimes 1_A \ .
\label{eq:sat_operator}
\end{align}
which projects the ancilla qubit into the $\ket{0}_A$ state for any state containing the unsatisfying bitstring. 
Otherwise it leaves the ancilla unchanged.
Here $P^0_i = \ket{0}\bra{0}$ and $P^1_i = \ket{1}\bra{1}$ are projectors onto the $\ket{0}$ or $\ket{1}$ states for qubit $i$.

\begin{figure}[t]
    \centering
    \includegraphics[width=\columnwidth]{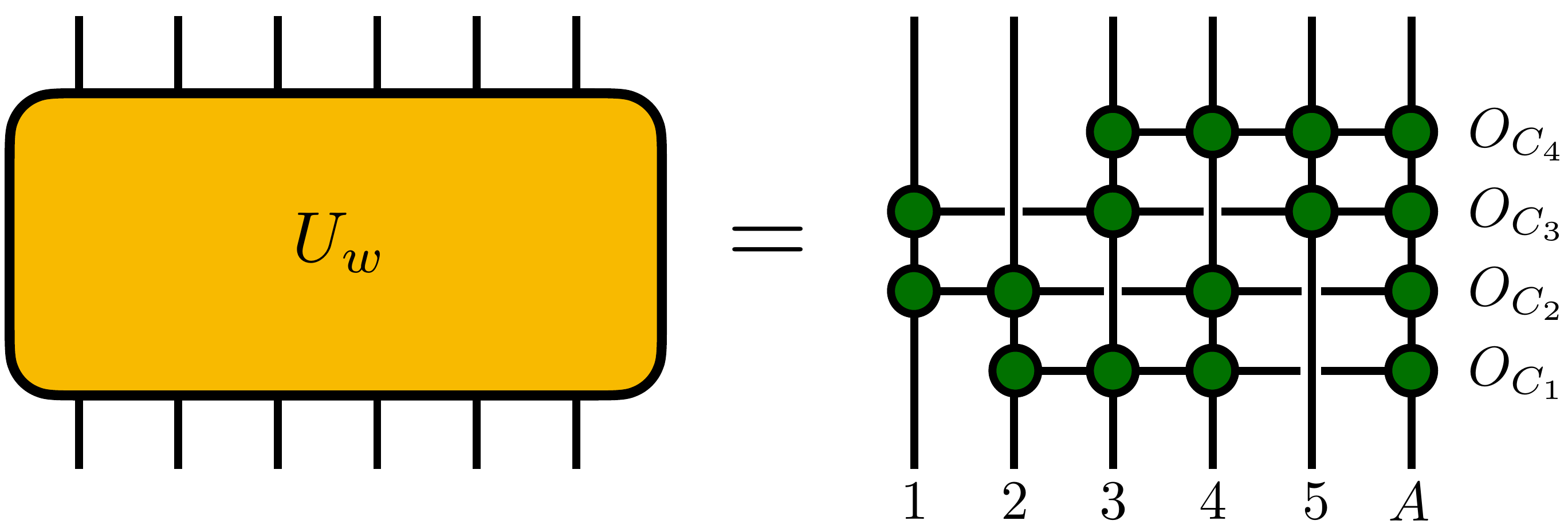}
    \caption{An oracle operator $U_w$ for a random 3-SAT instance expressed as a tensor network, 
    specifically a product of MPOs. Each MPO represents
    an operator $O_{C_j}$ of the type described in Eq.~(\ref{eq:sat_operator}) which acts non-trivially on three of the input qubits as well as the
    ancilla (last) qubit.
     \label{fig:sat_oracle}
    }
\end{figure}

In our classical implementation, the operator $O_C$ can be applied in its above form using straightforward tensor network methods.
We used the approach of implementing each $O_C$ as an MPO and applying these MPOs to the quantum state represented as an MPS. 
As an MPO, $O_C$ has a rank $\chi=3$, which can be understood from the fact that when one expands all the terms it is the sum of three product operators \cite{McCulloch}. Figure~\ref{fig:sat_oracle} shows an example 3-SAT oracle written as a product of such MPOs.

After applying the $O_C$ operators for every SAT clause, the state of the system becomes 
\begin{align}
\frac{1}{\sqrt{2^{n}}} \sum_{\alpha=1}^S \ket{w^\alpha} \ket{1}_A + \frac{1}{\sqrt{2^{n}}}  \sum_{\beta=1}^{2^n-S} \ket{\tilde{w}^\beta} \ket{0}_A
\label{eq:sat_mps}
\end{align}
where the $\{w^\alpha\}$ are the satisfying bitstrings and the $\{\tilde{w}^\beta\}$ are the unsatisfying ones. To convert this state
to a conventional Grover's post-oracle state Eq.~(\ref{eq:sum_of_products}), one can perform a post-selected, or forced, measurement of 
the ancilla qubit to be in the $\ket{-} = H\ket{1}$ state.  (Note that for a tensor network such a forced measurement always succeeds on the first try
and can be done by just acting with a single-qubit projection operator.) 
After dropping the now unentangled ancilla qubit, the
state will take the form of Eq.~(\ref{eq:sum_of_products}). If one is only interested in solving the Grover problem rather than constructing the post-oracle state, one simply projects the ancilla of Eq.\eqref{eq:sat_mps} onto the state $\ket{1}_A$. 

\subsection{Random 3-SAT Experiments (Moderately Hard Instance)\label{sec:randomSAT}}

We have tested this classical oracle implementation on fully random SAT clauses (variables $b_{i_p}$ chosen by drawing $i_p$ randomly from $1,2,..,n$ and with each variable negated with probability $1/2$) for up to $n=40$ variables or qubits, using the
ITensor software \cite{itensor,itensor-r0.3} running on a single workstation with four Xeon 3.6 GHz processors and 256 Gb of RAM.
For all experiments we used a ratio of clauses to variables of $p/n = 4.2$. This value is intentionally chosen close to the SAT-UNSAT transition ($p_c/n=4.267$) so that the number of satisfying $S$ assignments is typically very few, making these instances as challenging as possible for both classical and quantum algorithms \cite{Braunstein_Survey}.
The results shown in Table~\ref{sat_results} are for various experiments over a range of $n$ and different random instances for the same $n$ with different
numbers $S$ of satisfying bitstrings. We report both the maximum MPS rank $\chi_\text{max}$, which was the largest rank 
encountered during application of the $O_C$ operators, and the total time required to apply all of the operators and construct the post-oracle state.

\begin{table}
\begin{tabular}{| c c l l |}
\hline
n & S & $\chi_\text{max}$ & time  \\
\hline
30  &  4  &  467   & 21 s   \\
32   &  2  &  954  &  1.8 minutes  \\
34   &  48 &  1162 &  3.2 minutes \\
36   & 16  & 1994 &  8.3 minutes  \\
38   &  8  &  5867 &  1.6 hours \\
40   &  0  &  1402 &  4.2 minutes\\
40   &  28 &  2926  & 21 minutes  \\
40   &  161 & 5690 &  1.65 hours \\
40   &  174  & 10374  & 6.5 hours \\
\hline
\end{tabular}
\caption{Maximum MPS ranks $\chi_\text{max}$ and execution times to compute the post-oracle state corresponding to random 3-SAT problem
instances involving $n$ variables ($n$ qubits). The table also shows the number of satisfying assignments or bitstrings $S$ for each problem instance.
\label{sat_results}
}
\end{table}

After each post-oracle state was prepared, its quality was checked by projecting (post-selecting) the ancilla qubit into the state $\ket{1}_A$ then sampling 5000 bitstrings from the other $n$ qubits to verify that all samples satisfied the SAT clauses.
To count the number $S$ of satisfying bitstrings (\#SAT problem) we applied the MPOs $O_C$ to an \emph{unnormalized} state with each qubit (except the ancilla) initialized to $(\ket{0}+\ket{1})$. Afterward, we projected the ancilla into the $\ket{1}_A$ state and computed the norm of the resulting state which is equal to $S$. For smaller systems these counts were verified to be correct by checking with exact enumeration. The resulting counts $S$ are shown in the second column of Table~\ref{sat_results}, 

These results indicate the post-Grover's-oracle state can be prepared classically for typical 3-SAT instances for at least $n=40$ qubits 
on a single workstation. For problems of this size, the optimal number of iterations of Grover's algorithm 
would be \mbox{$r=823,500$} in contrast to the single application of the oracle used here.
The largest MPS rank encountered across the experiments we performed was $\chi=10,374$ which is a small fraction ($1\%$) of the 
maximum possible rank $2^{40/2}\approx 10^6$ (and $0.01\%$ in terms of
the maximum memory footprint). 
The entanglement barrier in the random 3-SAT problem is not only relatively modest for finite problem sizes, but was typically observed to be lower for the case of fewer satisfying solutions $S$. 
Hence \QIGA appears to perform better on problems with few solutions. 

It is important to note that the approach above can be straightforwardly parallelized through a ``slicing'' approach, similar to other recent high-performance quantum circuit simulations \cite{chen2018classical, Gray2021hyper, Pan:2022}. Instead of initializing all the $n$ input qubits to the $\ket{+}$ state, a subset of $p$ of these qubits can be initialized to either the $\ket{0}$ or $\ket{1}$ states. By running $2^p$ separate calculations for each setting of these computational-basis qubits one can obtain the same final result by combining the solutions obtained from $2^p$ computers working with no communication overhead. When trying this in practice, we found that the total computational effort (sum of running times of each of the parallel computers) was comparable to the original (serial) running time, while the maximum time for any one computer was significantly less than the original time.
Because the oracle acts classically on the computational basis qubits, the maximum entanglement during the oracle application is generally much lower for each sliced input so that the parallel approach  results in a faster time to solution.

Note that our implementation of a SAT solver as a Grover oracle operator is nevertheless far slower than the most efficient classical SAT solving 
algorithms, some of which also happen to use tensor networks \cite{Kourtis2019} and can solve
typical $n=40$ instances of 3-SAT in under one second. The appeal of using \QIGA lies
in the fact that it is not specific. It exists for {\it all} Grover problems.

\subsection{Quasi-One-Dimensional SAT Experiments (Easy Instance)\label{sec:1dSAT}}

In this section, we design instances of the 3-SAT problem where the QiGA approach has \emph{provably linear scaling} in the number of 
qubits $n$, that is a logarithmic scaling with problem size $\log(N)$. 
The goal of the construction
and associated experiments we perform below is to illustrate two points. First, it shows that there {\it are} classes of problems for which QiGA is always exponentially faster than GA. Second, the underlying structure that makes the problem easier for QiGA {\it needs not} be known a priori: QiGA discovers this structure and takes advantage of it automatically.

We consider a quasi-1D case that involves grouping variables into blocks of size $B$ along a 1D path, with the first block $(1,2,\ldots,B)$, second block $(B+1,\ldots, 2B)$, etc. The SAT problem is then defined by two sets of SAT clauses required to be satisfied altogether:
\begin{enumerate}
\item $N_\text{intra}$ fully random SAT clauses where variables in each clause only act within each block
\item $L_\text{inter}$ layers of random SAT clauses where variables span across two neighboring blocks
\end{enumerate}
The cases we consider will allow $N_\text{intra}$ to be any size while $L_\text{inter}$ is fixed to a small value such as $L_\text{inter}= 1,2,3$.

\begin{figure}[t]
    \centering
    \includegraphics[width=\columnwidth]{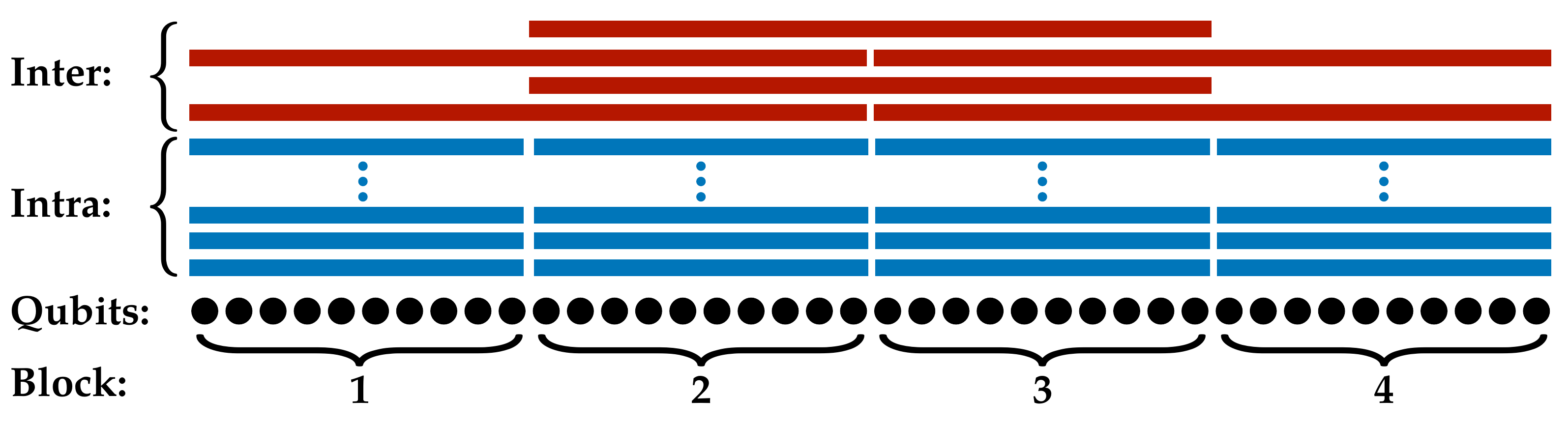}
    \caption{Schematic of the process of applying the operators (MPOs) $O_{C_p}$ enforcing the SAT clauses defining the quasi-1D SAT problem.
     The qubits are organized into blocks of size $B$ and a number of random, intra-block SAT clauses are enforced which act only inside each block. 
     Then a fixed number of layers of inter-block clauses are enforced which act between neighboring blocks.}  \label{fig:1dSAT}
\end{figure}

The proof of linear scaling consists of bounding the cost of applying the constraints of clauses in sets (1) and (2) above separately. 
We will use a similar approach as in Section~\ref{sec:sat_approach} above, with the slight modification that we will 
project out any unsatisfying bitstrings for a clause $C_p =(b_{i_p} \lor b_{j_p} \lor b_{k_p})$ by acting with an operator
\begin{align}
O_{C_p} = (1- P^{b_{i_p}}_{i_p} P^{b_{j_p}}_{j_p} P^{b_{k_p}}_{k_p})
\end{align}
that sets the state containing the single bitstring not satisfying $C_p$ to zero. There is no ancilla qubit in this approach.
The process of applying the MPOs $O_{C_p}$ is depicted in Fig.~\ref{fig:1dSAT} and explained in detail below.

Starting from the product superposition state
\begin{align}
\ket{s} = \ket{+}_1 \ket{+}_2 \ket{+}_3 \cdots \ket{+}_n 
\end{align}
and applying the operators for the set (1) of intra-block clauses, the cost scales as $(N_\text{intra}\ n \ 2^{ 3 B/2})$ in the worst case.
To see why, note that the state after applying the operators will be a product of MPS for each block, and the maximum bond dimension 
of an MPS with $B$ sites is $\chi = 2^{B/2}$..
Algorithms for applying operators to MPS of this size scale as $B \chi^{3} = B\, 2^{ 3 B/2}$. One has to apply each of the operators $O_{C_p}$ and there are $N_\text{intra}$ of these. Finally there are $n/B$ blocks. Multiplying each of these costs gives the above scaling. Thus the cost of enforcing the set (1) clauses scales only linearly with number of qubits $n$.

Next one enforces the $L_\text{inter}$ layers inter-block clauses in set (2). For each layer of such clauses, one can group the clauses into two sets, one acting across blocks 1 and 2, then 3 and 4, then 5 and 6, etc. and the next set acting across blocks 2 and 3, 4 and 5, 6 and 7, etc. The cost of applying the first layer scales at most as $n\ 2^{3 B/2}$ and doubles the bond dimension between blocks in the worst case, since the bond dimension of the $O_{C_p}$ MPOs is $2$. The cost of applying the next layer will then scale at most as $(n\ 8\ 2^{3 B/2})$, the extra factor of $8=2^3$ coming from the doubling of bond dimension due to the first layer. In general the cost of enforcing the $(2 L_\text{inter})$ layers of inter-block clauses will be 
$(n\ 2^{2 L_\text{inter}-1}\ 2^{3 B/2})$. 

Therefore the overall cost of enforcing all of the 1D 3-SAT clauses scales as
\begin{align}
(n\ N_\text{intra}\ \ 2^{ 3 B/2}) + (n\ 2^{2 L_\text{inter}-1} 2^{3 B/2})
\end{align}
which is manifestly linear in $n$, assuming $B$ and $L_\text{inter}$ are held constant, and that $N_\text{intra}$ depends only on $B$ (is chosen proportional to $B$).

\begin{table}
\begin{tabular}{| c c l l |}
\hline
n & S & $\chi_\text{max}$ & time  \\
\hline
40   &  99  &  21 &  0.973s\\
40   &  50  &  22 &  0.973s\\
40   &  108  &  16 &  0.989s\\
40   &  0 &  19 &  0.985s\\
60   &  4530 & 22 & 2.00s \\
60   &  0 & 19 & 1.98s \\
60   &  17920 & 19 & 1.96s \\
\hline
\end{tabular}
\caption{Maximum MPS ranks $\chi_\text{max}$ and execution times to compute the post-oracle state corresponding to blocked 1D 3-SAT problem
instances involving $n$ variables ($n$ qubits). In all cases the block size was chosen as $B=10$ and $N_\text{intra}=37$ random clauses were applied within each block while $N_\text{inter}=2$ layers of clauses were applied between neighboring blocks.
The table also shows the number of satisfying assignments or bitstrings $S$ for each problem instance.
\label{1d_sat_results}
}
\end{table}

In practice, when implementing the above method on test systems of $n=40$ qubits, using a fixed block size of $B=10$, taking $L_\text{inter}=2$ and choosing \mbox{$N_\text{intra} = 3.7 \cdot B$}, we find that all the clauses can be enforced in a running time of under 1 second. The maximum MPS bond dimension  observed is $\chi=22$. Systems with $n=60$ qubits are just as easy for the same block size and number of clauses per block, with similar maximum bond dimension and running times just under 2 seconds. See Table~\ref{1d_sat_results} for detailed results.

\subsection{Discussion of potentially hard instances\label{sec:hashing}}

Because arbitrary classical boolean logic can be encoded into quantum circuits, one cannot expect the \QIGA algorithm or
its variants to solve arbitrary inverse problems (finding $b$ such that $f(b)=1$) without $2^n$ operations in the worst case. For example, oracle quantum circuits which compute satisfying inputs for the \mbox{circuitSAT} family of problems are likely to require classical effort scaling as $2^n$ for worst-case instances \cite{Aaronson_QCLec18, Paturi_Complexity}. 
A more familiar example of a problem also likely to require close to $2^n$ effort for \QIGA would be inverting a 
cryptographic hashing function such as SHA-256, which can be used for secure digital transactions. Indeed, previous work using tensor networks
to simulate entire runs of Grover's algorithm (not using QiGA) to invert simplified versions of such hashing functions found that the entanglement during the oracle step grew to nearly its maximum value \cite{Ramos}. 

A discussion of the application of Grover's algorithm in the context of cryptography can be found in \cite{Bernstein,Gheorghiu} where it is argued that the limited quadratic speed-up of Grover is unlikely to impact security. In the context of large keys where massive parallelism would be needed, GA suffers from the fact that the quadratic speed-up applies only to the serial part of the calculation: if $K$ quantum computers are available, the time to solution is only reduced to $\sqrt{2^n/K}$ (independent parallel GA searches on the different quantum computers with $\log_2 (K)$ frozen bits) \cite{Gheorghiu}.
One should also keep in mind that, in the context of crypto-security, it is not sufficient to beat classical algorithms for GA to be useful. Even if, in a remote future, GA on a cluster of quantum computers could break a $128$ bits key while a classical computer could only break a $64$ bits key (this situation is unlikely as argued below), the algorithm would still not be useful in practice as one could switch to $256$ bits keys (something already implemented in some post-quantum cryptography proposals).

QiGA or its variants (such as the QiGA-2 variant discussed next) would not scale any better than brute-force searching or random guessing for
the kinds of worst-case instances described above. These are the kinds of problems for which it is reasonable to begin estimating the resources 
needed to solve the problem using classical versus quantum devices. 
However, we note that these problems form a small fraction of the all the problems
for which Grover could be applied. Moreover, as we will describe in detail in Section~\ref{sec:advantage},
the practical crossover point for which classical worst-case strategies exhaust modern resources is around $n=80$ variables or qubits.
The corresponding time scales and error rates required for quantum devices, even under extremely optimistic assumptions about hardware
quality, turn out to be very unfavorable to practically running Grover's algorithm.

\section{An alternative \QIGAbis algorithm for closed simulations}
\label{sec:quiga2}

In this section we propose an alternative classical algorithm to solve the Grover problem, \QIGAbis.  This algorithm does not necessarily require MPO/MPS tensor network technology so this section can be read separately and without previous knowledge of these techniques. In contrast to the previous QiGA, here we do not compute the full state (in MPS form) after a call to the oracle but rather compute individual amplitudes  
 $\bra{b} U_w \ket{s}$ for certain bitstrings $\ket{b}$.
The solution is found in $O(n)$ of such amplitude calculations. Such individual 
amplitude calculations are known as ``closed simulations'' and are easier than
the open simulations that provide the full state $\ket{\Psi_w}$. In particular there are efficient graph analysis heuristics to find the best contraction strategy \cite{Pan:2022}. We refer to \cite{Ayral} for a recent review. \QIGAbis contains some common features with a previously proposed algorithm that used a probabilistic (not quantum) formulation of MPS to solve the Grover's search problem \cite{Chamon}.

\subsection{Finding the solutions iteratively using closed circuit simulations}

\begin{figure}[t]
    \centering
    \includegraphics[width=0.8\columnwidth]{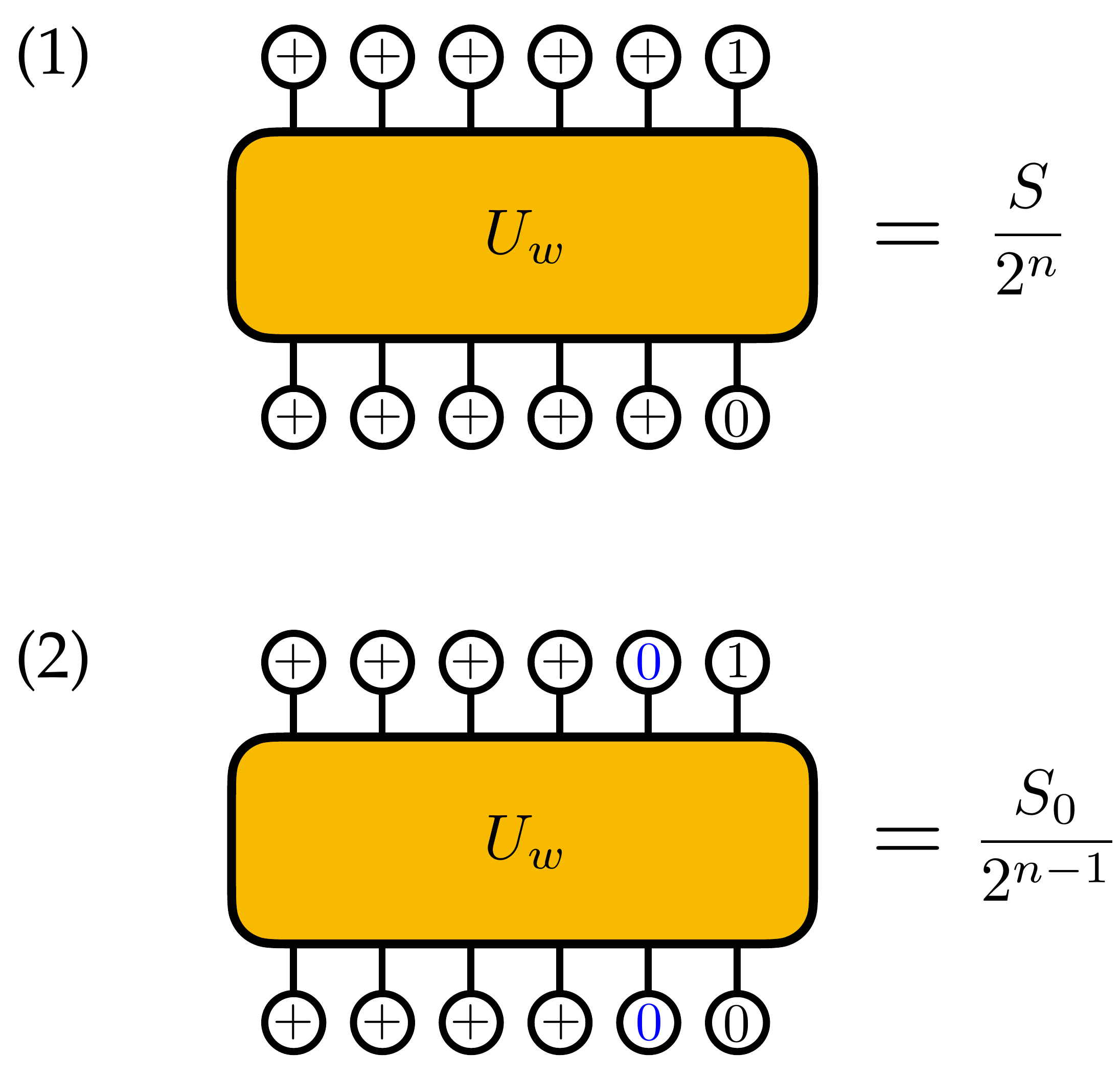}
    \caption{First two steps of the second quantum-inspired Grover's algorithm (QiGA-2) involving \emph{closed} simulations 
      of the oracle $U_w$. Each step involves computing only a single amplitude.
     The amplitude in step (1) counts the number of solutions up to a normalization factor (which can be avoided by using unnormalized states). 
     The amplitude in step (2) counts the number of solutions whose last bit is $0$. If this number is zero, then the last bit must be $1$. Proceeding
     through $n$ such steps gives one of the solutions, revealing it one bit at a time. \label{fig:qiga_closed}
    }
\end{figure}

We start by switching to a slightly different definition of the oracle $\tilde U_w$ that is more convenient for the present discussion. $\tilde U_w$ is fully equivalent to the previous one $U_w$ and one can switch from one to the other with simple circuits. Introducing an extra ancilla qubit, we define the oracle with a circuit that implements $\tilde U_w \ket{b}\ket{0} = \ket{b}\ket{f(b)}$ 
which leads to
\begin{equation}
\tilde U_w \ket{b}\ket{0} = 
\left\{
\begin{array}{l}
\ket{b}\ket{0}  \ {\rm if }\ b\ne w^\alpha  \\
\ket{b}\ket{1}  \ {\rm if }\ b = w^\alpha 
\end{array}
\right.
\end{equation}
We will be looking for the settings of the qubits $b_{n-1} \ldots b_2 b_1 b_0$ that correspond to one of the targets $b=w^\alpha$.

The first step is to calculate the amplitude
$\bra{+++...+}\bra{1} \tilde U_w \ket{+++...+}\ket{0}$. It is easy to see that
this amplitude simply counts the number of solutions $S$
\begin{equation}
\bra{+++...+}\bra{1} \tilde U_w \ket{+++...+}\ket{0}
= \frac{S}{2^n}
\end{equation}
Typically this is the most expensive step, with remaining steps contributing only a sub-leading cost.
In a practical calculation the normalization can be omitted to avoid dealing with exponentially small numbers by replacing $\ket{+}$ with
the un-normalized sum $\ket{0}+\ket{1}$.

In the second step, one changes the last input ($+$) qubit to the state $\ket{0}$ and
computes $\bra{+...++0}\bra{1} \tilde U_w \ket{+...++0}\ket{0}$. This amplitude is simply given by 
\begin{equation}
\bra{+...++0}\bra{1} \tilde U_w \ket{+...++0}\ket{0}
= \frac{S_0}{2^{n-1}}
\end{equation}
where $S_0$ is the number of solutions where the last bit is zero. We note $S_1$ the number of solutions where the last bit is one.
If $S_0 > 0$, we set $b_0=0$ otherwise we set $b_0=1$. Since $S_0+S_1=S$ we are guaranteed that one of the two choices will provide a solution if $S>0$. 

We continue iteratively
by setting the second-from-last bit $b_1$ to zero and computing
\begin{equation}
\bra{+...+0b_0}\bra{1} \tilde U_w \ket{+...+0b_0}\ket{0}
= \frac{S_{0b_0}}{2^{n-2}}
\end{equation}
where $S_{b_1b_0}$ is the number of solutions ending with the bits $b_1 b_0$. Again, If $S_{0b_0}>0$, we set $b_1=0$ otherwise we set $b_1=1$ knowing that $S_{0b_0}+S_{1b_0}= S_{b_0}>0$. This process is continued until all the bits have been calculated
and we have obtained a solution $w=b_{n-1}...b_2b_1b_0$. What is crucial here, and what a quantum computer cannot do, is to 
set the value of the ancilla, that is force the measurement to select the solution.
The \QIGAbis algorithm is totally agnostic to the method used (e.g. \cite{Crosswhite,McCulloch,Pirvu:2010,Zaletel} ) to perform the amplitude calculations.

\subsection{\QIGAbis for almost-Clifford quantum circuits}
\label{sec:clifford}

The computational cost for using \QIGA is directly linked to the level
of entanglement present in the application of a single oracle: the entanglement barrier. In the case of QiGA-2, it depends. If one uses tensor network contraction techniques such as \cite{Pan:2022}, then QiGA-2 faces similar difficulties as QiGA. However, there exist simulation techniques for amplitudes that can handle highly entangled states. A well-known example arises from the techniques that were derived from the Gottesman-Knill theorem \cite{Nielsen} for quantum circuits
that are {\it almost} uniquely composed of Clifford gates \cite{Aaronson:2004}.

Such a circuit is made uniquely of three Clifford gates, control-not $C_X$, Hadamard $H$ and phase $S$ gates, plus a small number of non-Clifford gates such as the $T$ gate or the Toffoli gate. These circuits are interesting in part because they arise naturally in fault tolerant quantum computing. In a nutshell, the idea of these simulations is to describe a ``stabilizer'' state $\ket{\Psi}$ of the quantum computer in terms of $n$ stabilizers $g_i$, each of them being a product of Pauli matrices (e.g.\ $g_1=X_1 Z_5 Z_6 Y_9$). The stabilizers (i) commute with each other $[g_i,g_j]=0$ and (ii)  $\ket{\Psi}$  is an eigenstate of $g_i$ with eigenvalue $\eta_i=\pm 1$. It is easy to see that the $g_i$ entirely characterize a state up to a global phase, that the initial state $\ket{0...0}$ is characterized by $g_i=Z_i$, $\eta_i=1$. Applying Clifford gates to a stabilizer state provides another stabilizer state with simple rules to update the $g_i$ and $\eta_i$. For instance applying $H_i$ amounts to replacing all the $Z_i$ by $X_i$ (and vice versa) in all the stabilizers. Such a simulation requires
only $O(n^2)$ bits of storage, $O(n)$ operations per gate, and $O(n^2)$ operations per measurement.

To account for the presence of $n_T$ non-Clifford gates, say $T$ gates, one simply decomposes the latter as a sum of Clifford gates $T_i = \cos(\pi/8) I_i +i\sin(\pi/8) Z_i$ and keeps track of the {\it at most} $2^{n_T}$ generated stabilizer states (actually only the $\eta_i$ must be kept for a gate that is decomposed onto Pauli matrices) so that the computing time scales as $\sim 2^{n_T}$ \cite{Garcia_2015}. A first version of this type of algorithm scaled as $4^{n_T}$
\cite{Aaronson:2004}. It was improved to 
$2^{n_T}$ \cite{Garcia_2015} and then to $2^{0.94 n_T}$ \cite{Bravyi_2016b}. 
There are also Monte-Carlo variations of these algorithms with scaling as low as $\sqrt{2^{n_T}}$ \cite{Bravyi_2016}.

Going back to the Grover problem, the \QIGAbis algorithm can be combined with stabilizer simulations. The computing time on a quantum computer will scale at least as $n_T \sqrt{2^{n}}$ while the computing time on the classical computer
scales at most as $n^2 2^{n_T}$ \cite{Garcia_2015} or even as $n^2 \sqrt{2^{n_T}}$ according to \cite{Bravyi_2016}. \QIGAbis combined with these techniques therefore provides another necessary condition that a quantum circuit must possess: one finds that a possible asymptotic quantum advantage requires $n_T$ to grow faster than $n$. The possibility of a practical advantage will be discussed at the end of this article. 
Interestingly, in this context the exponential scaling of \QIGAbis  appears when one ties up the scaling of $n_T$ with $n$. This connection strongly depends on the structure of the oracle. This observation demonstrates another example of the difficulty of establishing a quantum speed-up with Grover's algorithm.

\subsection{A connection between the closed circuit algorithm and the MPS algorithm using tensor cross interpolation}
Here, we make a connection between \QIGA and \QIGAbis
and show how the explicit MPS form of $\ket{\Psi_w} = U_w \ket{s}$ (we are back to the original definition of the ancilla) can be constructed from $O(n)$ amplitude calculations of the type $\bracket{b}{\Psi_w}$.

There has been recent progress in general algorithms able to 
construct a MPS representation of the state $\ket{\Psi_w}$ in $O(\chi^2 n)$ calls to a 
routine that calculates $\bracket{b}{\Psi_w}$.
Here, we make use of tensor cross interpolation \cite{Oseledets:2010,Savostyanov:2011,Savostyanov:2014,Dolgov:2020,Nunez-Fernandez} 
following the implementation described in \cite{Nunez-Fernandez}. The advantage of tensor cross interpolation is that it is
agnostic to the details of the quantum circuit $U_w$ and only requires an external  classical subroutine that computes
$\bracket{b}{\Psi_w}$. The algorithm requests the value of $\bracket{b}{\Psi_w}$
for certain values of $b$ using an active learning algorithm. It follows that it is directly compatible with the most advanced methods that have been
developed to calculate amplitudes of quantum circuits (including those that leverage on the high level of 
parallelism available in supercomputers).

Before we can use tensor cross interpolation effectively, a small adjustment must be performed. A direct calculation
of $\bracket{b}{\Psi_w}$ provides
\begin{align}
\label{eq:amplitude1}
\bracket{b}{\Psi_w} = \frac{1}{2^{n/2}} \left[ 1 - 2 \sum_{\alpha=1}^S \delta_{b,w^\alpha} \right] \ \ .
\end{align}
For a random bitstring $b$, one has $\bracket{b}{\Psi_w} = 1/\sqrt{2^{n}}$ since it is exponentially unlikely that
$b$ matches one of the solutions $w^\alpha$. It follows that the tensor cross interpolation algorithm would fail
to reconstruct the MPS as its exploration of the $\bracket{b}{\Psi_w}$ function would have an exponentially low probability of finding the relevant 
part (second half of the right hand side of Eq.\eqref{eq:amplitude1}).  Another way to see the same problem is to write $\bracket{b}{\Psi_w}$ in terms of the calls to the function $f(b)$. It reads
\begin{align}
\label{eq:amplitude2}
\bracket{b}{\Psi_w} = \frac{1}{2^{n/2}} (-1)^{f(b)}
\end{align}
i.e.\ the amplitudes can be calculated in a single call to $f(b)$. Hence, if $\ket{\Psi_w}$ MPS could be reconstructed
from $O(n)=O(\log N)$ calls to $\bracket{b}{\Psi_w}$, it would mean that the original problem could be solved in 
$O(n)$ calls to the function $f(b)$ hence that NP-complete problems could be solved in polynomial time,
an unlikely situation.

To solve this issue, we turn to the $\ket{\pm}$ basis and calculate $\bracket{\beta}{\Psi_w}$ instead of
$\bracket{b}{\Psi_w}$ where $\ket{\beta}$ is a product of $\ket{\pm}$ states (e.g.\ $\ket{+-+-...-+}$). 
Denoting the binary representation of $\beta$ as
$\beta_0,\beta_1...\beta_n$ with $\beta_i=0$ for a state $\ket{+}$
and $\beta_i=1$ for a state $\ket{-}$, we find
\begin{align}
\label{eq:amplitude3}
\bracket{\beta}{\Psi_w} =  \delta_{\beta,0} - 
\frac{2}{2^{n}} \sum_{\alpha=1}^S (-1)^{\sum_{i=0}^{n-1} \beta_i w_i^\alpha}
\end{align}
This form is directly suitable for tensor cross interpolation since information about the solutions $w^\alpha$
is now present for any bitstring $\beta$. We emphasize that \QIGA itself knows nothing about the solution $w^\alpha$
and only uses the amplitudes $\bracket{\beta}{\Psi_w}$. 
Calculating these amplitudes
$\bracket{\beta}{\Psi_w}$ has the same complexity as calculating $\bracket{b}{\Psi_w}$ since the two quantum circuits only differ by a layer of Hadamard gates at the end. Similarly, when the MPS is known in the $\ket{\beta}$ basis,
it is simply a matter of applying local Hadamard gates to get it back in the $\ket{b}$ basis. We have checked in explicit numerical calculations that our implementation of tensor cross interpolation can reconstruct the MPS in
$O(n)$ calls to the $\bracket{\beta}{\Psi_w}$ subroutine up to at least $n=1000$.

In terms of call the the function $f(b)$, the amplitudes $\bracket{\beta}{\Psi_w}$ take the form,
\begin{align}
\label{eq:amplitude4}
\bracket{\beta}{\Psi_w} =  
\frac{1}{2^{n}}\sum_{b=0}^{2^n-1} (-1)^{f(b) +\sum_{i=0}^{n-1} b_i \beta_i} 
\end{align}
which takes $O(2^n)$ calls to the classical function, if one does not take advantage of its quantum circuit form. 
Hence we recover the expected $O(N)$ classical scaling to solve the search problem if one does not use insights about the function $f$. 
When using the quantum circuit to compute the oracle amplitudes, the \QIGA complexity will depend on the entanglement barrier present in a single application of the quantum oracle as illustrated in Fig.\ref{fig:entropy}.

We  emphasize  that the approach outlined above is only feasible for a closed, classical simulation of the oracle circuit; it cannot be attempted on a quantum computer. 
Indeed, a quantum computer only provides bitstrings $\beta$ distributed according to the probability 
$|\bracket{\beta}{\Psi_w}|^2$ but it does not provide the actual value $\bracket{\beta}{\Psi_w}$ (nor can one choose the value of $\beta$).

\section{\label{sec:conclusions1} Conclusion of Part I and Introduction to Part II}

We have seen that for Grover's algorithm to be useful, it is not enough to  have a quantum computer, even an ideal one.
The advertised quadratic speedup  is only relative to a random guessing approach. In contrast, for problems where the oracle circuit encoding $f(b)$ has low entanglement or other structure such as low T-gate count, \QIGA can solve the problem on classical hardware with exponentially better scaling. In contrast to such low-difficulty problems, there are many medium-difficulty problems where \QIGA scales exponentially, but much better than $2^n$. Only for the hardest problems---most likely cryptographic in nature---could Grover's conceivably help.

Therefore in the remainder of this article, we narrow our focus to the task of running these hardest problems on quantum hardware and ask what resources would be needed.
This question has been tackled in part by others. References~\onlinecite{Babbush:2021,TroyerDisentangling} in particular consider the running times of algorithms offering quadratic speedups, such as Grover's. The conclusion is that  one would encounter prohibitively long running times of years or more, even for modest-sized problems, unless gate operation speeds were to improve by orders of magnitude beyond today's hardware. To this body of work, we discuss two additional  obstacles to running Grover's algorithm on actual hardware. First, not only must the gates operate at high speeds, but the error per step must be reduced exponentially with problem size. This puts additional demands on error-correcting codes. In the surface code, for example, it implies a quadratic growth of the code distance with problem size. Second, \QIGA provides lower bounds for the problem to be out of reach classically, in terms of T gate count or circuit depth. This provides us with more accurate estimates of the minimum problem sizes for which Grover could be competitive with classical approaches. Taken together, these different elements give us a general view of what it would take for Grover to thrive in practical use cases.

\section{\label{sec:scaling} Scaling of Errors in Grover's Algorithm}

Assuming one has taken on the task of running Grover's algorithm (GA) on a quantum computer, we now  
study the sensitivity of GA to the presence of imperfections such as gate errors
or decoherence. 
There has been some previous literature on this subject \cite{Koch,Regev,Long}.
The advantage of our tensor network approach is the capability to study relatively large system from which we can extract the asymptotic scaling of the errors in GA.

At the end of GA, the quantum computer is supposed to be  in the state $\ket{w}$ so that upon measuring the qubits one finds the solution $w$ almost certainly (up to negligible exponentially small $1/N$ corrections). However, due to imperfection in the gates or environmental coupling and decoherence 
the system will be in some other state
 instead, described by the density matrix $\rho$. The probability to get the correct  output $w$ when one performs the measurement of the qubits is $F = \bra{w}\rho\ket{w}$. $F$ is known as the {\it fidelity} of the calculation which 
 matches the success probability of the algorithm for GA. 
Generically, the fidelity decays exponentially with the number of gates applied in the calculation as well as with any idle time (decoherence). This exponential behavior has been observed ubiquitously in experiments and was established on a large scale in the seminal ``quantum supremacy'' experiment of Google \cite{Arute2019}. 
Denoting by $\epsilon$ the typical error per gate and $N_g$ the total number of gates used,
we therefore expect the decay rate to be $F \approx e^{-\epsilon N_g}$ with $N_g = r (N_o + N_d)$
where $N_o$ (resp. $N_d$) is the number of gates in the quantum circuit of the oracle (resp. diffusion operator) and $r \equiv  \frac{\pi}{4} 2^{n/2}$ is the optimum number of calls to the oracle.
In other words, for a noisy quantum computer the success probability of GA is expected to follow a
rather unfavorable {\it double exponential} decay with $n$,
\begin{align}
\label{eq:doubleexponential}
F \approx \exp\big[-\frac{\pi\epsilon}{4} (\sqrt{2})^n\, (N_o + N_d)\big].
\end{align}
  
\begin{figure}[t]
    \centering
    \includegraphics[width=0.95\columnwidth]{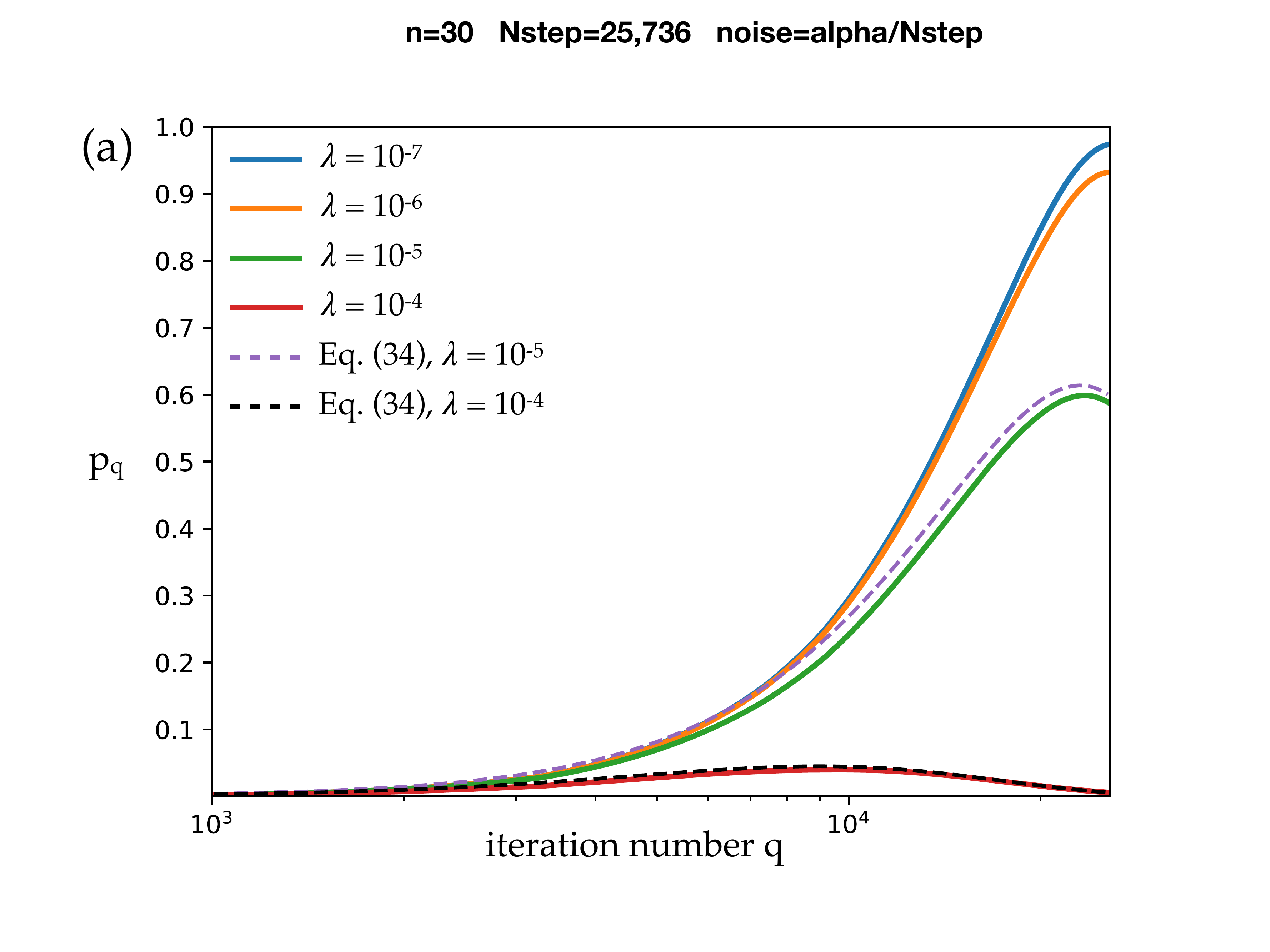}
    \includegraphics[width=0.95\columnwidth]{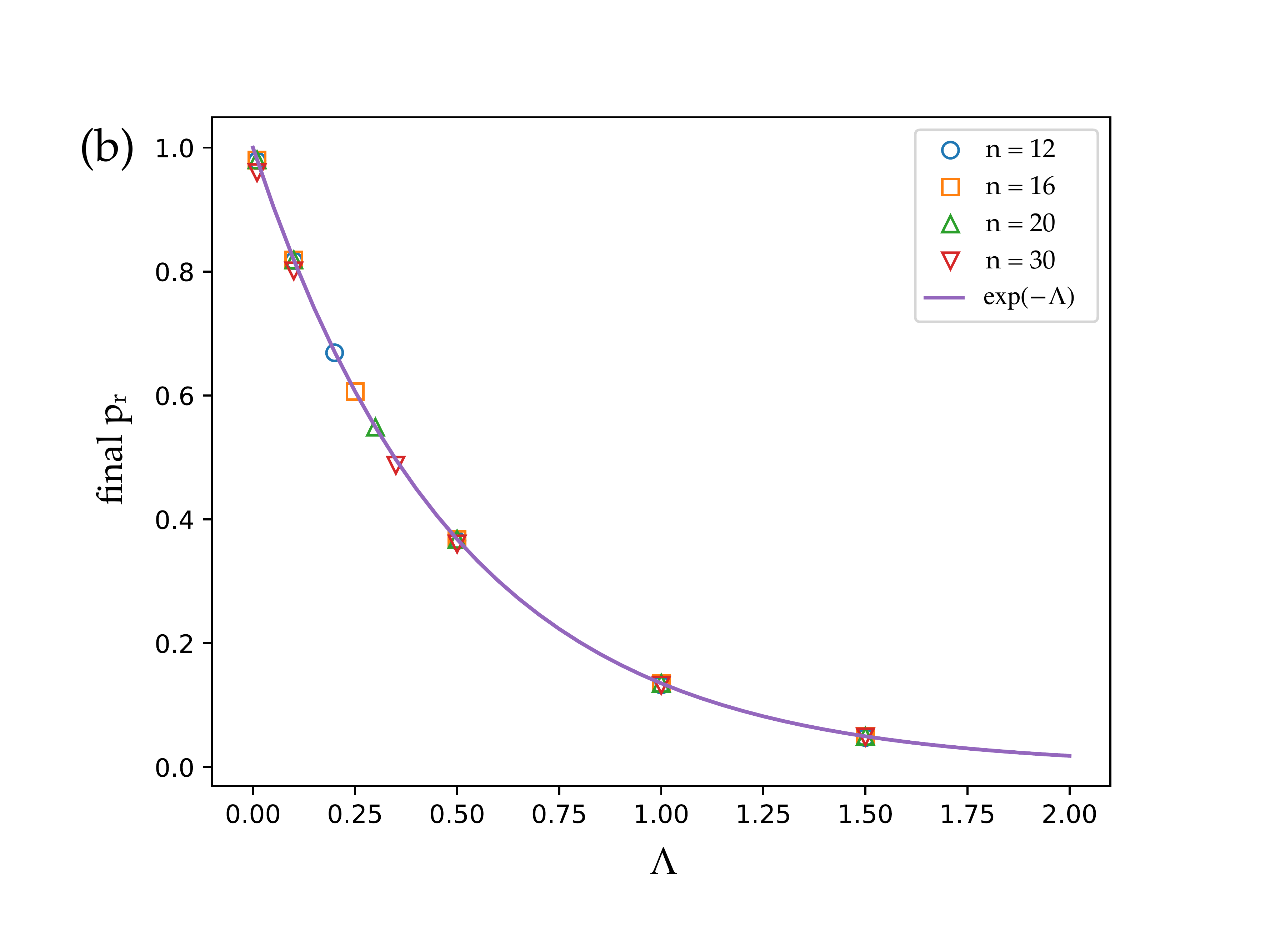}
    \caption{(a) Probability of success versus number of iterations of Grover's algorithm for $n=30$ qubits and different levels of noise $\lambda$.
    Also shown are theoretical fits to the cases $\lambda=10^{-5}, 10^{-4}$.
    For $n=30$ the theoretically optimal number of iterations is $r=25,736$. The final success probability 
    reaches a reasonable value for noise $\lambda < 5\times 10^{-5}$, but once the noise becomes as large as $\lambda = 10^{-4}$,
    the success probability reaches a maximum of only 0.04 at the 10,000th iteration then falls to about 0.006 by the 
    final iteration.
    (b) Final probability of success after the optimal number of iterations $r$ as a function of total noise $\Lambda=\lambda r$ where $\lambda$
    is the amount of depolarizing noise per iteration. }  \label{fig:noise_plot}
\end{figure}

To substantiate the scaling of GA with noise and number of qubits, we have performed classical simulations 
using an MPO state representation capable of representing mixed states. 
For these simulations we also implement the oracle and diffusion operators as MPO tensor
networks as described in Section~\ref{sec:grover_mpo}. As discussed earlier, without noise the MPS bond dimension $\chi$ of the state in between
the oracle and diffusion steps (so at every step of these simulations) is $\chi=2$, leading to an MPO-mixed-state representation of only $\chi=4$ 
so that each step can be performed rather quickly and it is possible to run the whole algorithm up to about $n=30$ qubits in under an hour. 
Adding noise to the simulations only modifies the bond dimension of the state very slightly and observed bond dimensions always remain less
than $\chi \lesssim 10$. 

To model the effects of noise, we apply a depolarizing noise channel $\Delta_\lambda(\rho) = (1-\lambda) \rho + \frac{\lambda}{2^n} I$ 
in between Grover iterations, that is once per iteration. 
By not including any noise during the application of the oracle or the diffuser our simulations 
are being very generous in favor of the success of GA. The noise per iteration $\lambda$ relates simply to the noise per gate $\lambda = \epsilon (N_o + N_d)$.

The results of our simulations for $n=30$ qubits and various levels of noise $\lambda$ are shown in Fig.~\ref{fig:noise_plot}(a). 
As long as the noise $\lambda \lesssim 5\times 10^{-5}$ the final probability of success reaches a reasonable value after
the optimal number of iterations, which for $n=30$ is $r = 25,736$. 
However, the probability of success after $r$ iterations falls below $1\%$ once the noise becomes larger than about $10^{-4}$. 
Note that non-zero noise leads to a maximum in the success probability at an earlier iteration $q^* < r$. We analyze the height and location 
of this maximum further below. 

Due to the transparent nature of GA and the depolarizing noise model, one can actually work out an exact result for the state 
after $q$ steps of the algorithm. By a recursive procedure, one finds:
\begin{align}
\rho_q = (1-\lambda)^{q} \ket{\Psi_q} \bra{\Psi_q} +  (1-(1-\lambda)^{q}) I \frac{1}{2^n}
\end{align}
where $\ket{\Psi_q}$ is the ideal pure state after $q$ noiseless Grover iterations. 
Using a well-known result for the probability of success of the noiseless GA after $q$ iterations \cite{Zhang}
\begin{align}
|\bracket{w}{\Psi_q}|^2 & = \sin^2((2q+1) \theta) \\
\theta & = \arcsin\Big(\frac{1}{\sqrt{2^n}}\Big)
\end{align}
it follows that the probability of success after iteration $q$ is given by
\begin{align}
p_q =  \bra{w} \rho_q \ket{w} = (1-\lambda)^{q} \sin^2((2q+1) \theta) \ \ .
\end{align}
Since $q\gg 1$, ignoring exponentially small corrections, we have:
\begin{align}
\label{eq:opt_prob}
p_q =  e^{-\lambda q} \sin^2(2q \theta) \ \ .
\end{align}
We show in Fig.~\ref{fig:noise_plot} that this theoretical fit works well, though there is a slight disagreement which 
we attribute not to the fit, but to a loss of precision in the early iterations of the numerical simulations
due to the use of double-precision floating point numbers and the very small signal of GA through the early 
iterations \cite{SaiToh}.

Interestingly, the form of $p_q$ Eq.~(\ref{eq:opt_prob}) means that if one defines the noise level in terms of
a parameter $\Lambda$ such that $\lambda = \Lambda/r$ then the final success probability is
\begin{align}
p_r = e^{-\Lambda}
\end{align}
regardless of the number of qubits, using the fact that $\sin^2(2r\theta)\approx 1$ up to exponentially small corrections. 
One can interpret $\Lambda=\lambda \cdot r$ as the total accumulated noise throughout a complete run of GA.
In Fig.~\ref{fig:noise_plot} we show how the 
final success probability $p_r$ observed  in noisy simulations fits very well to $e^{-\Lambda}$.

In the presence of large noise, due to a maximum that develops in the fidelity curve, it is advantageous for GA to stop the iteration
at a certain smaller value of $q=q^*$. An explicit calculation of the optimum of Eq.\eqref{eq:opt_prob}
provides
\begin{align}
\tan[2(r-q^*)\theta] = \frac{\Lambda}{\pi}
\end{align}
from which we arrive at the optimum success probability,
\begin{align}
\label{eq:opt_prob2}
p_{\rm success} =  \frac{e^{\frac{2\Lambda}{\pi}\arctan(\frac{\Lambda}{\pi})}}{1+(\Lambda/\pi)^2} 
e^{-\Lambda}. 
\end{align}
This formula behaves as $p_{\rm success} \approx e^{-\Lambda}$ for small $\Lambda$ (approaching 1.0 as $\Lambda \rightarrow 0$) 
and behaves as $p_{\rm success} \approx e^{-2}\cdot(\pi/\Lambda)^2$ for large $\Lambda$. 

Because $p_{\rm success}$ depends only on $\Lambda$,
an important conclusion is that for the GA success probability to scale successfully to large values of $n$, the total noise
$\Lambda$ must be held independent of $n$. The noise per iteration $\lambda$ must therefore scale as 
\begin{align}
\lambda = \frac{\Lambda}{r} \propto \frac{\Lambda}{(\sqrt{2})^n}
\end{align}
showing the noise per iteration (proportional to the noise per gate) must be reduced exponentially with $n$. 

\section{\label{sec:advantage} On the possibility of quantum advantage in Grover's algorithm}

There are three kinds of quantum advantage. 
The first is the theoretical kind, namely the possibility that in an idealized world
a perfect quantum computer could perform parametrically better than a classical one for some task. 
The second is the practical kind, namely the possibility that an actual device provides a useful result faster than a classical machine. Once again, in this article we assume that the quantum computer and the classical computer share the same information, namely the quantum circuit used to program the oracle of the former. 
The third kind of speed-up, the abstract theoretical speed-up measured only by the number of calls to the (respectively quantum and classical) oracles has already been discussed in Section~\ref{sec:point_of_view} and Fig.~\ref{fig:QigaContext}. In such a purely abstract context the quantum computer is always faster, but it is a context not relevant to our assumptions and we shall not come back to this point.

With respect to the first kind of quantum advantage our \QIGA merely reverses the charge of the proof.
We have shown that \QIGA and \QIGAbis automatically provide a classical algorithm that may (and in many cases will) be asymptotically faster than Grover.  In other words, there is no theoretical quantum advantage unless proven otherwise and quantum advantage has to be decided in a case-by-case manner. This considerably reduces the range of {\it theoretical} potential applications of Grover algorithm.

With respect to the second kind of quantum advantage involving an actual machine solving a practical problem of a fixed size, 
the existence of \QIGA and the demands for implementing GA 
place drastic bounds on the hardware needed which we will argue are simply out of reach.
When discussing hardware, there is long list of specifications that we could consider including heat management, control electronics, classical data bandwidth for e.g.\ syndrome analysis, device variability, power consumption, etc. 
In the sections on practical quantum advantage below, we will limit ourselves to discussing only following aspects: the total qubit count, the error budget per gate, and the time to solution.

\subsection{Absence of generic theoretical quantum advantage}
\label{sec:advantage_theo}

\QIGA implies that if the quantum circuit for the oracle can be simulated classically, then at most
$O(n)$ such calculations are sufficient to solve the problem while a quantum computer 
needs $O(2^{n/2})$ calls to the oracle. An immediate consequence is that no theoretical quantum advantage can be
claimed generically, that is irrespective of the nature of the underlying quantum circuit for the oracle. 
 This is an important point to make with respect to the large literature which assumes, implicitly or explicitly, the existence of a quantum speed-up every time GA replaces its classical counterpart \cite{Montanaro:2016}.

If the complexity for calculating one amplitude of the oracle is smaller than $(\sqrt{2})^n$, then \QIGA (and/or QiGA-2) is parametrically faster than its quantum counterpart. 
There are clear cases where classical algorithms win. For instance, if the oracle can be simulated with a fixed depth using local gates in 1D (respectively 2D for many systems \cite{Napp:2022,Tindall:2024}), then the problem can be solved in linear time using MPS or PEPS technology while GA would require exponential time. 
The quasi-1D SAT (Section~\ref{sec:1dSAT}) is another example.
Conversely, constructing an oracle whose classical simulation is provably harder than $(\sqrt{2})^n$ can most likely be done. Indeed, in the large depth limit, classical simulations of quantum circuits have a generic complexity of $2^n$. We surmise that for some specific tasks such as inverting a hashing function, the classical complexity might come close to the $2^n$ limit. However, \emph{proving} this classical complexity is a difficult task and we are not
aware of such a construction for an actual oracle (i.e.\ a circuit whose output amplitudes are only $\pm 1$).

A corollary of the existence of \QIGA is that the quantum circuit associated to the oracle of an NP complete problem 
must be exponentially difficult to simulate in the general case, i.e.\ presents an exponentially high entanglement barrier. Indeed, otherwise, one could simulate it in polynomial time which would prove $P=NP$ a statement widely believed to be untrue. 
Hence \QIGA provides a direct connection between classical complexity theory and the entanglement level of a quantum circuit.

Lastly, we would like to discuss the relation of this work to amplitude amplification \cite{Brassard}, a direct generalization of GA. In some cases, there exist  classical heuristic algorithms that can solve NP-hard problems faster than GA, though still scaling exponentially.
For instance, there exist very fast classical algorithms for the 3-SAT problem which we considered earlier in our  \QIGA  demonstrations
(incidentally, among the best are tensor network approaches \cite{Kourtis2019}). 
Amplitude amplification is claimed to recover the quadratic Grover speedup over such fast classical algorithms 
by combining these algorithms with a slight modification of GA. 
Below, we show that \QIGA applies in this context as well.
We again argue that the question of whether the oracle can be simulated classically is crucial.

Following the amplitude amplification framework of Ref.~\onlinecite{Brassard},
a classical heuristic algorithm takes the form of a function $b=G(r)$ that proposes
a bitstring $b$ as a possible solution. Here the variable $r$ captures the probabilistic nature of the heuristic, e.g.\ it can be the seed of the pseudo random number generator used in the heuristic. In a good heuristic, $r$ spans much fewer values than $b$. One example of a classical heuristic is Sch\"oning's algorithm \cite{Schoning} which solves 3-SAT problems with a complexity $(4/3)^n$ by an iterative process over bitstrings where at each stage an unsatisfied clause is chosen randomly and then a bit participating in the clause is flipped in order for the clause to become satisfied. 
To transform this heuristic into a quantum algorithm, amplitude amplification does not search for the bitstring $b$ that satisfies $f(b)=1$ but instead uses GA to search for the value of $r$ that satisfies
$f(G(r))=1$ (see the discussion around Theorem 6 of  \cite{Brassard}). Thus one needs to use GA with the oracle $U_w \ket{r} = (-1)^{f(G(r))} \ket{r}$. 
In the simplest setting of the amplitude amplification algorithm, the search space of $r$ requires a smaller number $n' < n$ qubits than the search space of inputs to $f(b)$ requiring $n$ qubits. Let $\kappa = n'/n < 1$. The number of iterations needed for GA in the new space scales as 
\begin{align}
2^{n'/2} = 2^{\kappa \, n/2} = (2^\kappa)^{n/2} = \sqrt{a^{n}}
\end{align}
where $a = 2^\kappa < 2$. This is to be contrasted with $2^{n'} = a^n$ iterations for the worst-case classical approach for finding a good input $r$. This result is often considered
as evidence that amplitude amplification leads to a quadratic speedup over the classical heuristic.
However, despite this relative speedup, the broader point is that \emph{amplitude amplification is just GA in a smaller search space}. It follows that our \QIGA approach applies directly to the amplitude amplification of these heuristics: one only needs to modify the oracle in the same way it would be modified for the quantum computer. Therefore one returns to the question of whether one can simulate the oracle circuit in this new space (which is smaller, hence easier to simulate) or if not, whether GA for a space this size can be run in practice on a quantum computer.

Anticipating the discussion of the next section, the existence of fast heuristics makes classical problems computable up to very large values of $n$. For instance, 3-SAT problems with $n\sim 10,000$ are well within range \cite{GomesSatisfiability}. It follows that the entrance level for GA for problems where good classical heuristics are available are much higher than for the problems for which no heuristic is available, as the circuits to compute the oracle $f(G(r))$ will require at least
$O(n)$ qubits (including ancillas) even if the seed space for $r$ is much smaller.  As we shall see in the estimates below, this would translate into inaccessible number of required qubits and astronomically large times to solution.

\subsection{Absence of practical quantum advantage on a noisy quantum computer}

We now turn to the question of a practical advantage and provide some estimates about the specifications
a quantum hardware must fullfill to solve a task better than what one can do classically. We start by estimating the total gate count $N_g$. The diffusion operator typically requires $2n$ Toffoli gates and the oracle at least the same
(a simpler oracle would likely be rather easy to solve classically). Each Toffoli gate must be decomposed into 15 gates (including $6$ Control-NOT and $7$ T gates). We arrive at a total gate count for GA of $N_g \ge 60\, n\, 2^{n/2}$ assuming perfect connectivity (i.e.\ that the two-qubit gates can be applied between any pairs of qubits). In order for the final success probability to be of order unity (here we choose $p_{\rm success}=1/2$) we need $\Lambda \approx 0.8$
which translates into $\epsilon \le 1/(60\, n\, 2^{n/2})$.

It follows that in order to apply GA on just $5$ qubits with $p_{\rm success}\ge 1/2$, one needs $\epsilon \le 5.10^{-4}$ which is better than any existing hardware. In fact, in a recent state-of-the-art experimental implementation of GA on $5$ qubits, sophisticated error mitigation techniques were needed to reach a success rate of 15\%, illustrating the inherent difficulty of implementing GA~\cite{Pokharel}. 
Indeed, the experimental value of the error per gate $\epsilon$ has been mostly stable in the last ten years, typically around $\epsilon\approx 0.01$ for state of the art superconducting qubits \cite{Arute2019} and around $\epsilon\approx 0.001$ for  trapped ion systems, which are however several orders of magnitude slower. Previous applications of GA for a few qubits used a much smaller gate count in order to retain a large enough fidelity. Or alternatively, they used examples where one uses Eq.~\eqref{eq:oracle_action} instead
of Eq.~\eqref{eq:oracle_action_basic} i.e.\ one explicitly uses information about the solution $w$ instead of performing the classical logic of computing $f(b)$. While all this is perfectly acceptable for proof-of-principle experiments, it does not correspond to the full application of GA to actually solve a problem. 
Going to $n=40$, which we can solve on desktop-like computers using QiGA, leads to $\epsilon \le 4.10^{-10}$. Manipulating tens of qubits with precisions better than one part in a billion for a total of billions of gates is in our view completely unrealistic. 

Existing exascale supercomputers are capable of performing $\approx 70 \times 10^{21} \approx 70 \times 2^{70}$ operations in one day of computing time so that in a worst case scenario, any $n=70$ problem can be solved in a day using  brute force search.
Note that these problems do not need to leverage the large memory available on those machines, since the functions one is interested in inverting are efficient to compute on individual classical or bitstring inputs. 
Since any problem of interest has at least some structure, we estimate that $n=80$ problems are
accessible on supercomputers (most probably $n\ge 100$) using the best quantum circuit simulators available (see \cite{Ayral} for a recent review) to perform QiGA. Solving such a problem on a quantum computer would require $\epsilon \le 2.10^{-16}$
as well as a time to solution (assuming very fast $10$ns gates) of more than one year of uninterrupted running time. This estimate assumes a linear scaling with $n$ of a single  oracle call/computation. For other harder scalings the ratio between quantum and classical computing time remains unchanged. 

\subsection{Absence of practical quantum advantage on a fault tolerant quantum computer}
 
The problem of limited qubit fidelity can in principle be solved by quantum error correction which should allow
to lower the effective error level per gate $\epsilon_L$ by constructing robust logical qubits out of several physical qubits.
However, quantum error correction trades a better effective $\epsilon_L$ with a much higher qubit count $n_L$
as well as a much higher time to solution since logical qubits require many physical qubit operations in order to 
make a single logical one. Hence, we can already anticipate that quantum error correction is unlikely to help given the above already high estimate of the time to solution.

To make quantitative estimates, we focus on the example of the surface code, one of the most robust QEC code with a clear path to implementation \cite{Fowler}. The construction of such a fault-tolerant quantum computer is a formidable task that would require controlling millions or billions of qubits with accuracy much better than the states of the art. Here, we focus on a few specific difficulties. In particular, we ignore problems such as non-correctable errors \cite{Waintal2019}, syndrome analysis
(that would require petabits per second of classical data to be extracted from the quantum computer and analyzed in real time), and other significant engineering problems.

In the theory of the surface code, the error $\epsilon_{\rm L}$ per logical gate on a logical qubit  scales as
\begin{align}
\epsilon_{\rm L} \propto \epsilon_{\rm ph}
\left(
\frac{\epsilon_{\rm ph}}{\epsilon_{\rm th}}
\right)^{\sqrt{N_c}/2}
\end{align}
where $N_c$ is the number of physical qubit per logical qubit,
$\epsilon_{\rm ph}$ is the error per gate on the physical qubits, and
$\epsilon_{\rm th}$ is the threshold of the code, around $\epsilon_{\rm th}\approx 0.01$ for the surface code \cite{Fowler}. Ignoring logarithmic corrections, we immediately observe that the exponentially large running time which implies that one must have $\epsilon_{\rm L} (\sqrt{2})^n \le 1$ translates into 
\begin{align}
N_c \propto n^2.
\end{align}
i.e.\ the number of physical qubits per logical qubit increases quadratically with $n$
in sharp contrast to e.g.\ Shor's algorithm where $N_c \propto (\log n)^2$ has a weak logarithmic increase. The total ``space'' constraint therefore scales as $~n^3$ qubits,
which is far more drastic than the space estimate for e.g. chemistry calculations
which was already estimated to involve millions, or more realistically billions, of very high quality physical qubits to address $n=100$ \cite{Reiher}.

Let us now turn to a \emph{time} estimate.
In the surface code, the time scale is no longer limited by the time it takes to execute one gate but by the time to make a projective measurement of the so-called stabilizers. For instance in superconducting transmon qubit, the former can be as low as $10$ ns while the later takes around $1 \mu$s. 
Systems based on atomic physics such as cold ions are typically much slower. Also, 
the error per gate $\epsilon$ is likely to be limited by the measurement errors that are typically much worse than the error per gate. 
There are different techniques for implementing a quantum algorithm on a surface code,
with different trade-off between ``space'' and ``time''  (number and speed of qubits). The fastest technique (that will have an important overhead of ancilla qubits) uses gate teleportation \cite{Fowler_2013} and its run time is limited by the number $n_T$ of non-Clifford gates in the quantum circuit. For such a circuit \QIGAbis
discussed in Section \ref{sec:clifford} provides a lower bound on the number $n_T$ of non-Clifford gates that the oracle must contain in order for the simulation to be inaccessible classically. The computing time of \QIGAbis coupled to an ``almost Clifford'' quantum circuit simulator is $\sim 2^{n_T}$ or
$\sim \sqrt{2^{n_T}}$ with circuits up to $n_T=48$ already demonstrated in \cite{Bravyi_2016} on a laptop computer in a few hours. We can surmise that $n_T\approx 80$ or slightly larger should be accessible to a supercomputer almost irrespectively of $n$ in the regime of interest. On the quantum computer side, a resource estimate analysis of the computing time for the surface code has been done recently in \cite{Babbush:2021}.
It includes the latest developments in magic state distillation and lattice surgery techniques for handling surface code \cite{Gidney_2019}.
Using these rather optimistic estimates, one finds a running time for $n=80$ of hundreds of years (for a single solution to a single problem) and a space-time footprint of 
millions, in units of number of qubits times number of years of computation.
Going from $n=80$ to a $n=100$ problem---a size that could be out of reach of classical computers for {\it some} problems that remain to be identified---one needs to multiply the above computing time by a factor $>1000$. We need not to elaborate further, it should be clear at this stage that the fate of quantum error correction for the implementation of GA is extremely doubtful.

\section{Conclusion}

Grover's algorithm is an elegant intellectual construction. Unfortunately our analysis indicates that it will remain so for the foreseeable future.

We have constructed a quantum inspired version of Grover's algorithm that can solve quantum problem in exponentially fewer executions of the oracle circuit than Grover's algorithm, provided one is able to compute $\log(N)$ individual amplitudes of the oracle or simulate the oracle exactly once. 
We have also provided specific cases where this ``classical advantage'' can be realized.

Since our classical algorithm is fully general, it provides a clear benchmark against which one can evaluate the potential speed-up of Grover algorithm both theoretically and practically. While we cannot exclude a \emph{theoretical} quantum speed-up for \emph{every} problem, assuming an idealized quantum implementation, we estimate that practical quantum implementations will be associated with astronomically large computing times. On the other hand, problems for which a quantum implementation may seem necessary could have hidden structure revealed by our classical algorithm in the form of low entanglement barriers on the way to a solution. 

A work which has some overlap with ours is the pioneering proposal of Chamon and Mucciolo \cite{Chamon} to use MPS tensor networks as a method for solving inverse problems. In contrast to our approach, their work employs a classical probability formalism and considers mainly worst-case bounds for entanglement growth.

A separate line of work that casts some doubt on practical quadratic quantum speedups, of the type offered by Grover's algorithm, focuses primarily on the problem size needed for a quantum/classical scaling crossover \cite{Babbush:2021,TroyerDisentangling}. Our work
complements and adds to these important observations by demonstrating an additional difficulty in scaling down error rates with problem size,
and the consequences for fault-tolerant quantum computing.

Beyond the above rather negative results, our quantum inspired algorithm 
may also lead to more positive results.
For instance, while we have focused on exact calculations of the the quantum circuit amplitudes, 
an interesting possibility would be to solve inverse problems from {\it approximate} closed calculations of 
amplitudes using tensor network or other classical techniques. It is unclear if the MPS resulting from such an approach 
would provide an efficient heuristic for solving the Grover problem.

\begin{acknowledgments}

EMS acknowledges illuminating conversations with Mazen Ali, Jielun Chris Chen, Antoine Georges, Matthew Fishman, Andrew Millis, Roger Mong, Vadim Oganesyan, Nicola Pancotti, Olivier Parcollet, Dries Sels, Jesko Sirker, Simon Trebst, Norm Tubman, and Steve White. The Flatiron Institute is a division of the Simons Foundation. XW acknowledges funding from the French ANR QPEG and the Plan France 2030 ANR-22-PETQ-0007 ``EPIQ''.  We thank Craig Gidney for  constructive remarks on current resource estimates for the surface code.
\end{acknowledgments}

\nocite{*}

\bibliography{main}

\providecommand{\noopsort}[1]{}\providecommand{\singleletter}[1]{#1}%
\begin{thebibliography}{94}%
\makeatletter
\providecommand \@ifxundefined [1]{%
 \@ifx{#1\undefined}
}%
\providecommand \@ifnum [1]{%
 \ifnum #1\expandafter \@firstoftwo
 \else \expandafter \@secondoftwo
 \fi
}%
\providecommand \@ifx [1]{%
 \ifx #1\expandafter \@firstoftwo
 \else \expandafter \@secondoftwo
 \fi
}%
\providecommand \natexlab [1]{#1}%
\providecommand \enquote  [1]{``#1''}%
\providecommand \bibnamefont  [1]{#1}%
\providecommand \bibfnamefont [1]{#1}%
\providecommand \citenamefont [1]{#1}%
\providecommand \href@noop [0]{\@secondoftwo}%
\providecommand \href [0]{\begingroup \@sanitize@url \@href}%
\providecommand \@href[1]{\@@startlink{#1}\@@href}%
\providecommand \@@href[1]{\endgroup#1\@@endlink}%
\providecommand \@sanitize@url [0]{\catcode `\\12\catcode `\$12\catcode
  `\&12\catcode `\#12\catcode `\^12\catcode `\_12\catcode `\%12\relax}%
\providecommand \@@startlink[1]{}%
\providecommand \@@endlink[0]{}%
\providecommand \url  [0]{\begingroup\@sanitize@url \@url }%
\providecommand \@url [1]{\endgroup\@href {#1}{\urlprefix }}%
\providecommand \urlprefix  [0]{URL }%
\providecommand \Eprint [0]{\href }%
\providecommand \doibase [0]{https://doi.org/}%
\providecommand \selectlanguage [0]{\@gobble}%
\providecommand \bibinfo  [0]{\@secondoftwo}%
\providecommand \bibfield  [0]{\@secondoftwo}%
\providecommand \translation [1]{[#1]}%
\providecommand \BibitemOpen [0]{}%
\providecommand \bibitemStop [0]{}%
\providecommand \bibitemNoStop [0]{.\EOS\space}%
\providecommand \EOS [0]{\spacefactor3000\relax}%
\providecommand \BibitemShut  [1]{\csname bibitem#1\endcsname}%
\let\auto@bib@innerbib\@empty
\bibitem [{\citenamefont {Shor}(1994)}]{Shor:1994}%
  \BibitemOpen
  \bibfield  {author} {\bibinfo {author} {\bibfnamefont {P.}~\bibnamefont
  {Shor}},\ }\bibfield  {title} {\bibinfo {title} {Algorithms for quantum
  computation: discrete logarithms and factoring},\ }in\ \href
  {https://doi.org/10.1109/SFCS.1994.365700} {\emph {\bibinfo {booktitle}
  {Proceedings 35th Annual Symposium on Foundations of Computer Science}}}\
  (\bibinfo {year} {1994})\ pp.\ \bibinfo {pages} {124--134}\BibitemShut
  {NoStop}%
\bibitem [{\citenamefont {Beauregard}(2002)}]{Beauregard}%
  \BibitemOpen
  \bibfield  {author} {\bibinfo {author} {\bibfnamefont {S.}~\bibnamefont
  {Beauregard}},\ }\href {https://doi.org/10.48550/ARXIV.QUANT-PH/0205095}
  {\bibinfo {title} {Circuit for \mbox{Shor's} algorithm using 2n+3 qubits}}
  (\bibinfo {year} {2002})\BibitemShut {NoStop}%
\bibitem [{\citenamefont {Kitaev}(1995)}]{Kitaev}%
  \BibitemOpen
  \bibfield  {author} {\bibinfo {author} {\bibfnamefont {A.~Y.}\ \bibnamefont
  {Kitaev}},\ }\href {https://doi.org/10.48550/ARXIV.QUANT-PH/9511026}
  {\bibinfo {title} {Quantum measurements and the \mbox{Abelian} stabilizer
  problem}} (\bibinfo {year} {1995})\BibitemShut {NoStop}%
\bibitem [{\citenamefont {Harrow}\ \emph {et~al.}(2009)\citenamefont {Harrow},
  \citenamefont {Hassidim},\ and\ \citenamefont {Lloyd}}]{HHL}%
  \BibitemOpen
  \bibfield  {author} {\bibinfo {author} {\bibfnamefont {A.~W.}\ \bibnamefont
  {Harrow}}, \bibinfo {author} {\bibfnamefont {A.}~\bibnamefont {Hassidim}},\
  and\ \bibinfo {author} {\bibfnamefont {S.}~\bibnamefont {Lloyd}},\ }\bibfield
   {title} {\bibinfo {title} {Quantum algorithm for linear systems of
  equations},\ }\href {https://doi.org/10.1103/PhysRevLett.103.150502}
  {\bibfield  {journal} {\bibinfo  {journal} {Phys. Rev. Lett.}\ }\textbf
  {\bibinfo {volume} {103}},\ \bibinfo {pages} {150502} (\bibinfo {year}
  {2009})}\BibitemShut {NoStop}%
\bibitem [{\citenamefont {Grover}(1997)}]{Grover}%
  \BibitemOpen
  \bibfield  {author} {\bibinfo {author} {\bibfnamefont {L.~K.}\ \bibnamefont
  {Grover}},\ }\bibfield  {title} {\bibinfo {title} {Quantum mechanics helps in
  searching for a needle in a haystack},\ }\href
  {https://doi.org/10.1103/PhysRevLett.79.325} {\bibfield  {journal} {\bibinfo
  {journal} {Phys. Rev. Lett.}\ }\textbf {\bibinfo {volume} {79}},\ \bibinfo
  {pages} {325} (\bibinfo {year} {1997})}\BibitemShut {NoStop}%
\bibitem [{\citenamefont {Grover}(2001)}]{GroverReview}%
  \BibitemOpen
  \bibfield  {author} {\bibinfo {author} {\bibfnamefont {L.~K.}\ \bibnamefont
  {Grover}},\ }\bibfield  {title} {\bibinfo {title} {{From
  \mbox{Schr\"odinger's} equation to the quantum search algorithm}},\ }\href
  {https://doi.org/10.1007/s12043-001-0128-3} {\bibfield  {journal} {\bibinfo
  {journal} {Pramana - J. Phys.}\ }\textbf {\bibinfo {volume} {56}},\ \bibinfo
  {pages} {333} (\bibinfo {year} {2001})}\BibitemShut {NoStop}%
\bibitem [{\citenamefont {Brassard}\ \emph {et~al.}(2002)\citenamefont
  {Brassard}, \citenamefont {Hoyer}, \citenamefont {Mosca},\ and\ \citenamefont
  {Tapp}}]{Brassard}%
  \BibitemOpen
  \bibfield  {author} {\bibinfo {author} {\bibfnamefont {G.}~\bibnamefont
  {Brassard}}, \bibinfo {author} {\bibfnamefont {P.}~\bibnamefont {Hoyer}},
  \bibinfo {author} {\bibfnamefont {M.}~\bibnamefont {Mosca}},\ and\ \bibinfo
  {author} {\bibfnamefont {A.}~\bibnamefont {Tapp}},\ }\bibfield  {title}
  {\bibinfo {title} {Quantum amplitude amplification and estimation},\
  }\href@noop {} {\bibfield  {journal} {\bibinfo  {journal} {Contemporary
  Mathematics}\ }\textbf {\bibinfo {volume} {305}},\ \bibinfo {pages} {53}
  (\bibinfo {year} {2002})}\BibitemShut {NoStop}%
\bibitem [{\citenamefont {Bennett}\ \emph {et~al.}(1997)\citenamefont
  {Bennett}, \citenamefont {Bernstein}, \citenamefont {Brassard},\ and\
  \citenamefont {Vazirani}}]{Bennett}%
  \BibitemOpen
  \bibfield  {author} {\bibinfo {author} {\bibfnamefont {C.~H.}\ \bibnamefont
  {Bennett}}, \bibinfo {author} {\bibfnamefont {E.}~\bibnamefont {Bernstein}},
  \bibinfo {author} {\bibfnamefont {G.}~\bibnamefont {Brassard}},\ and\
  \bibinfo {author} {\bibfnamefont {U.}~\bibnamefont {Vazirani}},\ }\bibfield
  {title} {\bibinfo {title} {Strengths and weaknesses of quantum computing},\
  }\href {https://doi.org/10.1137/S0097539796300933} {\bibfield  {journal}
  {\bibinfo  {journal} {SIAM Journal on Computing}\ }\textbf {\bibinfo {volume}
  {26}},\ \bibinfo {pages} {1510} (\bibinfo {year} {1997})},\ \Eprint
  {https://arxiv.org/abs/https://doi.org/10.1137/S0097539796300933}
  {https://doi.org/10.1137/S0097539796300933} \BibitemShut {NoStop}%
\bibitem [{NP_()}]{NP_footnote}%
  \BibitemOpen
  \href@noop {} {\bibinfo {title} {Non-deterministic polynomial (\mbox{NP})
  time is a formal com- plexity class of decision (yes or no) problems defined
  by having solutions whose validity can be checked in poly- nomial time, while
  finding a solution might take exponential time for some problem instances.
  \mbox{NP}-complete problems are a subset of \mbox{NP} problems for which, if
  a general-purpose polynomial-time solving algorithm could be found, it could
  also be used to solve all other problems in \mbox{NP}.}}\BibitemShut {Stop}%
\bibitem [{\citenamefont {Baritompa}\ \emph {et~al.}(2005)\citenamefont
  {Baritompa}, \citenamefont {Bulger},\ and\ \citenamefont {Wood}}]{Baritompa}%
  \BibitemOpen
  \bibfield  {author} {\bibinfo {author} {\bibfnamefont {W.~P.}\ \bibnamefont
  {Baritompa}}, \bibinfo {author} {\bibfnamefont {D.~W.}\ \bibnamefont
  {Bulger}},\ and\ \bibinfo {author} {\bibfnamefont {G.~R.}\ \bibnamefont
  {Wood}},\ }\bibfield  {title} {\bibinfo {title} {\mbox{Grover's} quantum
  algorithm applied to global optimization},\ }\href
  {https://doi.org/10.1137/040605072} {\bibfield  {journal} {\bibinfo
  {journal} {SIAM Journal on Optimization}\ }\textbf {\bibinfo {volume} {15}},\
  \bibinfo {pages} {1170} (\bibinfo {year} {2005})}\BibitemShut {NoStop}%
\bibitem [{\citenamefont {Wei}\ \emph {et~al.}(2020)\citenamefont {Wei},
  \citenamefont {Naik}, \citenamefont {Harrow},\ and\ \citenamefont
  {Thaler}}]{Wei}%
  \BibitemOpen
  \bibfield  {author} {\bibinfo {author} {\bibfnamefont {A.~Y.}\ \bibnamefont
  {Wei}}, \bibinfo {author} {\bibfnamefont {P.}~\bibnamefont {Naik}}, \bibinfo
  {author} {\bibfnamefont {A.~W.}\ \bibnamefont {Harrow}},\ and\ \bibinfo
  {author} {\bibfnamefont {J.}~\bibnamefont {Thaler}},\ }\bibfield  {title}
  {\bibinfo {title} {Quantum algorithms for jet clustering},\ }\href
  {https://doi.org/10.1103/PhysRevD.101.094015} {\bibfield  {journal} {\bibinfo
   {journal} {Phys. Rev. D}\ }\textbf {\bibinfo {volume} {101}},\ \bibinfo
  {pages} {094015} (\bibinfo {year} {2020})}\BibitemShut {NoStop}%
\bibitem [{\citenamefont {D\"{u}rr}\ \emph {et~al.}(2006)\citenamefont
  {D\"{u}rr}, \citenamefont {Heiligman}, \citenamefont {Hoyer},\ and\
  \citenamefont {Mhalla}}]{Durr}%
  \BibitemOpen
  \bibfield  {author} {\bibinfo {author} {\bibfnamefont {C.}~\bibnamefont
  {D\"{u}rr}}, \bibinfo {author} {\bibfnamefont {M.}~\bibnamefont {Heiligman}},
  \bibinfo {author} {\bibfnamefont {P.}~\bibnamefont {Hoyer}},\ and\ \bibinfo
  {author} {\bibfnamefont {M.}~\bibnamefont {Mhalla}},\ }\bibfield  {title}
  {\bibinfo {title} {Quantum query complexity of some graph problems},\ }\href
  {https://doi.org/10.1137/050644719} {\bibfield  {journal} {\bibinfo
  {journal} {SIAM Journal on Computing}\ }\textbf {\bibinfo {volume} {35}},\
  \bibinfo {pages} {1310} (\bibinfo {year} {2006})}\BibitemShut {NoStop}%
\bibitem [{\citenamefont {Stamatopoulos}\ \emph {et~al.}(2020)\citenamefont
  {Stamatopoulos}, \citenamefont {Egger}, \citenamefont {Sun}, \citenamefont
  {Zoufal}, \citenamefont {Iten}, \citenamefont {Shen},\ and\ \citenamefont
  {Woerner}}]{Stamatopoulos2020}%
  \BibitemOpen
  \bibfield  {author} {\bibinfo {author} {\bibfnamefont {N.}~\bibnamefont
  {Stamatopoulos}}, \bibinfo {author} {\bibfnamefont {D.~J.}\ \bibnamefont
  {Egger}}, \bibinfo {author} {\bibfnamefont {Y.}~\bibnamefont {Sun}}, \bibinfo
  {author} {\bibfnamefont {C.}~\bibnamefont {Zoufal}}, \bibinfo {author}
  {\bibfnamefont {R.}~\bibnamefont {Iten}}, \bibinfo {author} {\bibfnamefont
  {N.}~\bibnamefont {Shen}},\ and\ \bibinfo {author} {\bibfnamefont
  {S.}~\bibnamefont {Woerner}},\ }\bibfield  {title} {\bibinfo {title} {Option
  {P}ricing using {Q}uantum {C}omputers},\ }\href
  {https://doi.org/10.22331/q-2020-07-06-291} {\bibfield  {journal} {\bibinfo
  {journal} {{Quantum}}\ }\textbf {\bibinfo {volume} {4}},\ \bibinfo {pages}
  {291} (\bibinfo {year} {2020})}\BibitemShut {NoStop}%
\bibitem [{\citenamefont {Ramesh}\ and\ \citenamefont {Vinay}(2003)}]{Ramesh}%
  \BibitemOpen
  \bibfield  {author} {\bibinfo {author} {\bibfnamefont {H.}~\bibnamefont
  {Ramesh}}\ and\ \bibinfo {author} {\bibfnamefont {V.}~\bibnamefont {Vinay}},\
  }\bibfield  {title} {\bibinfo {title} {String matching in \mbox{O(n+m)}
  quantum time},\ }\href
  {https://doi.org/https://doi.org/10.1016/S1570-8667(03)00010-8} {\bibfield
  {journal} {\bibinfo  {journal} {Journal of Discrete Algorithms}\ }\textbf
  {\bibinfo {volume} {1}},\ \bibinfo {pages} {103} (\bibinfo {year} {2003})},\
  \bibinfo {note} {combinatorial Algorithms}\BibitemShut {NoStop}%
\bibitem [{\citenamefont {A{\"\i}meur}\ \emph {et~al.}(2013)\citenamefont
  {A{\"\i}meur}, \citenamefont {Brassard},\ and\ \citenamefont
  {Gambs}}]{Aimeur}%
  \BibitemOpen
  \bibfield  {author} {\bibinfo {author} {\bibfnamefont {E.}~\bibnamefont
  {A{\"\i}meur}}, \bibinfo {author} {\bibfnamefont {G.}~\bibnamefont
  {Brassard}},\ and\ \bibinfo {author} {\bibfnamefont {S.}~\bibnamefont
  {Gambs}},\ }\bibfield  {title} {\bibinfo {title} {Quantum speed-up for
  unsupervised learning},\ }\href@noop {} {\bibfield  {journal} {\bibinfo
  {journal} {Machine Learning}\ }\textbf {\bibinfo {volume} {90}},\ \bibinfo
  {pages} {261} (\bibinfo {year} {2013})}\BibitemShut {NoStop}%
\bibitem [{\citenamefont {Kapoor}\ \emph {et~al.}(2016)\citenamefont {Kapoor},
  \citenamefont {Wiebe},\ and\ \citenamefont {Svore}}]{Kapoor}%
  \BibitemOpen
  \bibfield  {author} {\bibinfo {author} {\bibfnamefont {A.}~\bibnamefont
  {Kapoor}}, \bibinfo {author} {\bibfnamefont {N.}~\bibnamefont {Wiebe}},\ and\
  \bibinfo {author} {\bibfnamefont {K.}~\bibnamefont {Svore}},\ }\bibfield
  {title} {\bibinfo {title} {Quantum perceptron models},\ }in\ \href
  {https://proceedings.neurips.cc/paper/2016/file/d47268e9db2e9aa3827bba3afb7ff94a-Paper.pdf}
  {\emph {\bibinfo {booktitle} {Advances in Neural Information Processing
  Systems}}},\ Vol.~\bibinfo {volume} {29},\ \bibinfo {editor} {edited by\
  \bibinfo {editor} {\bibfnamefont {D.}~\bibnamefont {Lee}}, \bibinfo {editor}
  {\bibfnamefont {M.}~\bibnamefont {Sugiyama}}, \bibinfo {editor}
  {\bibfnamefont {U.}~\bibnamefont {Luxburg}}, \bibinfo {editor} {\bibfnamefont
  {I.}~\bibnamefont {Guyon}},\ and\ \bibinfo {editor} {\bibfnamefont
  {R.}~\bibnamefont {Garnett}}}\ (\bibinfo  {publisher} {Curran Associates,
  Inc.},\ \bibinfo {year} {2016})\BibitemShut {NoStop}%
\bibitem [{\citenamefont {Dong}\ \emph {et~al.}(2008)\citenamefont {Dong},
  \citenamefont {Chen}, \citenamefont {Li},\ and\ \citenamefont {Tarn}}]{Dong}%
  \BibitemOpen
  \bibfield  {author} {\bibinfo {author} {\bibfnamefont {D.}~\bibnamefont
  {Dong}}, \bibinfo {author} {\bibfnamefont {C.}~\bibnamefont {Chen}}, \bibinfo
  {author} {\bibfnamefont {H.}~\bibnamefont {Li}},\ and\ \bibinfo {author}
  {\bibfnamefont {T.-J.}\ \bibnamefont {Tarn}},\ }\bibfield  {title} {\bibinfo
  {title} {Quantum reinforcement learning},\ }\href
  {https://doi.org/10.1109/TSMCB.2008.925743} {\bibfield  {journal} {\bibinfo
  {journal} {IEEE Transactions on Systems, Man, and Cybernetics, Part B
  (Cybernetics)}\ }\textbf {\bibinfo {volume} {38}},\ \bibinfo {pages} {1207}
  (\bibinfo {year} {2008})}\BibitemShut {NoStop}%
\bibitem [{\citenamefont {Dewes}\ \emph {et~al.}(2012)\citenamefont {Dewes},
  \citenamefont {Lauro}, \citenamefont {Ong}, \citenamefont {Schmitt},
  \citenamefont {Milman}, \citenamefont {Bertet}, \citenamefont {Vion},\ and\
  \citenamefont {Esteve}}]{Dewes}%
  \BibitemOpen
  \bibfield  {author} {\bibinfo {author} {\bibfnamefont {A.}~\bibnamefont
  {Dewes}}, \bibinfo {author} {\bibfnamefont {R.}~\bibnamefont {Lauro}},
  \bibinfo {author} {\bibfnamefont {F.~R.}\ \bibnamefont {Ong}}, \bibinfo
  {author} {\bibfnamefont {V.}~\bibnamefont {Schmitt}}, \bibinfo {author}
  {\bibfnamefont {P.}~\bibnamefont {Milman}}, \bibinfo {author} {\bibfnamefont
  {P.}~\bibnamefont {Bertet}}, \bibinfo {author} {\bibfnamefont
  {D.}~\bibnamefont {Vion}},\ and\ \bibinfo {author} {\bibfnamefont
  {D.}~\bibnamefont {Esteve}},\ }\bibfield  {title} {\bibinfo {title} {Quantum
  speeding-up of computation demonstrated in a superconducting two-qubit
  processor},\ }\href {https://doi.org/10.1103/PhysRevB.85.140503} {\bibfield
  {journal} {\bibinfo  {journal} {Phys. Rev. B}\ }\textbf {\bibinfo {volume}
  {85}},\ \bibinfo {pages} {140503(R)} (\bibinfo {year} {2012})}\BibitemShut
  {NoStop}%
\bibitem [{\citenamefont {Figgatt}\ \emph {et~al.}(2017)\citenamefont
  {Figgatt}, \citenamefont {Maslov}, \citenamefont {Landsman}, \citenamefont
  {Linke}, \citenamefont {Debnath},\ and\ \citenamefont {Monroe}}]{Figgatt}%
  \BibitemOpen
  \bibfield  {author} {\bibinfo {author} {\bibfnamefont {C.}~\bibnamefont
  {Figgatt}}, \bibinfo {author} {\bibfnamefont {D.}~\bibnamefont {Maslov}},
  \bibinfo {author} {\bibfnamefont {K.~A.}\ \bibnamefont {Landsman}}, \bibinfo
  {author} {\bibfnamefont {N.~M.}\ \bibnamefont {Linke}}, \bibinfo {author}
  {\bibfnamefont {S.}~\bibnamefont {Debnath}},\ and\ \bibinfo {author}
  {\bibfnamefont {C.}~\bibnamefont {Monroe}},\ }\bibfield  {title} {\bibinfo
  {title} {Complete 3-qubit \mbox{Grover} search on a programmable quantum
  computer},\ }\href {https://doi.org/10.1038/s41467-017-01904-7} {\bibfield
  {journal} {\bibinfo  {journal} {Nature communications}\ }\textbf {\bibinfo
  {volume} {8}},\ \bibinfo {pages} {1} (\bibinfo {year} {2017})}\BibitemShut
  {NoStop}%
\bibitem [{\citenamefont {Mandviwalla}\ \emph {et~al.}(2018)\citenamefont
  {Mandviwalla}, \citenamefont {Ohshiro},\ and\ \citenamefont
  {Ji}}]{Mandviwalla}%
  \BibitemOpen
  \bibfield  {author} {\bibinfo {author} {\bibfnamefont {A.}~\bibnamefont
  {Mandviwalla}}, \bibinfo {author} {\bibfnamefont {K.}~\bibnamefont
  {Ohshiro}},\ and\ \bibinfo {author} {\bibfnamefont {B.}~\bibnamefont {Ji}},\
  }\bibfield  {title} {\bibinfo {title} {{Implementing \mbox{Grover's}
  Algorithm on the IBM Quantum Computers}},\ }in\ \href
  {https://doi.org/10.1109/BigData.2018.8622457} {\emph {\bibinfo {booktitle}
  {2018 IEEE International Conference on Big Data (Big Data)}}}\ (\bibinfo
  {year} {2018})\ pp.\ \bibinfo {pages} {2531--2537}\BibitemShut {NoStop}%
\bibitem [{\citenamefont {Pokharel}\ and\ \citenamefont
  {Lidar}(2022)}]{Pokharel}%
  \BibitemOpen
  \bibfield  {author} {\bibinfo {author} {\bibfnamefont {B.}~\bibnamefont
  {Pokharel}}\ and\ \bibinfo {author} {\bibfnamefont {D.}~\bibnamefont
  {Lidar}},\ }\href {https://doi.org/10.48550/ARXIV.2211.04543} {\bibinfo
  {title} {Better-than-classical \mbox{Grover} search via quantum error
  detection and suppression}} (\bibinfo {year} {2022}),\ \Eprint
  {https://arxiv.org/abs/arxiv:2211.04543} {arxiv:2211.04543} \BibitemShut
  {NoStop}%
\bibitem [{\citenamefont {Nielsen}\ and\ \citenamefont
  {Chuang}(2016)}]{Nielsen}%
  \BibitemOpen
  \bibfield  {author} {\bibinfo {author} {\bibfnamefont {M.~A.}\ \bibnamefont
  {Nielsen}}\ and\ \bibinfo {author} {\bibfnamefont {I.~L.}\ \bibnamefont
  {Chuang}},\ }\href
  {https://www.cambridge.org/de/academic/subjects/physics/quantum-physics-quantum-information-and-quantum-computation/quantum-computation-and-quantum-information-10th-anniversary-edition?format=HB}
  {\emph {\bibinfo {title} {Quantum Computation and Quantum Information (10th
  Anniversary edition)}}}\ (\bibinfo  {publisher} {Cambridge University
  Press},\ \bibinfo {year} {2016})\BibitemShut {NoStop}%
\bibitem [{SAT()}]{SAT_footnote}%
  \BibitemOpen
  \href@noop {} {\bibinfo {title} {The acronym \mbox{SAT} is shorthand for
  ``satisfiability problem,'' a type of decision problem over binary strings.
  \mbox{Circuit SAT} is a problem about deciding if an input exists to a given
  \mbox{Boolean} circuit for which the output is true. \mbox{Another}
  prototypical \mbox{SAT} problem class is \mbox{3-SAT}, involving constraints
  on groups of three \mbox{Boolean} variables.}}\BibitemShut {Stop}%
\bibitem [{\citenamefont {Aaronson}(2018)}]{Aaronson_QCLec18}%
  \BibitemOpen
  \bibfield  {author} {\bibinfo {author} {\bibfnamefont {S.}~\bibnamefont
  {Aaronson}},\ }\href {https://www.scottaaronson.com/qclec.pdf} {\bibinfo
  {title} {Introduction to quantum information science}} (\bibinfo {year}
  {2018})\BibitemShut {NoStop}%
\bibitem [{\citenamefont {Ma}\ and\ \citenamefont {Yang}(2022)}]{Ma}%
  \BibitemOpen
  \bibfield  {author} {\bibinfo {author} {\bibfnamefont {L.}~\bibnamefont
  {Ma}}\ and\ \bibinfo {author} {\bibfnamefont {C.}~\bibnamefont {Yang}},\
  }\bibfield  {title} {\bibinfo {title} {Low rank approximation in simulations
  of quantum algorithms},\ }\href
  {https://doi.org/https://doi.org/10.1016/j.jocs.2022.101561} {\bibfield
  {journal} {\bibinfo  {journal} {Journal of Computational Science}\ }\textbf
  {\bibinfo {volume} {59}},\ \bibinfo {pages} {101561} (\bibinfo {year}
  {2022})}\BibitemShut {NoStop}%
\bibitem [{\citenamefont {\"Ostlund}\ and\ \citenamefont
  {Rommer}(1995)}]{Ostlund}%
  \BibitemOpen
  \bibfield  {author} {\bibinfo {author} {\bibfnamefont {S.}~\bibnamefont
  {\"Ostlund}}\ and\ \bibinfo {author} {\bibfnamefont {S.}~\bibnamefont
  {Rommer}},\ }\bibfield  {title} {\bibinfo {title} {Thermodynamic limit of
  density matrix renormalization},\ }\href
  {https://doi.org/10.1103/PhysRevLett.75.3537} {\bibfield  {journal} {\bibinfo
   {journal} {Phys. Rev. Lett.}\ }\textbf {\bibinfo {volume} {75}},\ \bibinfo
  {pages} {3537} (\bibinfo {year} {1995})}\BibitemShut {NoStop}%
\bibitem [{\citenamefont {Vidal}(2003{\natexlab{a}})}]{Vidal}%
  \BibitemOpen
  \bibfield  {author} {\bibinfo {author} {\bibfnamefont {G.}~\bibnamefont
  {Vidal}},\ }\bibfield  {title} {\bibinfo {title} {Efficient classical
  simulation of slightly entangled quantum computations},\ }\href
  {https://doi.org/10.1103/PhysRevLett.91.147902} {\bibfield  {journal}
  {\bibinfo  {journal} {Phys. Rev. Lett.}\ }\textbf {\bibinfo {volume} {91}},\
  \bibinfo {pages} {147902} (\bibinfo {year} {2003}{\natexlab{a}})}\BibitemShut
  {NoStop}%
\bibitem [{\citenamefont {Perez-Garcia}\ \emph {et~al.}(2007)\citenamefont
  {Perez-Garcia}, \citenamefont {Verstraete}, \citenamefont {Wolf},\ and\
  \citenamefont {Cirac}}]{Perez-Garcia}%
  \BibitemOpen
  \bibfield  {author} {\bibinfo {author} {\bibfnamefont {D.}~\bibnamefont
  {Perez-Garcia}}, \bibinfo {author} {\bibfnamefont {F.}~\bibnamefont
  {Verstraete}}, \bibinfo {author} {\bibfnamefont {M.~M.}\ \bibnamefont
  {Wolf}},\ and\ \bibinfo {author} {\bibfnamefont {J.~I.}\ \bibnamefont
  {Cirac}},\ }\bibfield  {title} {\bibinfo {title} {Matrix product state
  representations},\ }\href@noop {} {\bibfield  {journal} {\bibinfo  {journal}
  {Quantum Info. Comput.}\ }\textbf {\bibinfo {volume} {7}},\ \bibinfo {pages}
  {401–430} (\bibinfo {year} {2007})}\BibitemShut {NoStop}%
\bibitem [{\citenamefont {McCulloch}(2007)}]{McCulloch}%
  \BibitemOpen
  \bibfield  {author} {\bibinfo {author} {\bibfnamefont {I.~P.}\ \bibnamefont
  {McCulloch}},\ }\bibfield  {title} {\bibinfo {title} {From density-matrix
  renormalization group to matrix product states},\ }\href
  {https://doi.org/10.1088/1742-5468/2007/10/p10014} {\bibfield  {journal}
  {\bibinfo  {journal} {Journal of Statistical Mechanics: Theory and
  Experiment}\ }\textbf {\bibinfo {volume} {2007}},\ \bibinfo {pages} {P10014}
  (\bibinfo {year} {2007})}\BibitemShut {NoStop}%
\bibitem [{\citenamefont {Verstraete}\ \emph {et~al.}(2004)\citenamefont
  {Verstraete}, \citenamefont {Garc\'{\i}a-Ripoll},\ and\ \citenamefont
  {Cirac}}]{Verstraete:2004}%
  \BibitemOpen
  \bibfield  {author} {\bibinfo {author} {\bibfnamefont {F.}~\bibnamefont
  {Verstraete}}, \bibinfo {author} {\bibfnamefont {J.~J.}\ \bibnamefont
  {Garc\'{\i}a-Ripoll}},\ and\ \bibinfo {author} {\bibfnamefont {J.~I.}\
  \bibnamefont {Cirac}},\ }\bibfield  {title} {\bibinfo {title} {Matrix product
  density operators: Simulation of finite-temperature and dissipative
  systems},\ }\href {https://doi.org/10.1103/PhysRevLett.93.207204} {\bibfield
  {journal} {\bibinfo  {journal} {Phys. Rev. Lett.}\ }\textbf {\bibinfo
  {volume} {93}},\ \bibinfo {pages} {207204} (\bibinfo {year}
  {2004})}\BibitemShut {NoStop}%
\bibitem [{\citenamefont {Vidal}(2003{\natexlab{b}})}]{VidalEfficient}%
  \BibitemOpen
  \bibfield  {author} {\bibinfo {author} {\bibfnamefont {G.}~\bibnamefont
  {Vidal}},\ }\bibfield  {title} {\bibinfo {title} {Efficient classical
  simulation of slightly entangled quantum computations},\ }\href
  {https://doi.org/10.1103/PhysRevLett.91.147902} {\bibfield  {journal}
  {\bibinfo  {journal} {Phys. Rev. Lett.}\ }\textbf {\bibinfo {volume} {91}},\
  \bibinfo {pages} {147902} (\bibinfo {year} {2003}{\natexlab{b}})}\BibitemShut
  {NoStop}%
\bibitem [{\citenamefont {Zhou}\ \emph {et~al.}(2020)\citenamefont {Zhou},
  \citenamefont {Stoudenmire},\ and\ \citenamefont {Waintal}}]{Zhou}%
  \BibitemOpen
  \bibfield  {author} {\bibinfo {author} {\bibfnamefont {Y.}~\bibnamefont
  {Zhou}}, \bibinfo {author} {\bibfnamefont {E.~M.}\ \bibnamefont
  {Stoudenmire}},\ and\ \bibinfo {author} {\bibfnamefont {X.}~\bibnamefont
  {Waintal}},\ }\bibfield  {title} {\bibinfo {title} {What limits the
  simulation of quantum computers?},\ }\href
  {https://doi.org/10.1103/PhysRevX.10.041038} {\bibfield  {journal} {\bibinfo
  {journal} {Phys. Rev. X}\ }\textbf {\bibinfo {volume} {10}},\ \bibinfo
  {pages} {041038} (\bibinfo {year} {2020})}\BibitemShut {NoStop}%
\bibitem [{\citenamefont {Orús}(2014)}]{Orus_2014}%
  \BibitemOpen
  \bibfield  {author} {\bibinfo {author} {\bibfnamefont {R.}~\bibnamefont
  {Orús}},\ }\bibfield  {title} {\bibinfo {title} {A practical introduction to
  tensor networks: Matrix product states and projected entangled pair states},\
  }\href {https://doi.org/https://doi.org/10.1016/j.aop.2014.06.013} {\bibfield
   {journal} {\bibinfo  {journal} {Annals of Physics}\ }\textbf {\bibinfo
  {volume} {349}},\ \bibinfo {pages} {117} (\bibinfo {year}
  {2014})}\BibitemShut {NoStop}%
\bibitem [{\citenamefont {Bridgeman}\ and\ \citenamefont
  {Chubb}(2017)}]{Bridgeman_2017}%
  \BibitemOpen
  \bibfield  {author} {\bibinfo {author} {\bibfnamefont {J.~C.}\ \bibnamefont
  {Bridgeman}}\ and\ \bibinfo {author} {\bibfnamefont {C.~T.}\ \bibnamefont
  {Chubb}},\ }\bibfield  {title} {\bibinfo {title} {Hand-waving and
  interpretive dance: an introductory course on tensor networks},\ }\href
  {https://doi.org/10.1088/1751-8121/aa6dc3} {\bibfield  {journal} {\bibinfo
  {journal} {Journal of Physics A: Mathematical and Theoretical}\ }\textbf
  {\bibinfo {volume} {50}},\ \bibinfo {pages} {223001} (\bibinfo {year}
  {2017})}\BibitemShut {NoStop}%
\bibitem [{\citenamefont {Damme}\ \emph {et~al.}(2023)\citenamefont {Damme},
  \citenamefont {Haegeman}, \citenamefont {McCulloch},\ and\ \citenamefont
  {Vanderstraeten}}]{Vandamme2023efficient}%
  \BibitemOpen
  \bibfield  {author} {\bibinfo {author} {\bibfnamefont {M.~V.}\ \bibnamefont
  {Damme}}, \bibinfo {author} {\bibfnamefont {J.}~\bibnamefont {Haegeman}},
  \bibinfo {author} {\bibfnamefont {I.}~\bibnamefont {McCulloch}},\ and\
  \bibinfo {author} {\bibfnamefont {L.}~\bibnamefont {Vanderstraeten}},\ }\href
  {https://arxiv.org/abs/2302.14181} {\bibinfo {title} {Efficient higher-order
  matrix product operators for time evolution}} (\bibinfo {year} {2023}),\
  \Eprint {https://arxiv.org/abs/2302.14181} {arXiv:2302.14181
  [cond-mat.str-el]} \BibitemShut {NoStop}%
\bibitem [{\citenamefont {Ferris}\ and\ \citenamefont
  {Vidal}(2012)}]{Ferris:2012}%
  \BibitemOpen
  \bibfield  {author} {\bibinfo {author} {\bibfnamefont {A.~J.}\ \bibnamefont
  {Ferris}}\ and\ \bibinfo {author} {\bibfnamefont {G.}~\bibnamefont {Vidal}},\
  }\bibfield  {title} {\bibinfo {title} {Perfect sampling with unitary tensor
  networks},\ }\href {https://doi.org/10.1103/PhysRevB.85.165146} {\bibfield
  {journal} {\bibinfo  {journal} {Phys. Rev. B}\ }\textbf {\bibinfo {volume}
  {85}},\ \bibinfo {pages} {165146} (\bibinfo {year} {2012})}\BibitemShut
  {NoStop}%
\bibitem [{\citenamefont {Stoudenmire}\ and\ \citenamefont
  {White}(2010)}]{METTS}%
  \BibitemOpen
  \bibfield  {author} {\bibinfo {author} {\bibfnamefont {E.~M.}\ \bibnamefont
  {Stoudenmire}}\ and\ \bibinfo {author} {\bibfnamefont {S.~R.}\ \bibnamefont
  {White}},\ }\bibfield  {title} {\bibinfo {title} {Minimally entangled typical
  thermal state algorithms},\ }\href
  {https://doi.org/10.1088/1367-2630/12/5/055026} {\bibfield  {journal}
  {\bibinfo  {journal} {New Journal of Physics}\ }\textbf {\bibinfo {volume}
  {12}},\ \bibinfo {pages} {055026} (\bibinfo {year} {2010})}\BibitemShut
  {NoStop}%
\bibitem [{Note1()}]{Note1}%
  \BibitemOpen
  \bibinfo {note} {The $n \chi ^2$ scaling for MPS sampling assumes the MPS has
  been brought into an ``orthogonal form'' which is typically the case for MPS
  computed by circuit evolution algorithms. Otherwise an MPS can be brought
  into this form at a cost of $n \chi ^3$}\BibitemShut {NoStop}%
\bibitem [{\citenamefont {Or{\'u}s}(2019)}]{Orus_2019}%
  \BibitemOpen
  \bibfield  {author} {\bibinfo {author} {\bibfnamefont {R.}~\bibnamefont
  {Or{\'u}s}},\ }\bibfield  {title} {\bibinfo {title} {Tensor networks for
  complex quantum systems},\ }\href@noop {} {\bibfield  {journal} {\bibinfo
  {journal} {Nature Reviews Physics}\ }\textbf {\bibinfo {volume} {1}},\
  \bibinfo {pages} {538} (\bibinfo {year} {2019})}\BibitemShut {NoStop}%
\bibitem [{\citenamefont {Evenbly}\ and\ \citenamefont
  {Vidal}(2011)}]{Evenbly2011tensor}%
  \BibitemOpen
  \bibfield  {author} {\bibinfo {author} {\bibfnamefont {G.}~\bibnamefont
  {Evenbly}}\ and\ \bibinfo {author} {\bibfnamefont {G.}~\bibnamefont
  {Vidal}},\ }\bibfield  {title} {\bibinfo {title} {Tensor network states and
  geometry},\ }\href@noop {} {\bibfield  {journal} {\bibinfo  {journal}
  {Journal of Statistical Physics}\ }\textbf {\bibinfo {volume} {145}},\
  \bibinfo {pages} {891} (\bibinfo {year} {2011})}\BibitemShut {NoStop}%
\bibitem [{\citenamefont {Ayral}\ \emph {et~al.}(2023)\citenamefont {Ayral},
  \citenamefont {Louvet}, \citenamefont {Zhou}, \citenamefont {Lambert},
  \citenamefont {Stoudenmire},\ and\ \citenamefont {Waintal}}]{Ayral}%
  \BibitemOpen
  \bibfield  {author} {\bibinfo {author} {\bibfnamefont {T.}~\bibnamefont
  {Ayral}}, \bibinfo {author} {\bibfnamefont {T.}~\bibnamefont {Louvet}},
  \bibinfo {author} {\bibfnamefont {Y.}~\bibnamefont {Zhou}}, \bibinfo {author}
  {\bibfnamefont {C.}~\bibnamefont {Lambert}}, \bibinfo {author} {\bibfnamefont
  {E.~M.}\ \bibnamefont {Stoudenmire}},\ and\ \bibinfo {author} {\bibfnamefont
  {X.}~\bibnamefont {Waintal}},\ }\bibfield  {title} {\bibinfo {title}
  {Density-matrix renormalization group algorithm for simulating quantum
  circuits with a finite fidelity},\ }\href
  {https://doi.org/10.1103/PRXQuantum.4.020304} {\bibfield  {journal} {\bibinfo
   {journal} {PRX Quantum}\ }\textbf {\bibinfo {volume} {4}},\ \bibinfo {pages}
  {020304} (\bibinfo {year} {2023})}\BibitemShut {NoStop}%
\bibitem [{\citenamefont {White}\ \emph {et~al.}(2018)\citenamefont {White},
  \citenamefont {Zaletel}, \citenamefont {Mong},\ and\ \citenamefont
  {Refael}}]{White_DMT}%
  \BibitemOpen
  \bibfield  {author} {\bibinfo {author} {\bibfnamefont {C.~D.}\ \bibnamefont
  {White}}, \bibinfo {author} {\bibfnamefont {M.}~\bibnamefont {Zaletel}},
  \bibinfo {author} {\bibfnamefont {R.~S.~K.}\ \bibnamefont {Mong}},\ and\
  \bibinfo {author} {\bibfnamefont {G.}~\bibnamefont {Refael}},\ }\bibfield
  {title} {\bibinfo {title} {Quantum dynamics of thermalizing systems},\ }\href
  {https://doi.org/10.1103/PhysRevB.97.035127} {\bibfield  {journal} {\bibinfo
  {journal} {Phys. Rev. B}\ }\textbf {\bibinfo {volume} {97}},\ \bibinfo
  {pages} {035127} (\bibinfo {year} {2018})}\BibitemShut {NoStop}%
\bibitem [{\citenamefont {Rakovszky}\ \emph {et~al.}(2022)\citenamefont
  {Rakovszky}, \citenamefont {von Keyserlingk},\ and\ \citenamefont
  {Pollmann}}]{Rakovszky}%
  \BibitemOpen
  \bibfield  {author} {\bibinfo {author} {\bibfnamefont {T.}~\bibnamefont
  {Rakovszky}}, \bibinfo {author} {\bibfnamefont {C.~W.}\ \bibnamefont {von
  Keyserlingk}},\ and\ \bibinfo {author} {\bibfnamefont {F.}~\bibnamefont
  {Pollmann}},\ }\bibfield  {title} {\bibinfo {title} {Dissipation-assisted
  operator evolution method for capturing hydrodynamic transport},\ }\href
  {https://doi.org/10.1103/PhysRevB.105.075131} {\bibfield  {journal} {\bibinfo
   {journal} {Phys. Rev. B}\ }\textbf {\bibinfo {volume} {105}},\ \bibinfo
  {pages} {075131} (\bibinfo {year} {2022})}\BibitemShut {NoStop}%
\bibitem [{\citenamefont {Massacci}\ and\ \citenamefont
  {Marraro}(2000)}]{Massacci:2000}%
  \BibitemOpen
  \bibfield  {author} {\bibinfo {author} {\bibfnamefont {F.}~\bibnamefont
  {Massacci}}\ and\ \bibinfo {author} {\bibfnamefont {L.}~\bibnamefont
  {Marraro}},\ }\bibfield  {title} {\bibinfo {title} {Logical cryptanalysis as
  a \mbox{SAT} problem},\ }\href@noop {} {\bibfield  {journal} {\bibinfo
  {journal} {Journal of Automated Reasoning}\ }\textbf {\bibinfo {volume}
  {24}},\ \bibinfo {pages} {165} (\bibinfo {year} {2000})}\BibitemShut
  {NoStop}%
\bibitem [{\citenamefont {Mironov}\ and\ \citenamefont
  {Zhang}(2006)}]{Mironov:2006}%
  \BibitemOpen
  \bibfield  {author} {\bibinfo {author} {\bibfnamefont {I.}~\bibnamefont
  {Mironov}}\ and\ \bibinfo {author} {\bibfnamefont {L.}~\bibnamefont
  {Zhang}},\ }\bibfield  {title} {\bibinfo {title} {Applications of \mbox{SAT}
  solvers to cryptanalysis of hash functions},\ }in\ \href@noop {} {\emph
  {\bibinfo {booktitle} {International Conference on Theory and Applications of
  Satisfiability Testing}}}\ (\bibinfo {organization} {Springer},\ \bibinfo
  {year} {2006})\ pp.\ \bibinfo {pages} {102--115}\BibitemShut {NoStop}%
\bibitem [{\citenamefont {Perron}\ and\ \citenamefont
  {Furnon}(2022)}]{ortools}%
  \BibitemOpen
  \bibfield  {author} {\bibinfo {author} {\bibfnamefont {L.}~\bibnamefont
  {Perron}}\ and\ \bibinfo {author} {\bibfnamefont {V.}~\bibnamefont
  {Furnon}},\ }\href {https://developers.google.com/optimization/} {\bibinfo
  {title} {Or-tools}} (\bibinfo {year} {2022})\BibitemShut {NoStop}%
\bibitem [{\citenamefont {Corblin}\ \emph {et~al.}(2007)\citenamefont
  {Corblin}, \citenamefont {Bordeaux}, \citenamefont {Hamadi}, \citenamefont
  {Fanchon},\ and\ \citenamefont {Trilling}}]{corblin2007sat}%
  \BibitemOpen
  \bibfield  {author} {\bibinfo {author} {\bibfnamefont {F.}~\bibnamefont
  {Corblin}}, \bibinfo {author} {\bibfnamefont {L.}~\bibnamefont {Bordeaux}},
  \bibinfo {author} {\bibfnamefont {Y.}~\bibnamefont {Hamadi}}, \bibinfo
  {author} {\bibfnamefont {E.}~\bibnamefont {Fanchon}},\ and\ \bibinfo {author}
  {\bibfnamefont {L.}~\bibnamefont {Trilling}},\ }\bibfield  {title} {\bibinfo
  {title} {A \mbox{SAT}-based approach to decipher gene regulatory networks},\
  }\href@noop {} {\bibfield  {journal} {\bibinfo  {journal} {Integrative
  Post-Genomics, RIAMS, Lyon}\ } (\bibinfo {year} {2007})}\BibitemShut
  {NoStop}%
\bibitem [{\citenamefont {Fishman}\ \emph
  {et~al.}(2022{\natexlab{a}})\citenamefont {Fishman}, \citenamefont {White},\
  and\ \citenamefont {Stoudenmire}}]{itensor}%
  \BibitemOpen
  \bibfield  {author} {\bibinfo {author} {\bibfnamefont {M.}~\bibnamefont
  {Fishman}}, \bibinfo {author} {\bibfnamefont {S.~R.}\ \bibnamefont {White}},\
  and\ \bibinfo {author} {\bibfnamefont {E.~M.}\ \bibnamefont {Stoudenmire}},\
  }\bibfield  {title} {\bibinfo {title} {{The ITensor Software Library for
  Tensor Network Calculations}},\ }\href
  {https://doi.org/10.21468/SciPostPhysCodeb.4} {\bibfield  {journal} {\bibinfo
   {journal} {SciPost Phys. Codebases}\ ,\ \bibinfo {pages} {4}} (\bibinfo
  {year} {2022}{\natexlab{a}})}\BibitemShut {NoStop}%
\bibitem [{\citenamefont {Fishman}\ \emph
  {et~al.}(2022{\natexlab{b}})\citenamefont {Fishman}, \citenamefont {White},\
  and\ \citenamefont {Stoudenmire}}]{itensor-r0.3}%
  \BibitemOpen
  \bibfield  {author} {\bibinfo {author} {\bibfnamefont {M.}~\bibnamefont
  {Fishman}}, \bibinfo {author} {\bibfnamefont {S.~R.}\ \bibnamefont {White}},\
  and\ \bibinfo {author} {\bibfnamefont {E.~M.}\ \bibnamefont {Stoudenmire}},\
  }\bibfield  {title} {\bibinfo {title} {{Codebase release 0.3 for ITensor}},\
  }\href {https://doi.org/10.21468/SciPostPhysCodeb.4-r0.3} {\bibfield
  {journal} {\bibinfo  {journal} {SciPost Phys. Codebases}\ ,\ \bibinfo {pages}
  {4}} (\bibinfo {year} {2022}{\natexlab{b}})}\BibitemShut {NoStop}%
\bibitem [{\citenamefont {Braunstein}\ \emph {et~al.}(2005)\citenamefont
  {Braunstein}, \citenamefont {Mézard},\ and\ \citenamefont
  {Zecchina}}]{Braunstein_Survey}%
  \BibitemOpen
  \bibfield  {author} {\bibinfo {author} {\bibfnamefont {A.}~\bibnamefont
  {Braunstein}}, \bibinfo {author} {\bibfnamefont {M.}~\bibnamefont
  {Mézard}},\ and\ \bibinfo {author} {\bibfnamefont {R.}~\bibnamefont
  {Zecchina}},\ }\bibfield  {title} {\bibinfo {title} {Survey propagation: An
  algorithm for satisfiability},\ }\href
  {https://doi.org/https://doi.org/10.1002/rsa.20057} {\bibfield  {journal}
  {\bibinfo  {journal} {Random Structures \& Algorithms}\ }\textbf {\bibinfo
  {volume} {27}},\ \bibinfo {pages} {201} (\bibinfo {year} {2005})},\ \Eprint
  {https://arxiv.org/abs/https://onlinelibrary.wiley.com/doi/pdf/10.1002/rsa.20057}
  {https://onlinelibrary.wiley.com/doi/pdf/10.1002/rsa.20057} \BibitemShut
  {NoStop}%
\bibitem [{\citenamefont {Chen}\ \emph {et~al.}(2018)\citenamefont {Chen},
  \citenamefont {Zhang}, \citenamefont {Huang}, \citenamefont {Newman},\ and\
  \citenamefont {Shi}}]{chen2018classical}%
  \BibitemOpen
  \bibfield  {author} {\bibinfo {author} {\bibfnamefont {J.}~\bibnamefont
  {Chen}}, \bibinfo {author} {\bibfnamefont {F.}~\bibnamefont {Zhang}},
  \bibinfo {author} {\bibfnamefont {C.}~\bibnamefont {Huang}}, \bibinfo
  {author} {\bibfnamefont {M.}~\bibnamefont {Newman}},\ and\ \bibinfo {author}
  {\bibfnamefont {Y.}~\bibnamefont {Shi}},\ }\href@noop {} {\bibinfo {title}
  {Classical simulation of intermediate-size quantum circuits}} (\bibinfo
  {year} {2018}),\ \Eprint {https://arxiv.org/abs/1805.01450} {arXiv:1805.01450
  [quant-ph]} \BibitemShut {NoStop}%
\bibitem [{\citenamefont {Gray}\ and\ \citenamefont
  {Kourtis}(2021)}]{Gray2021hyper}%
  \BibitemOpen
  \bibfield  {author} {\bibinfo {author} {\bibfnamefont {J.}~\bibnamefont
  {Gray}}\ and\ \bibinfo {author} {\bibfnamefont {S.}~\bibnamefont {Kourtis}},\
  }\bibfield  {title} {\bibinfo {title} {Hyper-optimized tensor network
  contraction},\ }\href {https://quantum-journal.org/papers/q-2021-03-15-410/}
  {\bibfield  {journal} {\bibinfo  {journal} {Quantum}\ }\textbf {\bibinfo
  {volume} {5}},\ \bibinfo {pages} {410} (\bibinfo {year} {2021})}\BibitemShut
  {NoStop}%
\bibitem [{\citenamefont {Pan}\ \emph {et~al.}(2022)\citenamefont {Pan},
  \citenamefont {Chen},\ and\ \citenamefont {Zhang}}]{Pan:2022}%
  \BibitemOpen
  \bibfield  {author} {\bibinfo {author} {\bibfnamefont {F.}~\bibnamefont
  {Pan}}, \bibinfo {author} {\bibfnamefont {K.}~\bibnamefont {Chen}},\ and\
  \bibinfo {author} {\bibfnamefont {P.}~\bibnamefont {Zhang}},\ }\bibfield
  {title} {\bibinfo {title} {Solving the sampling problem of the sycamore
  quantum circuits},\ }\href {https://doi.org/10.1103/PhysRevLett.129.090502}
  {\bibfield  {journal} {\bibinfo  {journal} {Phys. Rev. Lett.}\ }\textbf
  {\bibinfo {volume} {129}},\ \bibinfo {pages} {090502} (\bibinfo {year}
  {2022})}\BibitemShut {NoStop}%
\bibitem [{\citenamefont {Kourtis}\ \emph {et~al.}(2019)\citenamefont
  {Kourtis}, \citenamefont {Chamon}, \citenamefont {Mucciolo},\ and\
  \citenamefont {Ruckenstein}}]{Kourtis2019}%
  \BibitemOpen
  \bibfield  {author} {\bibinfo {author} {\bibfnamefont {S.}~\bibnamefont
  {Kourtis}}, \bibinfo {author} {\bibfnamefont {C.}~\bibnamefont {Chamon}},
  \bibinfo {author} {\bibfnamefont {E.~R.}\ \bibnamefont {Mucciolo}},\ and\
  \bibinfo {author} {\bibfnamefont {A.~E.}\ \bibnamefont {Ruckenstein}},\
  }\bibfield  {title} {\bibinfo {title} {{Fast counting with tensor
  networks}},\ }\href {https://doi.org/10.21468/SciPostPhys.7.5.060} {\bibfield
   {journal} {\bibinfo  {journal} {SciPost Phys.}\ }\textbf {\bibinfo {volume}
  {7}},\ \bibinfo {pages} {060} (\bibinfo {year} {2019})}\BibitemShut {NoStop}%
\bibitem [{\citenamefont {Paturi}\ and\ \citenamefont
  {Pudlak}(2010)}]{Paturi_Complexity}%
  \BibitemOpen
  \bibfield  {author} {\bibinfo {author} {\bibfnamefont {R.}~\bibnamefont
  {Paturi}}\ and\ \bibinfo {author} {\bibfnamefont {P.}~\bibnamefont
  {Pudlak}},\ }\bibfield  {title} {\bibinfo {title} {On the complexity of
  circuit satisfiability},\ }in\ \href
  {https://doi.org/10.1145/1806689.1806724} {\emph {\bibinfo {booktitle}
  {Proceedings of the Forty-Second ACM Symposium on Theory of Computing}}},\
  \bibinfo {series and number} {STOC '10}\ (\bibinfo  {publisher} {Association
  for Computing Machinery},\ \bibinfo {address} {New York, NY, USA},\ \bibinfo
  {year} {2010})\ p.\ \bibinfo {pages} {241–250}\BibitemShut {NoStop}%
\bibitem [{\citenamefont {Ramos-Calderer}\ \emph {et~al.}(2021)\citenamefont
  {Ramos-Calderer}, \citenamefont {Bellini}, \citenamefont {Latorre},
  \citenamefont {Manzano},\ and\ \citenamefont {Mateu}}]{Ramos}%
  \BibitemOpen
  \bibfield  {author} {\bibinfo {author} {\bibfnamefont {S.}~\bibnamefont
  {Ramos-Calderer}}, \bibinfo {author} {\bibfnamefont {E.}~\bibnamefont
  {Bellini}}, \bibinfo {author} {\bibfnamefont {J.~I.}\ \bibnamefont
  {Latorre}}, \bibinfo {author} {\bibfnamefont {M.}~\bibnamefont {Manzano}},\
  and\ \bibinfo {author} {\bibfnamefont {V.}~\bibnamefont {Mateu}},\ }\bibfield
   {title} {\bibinfo {title} {Quantum search for scaled hash function
  preimages},\ }\href@noop {} {\bibfield  {journal} {\bibinfo  {journal}
  {Quantum Information Processing}\ }\textbf {\bibinfo {volume} {20}},\
  \bibinfo {pages} {180} (\bibinfo {year} {2021})}\BibitemShut {NoStop}%
\bibitem [{\citenamefont {Bernstein}\ and\ \citenamefont
  {Lange}(2017)}]{Bernstein}%
  \BibitemOpen
  \bibfield  {author} {\bibinfo {author} {\bibfnamefont {D.~J.}\ \bibnamefont
  {Bernstein}}\ and\ \bibinfo {author} {\bibfnamefont {T.}~\bibnamefont
  {Lange}},\ }\href {https://eprint.iacr.org/2017/314} {\bibinfo {title}
  {Post-quantum cryptography---dealing with the fallout of physics success}},\
  \bibinfo {howpublished} {Cryptology ePrint Archive, Paper 2017/314} (\bibinfo
  {year} {2017}),\ \bibinfo {note}
  {\url{https://eprint.iacr.org/2017/314}}\BibitemShut {NoStop}%
\bibitem [{\citenamefont {Gheorghiu}\ and\ \citenamefont
  {Mosca}(2019)}]{Gheorghiu}%
  \BibitemOpen
  \bibfield  {author} {\bibinfo {author} {\bibfnamefont {V.}~\bibnamefont
  {Gheorghiu}}\ and\ \bibinfo {author} {\bibfnamefont {M.}~\bibnamefont
  {Mosca}},\ }\href@noop {} {\bibinfo {title} {Benchmarking the quantum
  cryptanalysis of symmetric, public-key and hash-based cryptographic schemes}}
  (\bibinfo {year} {2019}),\ \Eprint {https://arxiv.org/abs/1902.02332}
  {arXiv:1902.02332 [quant-ph]} \BibitemShut {NoStop}%
\bibitem [{\citenamefont {Chamon}\ and\ \citenamefont
  {Mucciolo}(2012)}]{Chamon}%
  \BibitemOpen
  \bibfield  {author} {\bibinfo {author} {\bibfnamefont {C.}~\bibnamefont
  {Chamon}}\ and\ \bibinfo {author} {\bibfnamefont {E.~R.}\ \bibnamefont
  {Mucciolo}},\ }\bibfield  {title} {\bibinfo {title} {Virtual parallel
  computing and a search algorithm using matrix product states},\ }\href
  {https://doi.org/10.1103/PhysRevLett.109.030503} {\bibfield  {journal}
  {\bibinfo  {journal} {Phys. Rev. Lett.}\ }\textbf {\bibinfo {volume} {109}},\
  \bibinfo {pages} {030503} (\bibinfo {year} {2012})}\BibitemShut {NoStop}%
\bibitem [{\citenamefont {Crosswhite}\ and\ \citenamefont
  {Bacon}(2008)}]{Crosswhite}%
  \BibitemOpen
  \bibfield  {author} {\bibinfo {author} {\bibfnamefont {G.~M.}\ \bibnamefont
  {Crosswhite}}\ and\ \bibinfo {author} {\bibfnamefont {D.}~\bibnamefont
  {Bacon}},\ }\bibfield  {title} {\bibinfo {title} {Finite automata for caching
  in matrix product algorithms},\ }\href
  {https://doi.org/10.1103/PhysRevA.78.012356} {\bibfield  {journal} {\bibinfo
  {journal} {Phys. Rev. A}\ }\textbf {\bibinfo {volume} {78}},\ \bibinfo
  {pages} {012356} (\bibinfo {year} {2008})}\BibitemShut {NoStop}%
\bibitem [{\citenamefont {Pirvu}\ \emph {et~al.}(2010)\citenamefont {Pirvu},
  \citenamefont {Murg}, \citenamefont {Cirac},\ and\ \citenamefont
  {Verstraete}}]{Pirvu:2010}%
  \BibitemOpen
  \bibfield  {author} {\bibinfo {author} {\bibfnamefont {B.}~\bibnamefont
  {Pirvu}}, \bibinfo {author} {\bibfnamefont {V.}~\bibnamefont {Murg}},
  \bibinfo {author} {\bibfnamefont {J.~I.}\ \bibnamefont {Cirac}},\ and\
  \bibinfo {author} {\bibfnamefont {F.}~\bibnamefont {Verstraete}},\ }\bibfield
   {title} {\bibinfo {title} {Matrix product operator representations},\ }\href
  {https://dx.doi.org/10.1088/1367-2630/12/2/025012} {\bibfield  {journal}
  {\bibinfo  {journal} {New Journal of Physics}\ }\textbf {\bibinfo {volume}
  {12}},\ \bibinfo {pages} {025012} (\bibinfo {year} {2010})}\BibitemShut
  {NoStop}%
\bibitem [{\citenamefont {Zaletel}\ \emph {et~al.}(2015)\citenamefont
  {Zaletel}, \citenamefont {Mong}, \citenamefont {Karrasch}, \citenamefont
  {Moore},\ and\ \citenamefont {Pollmann}}]{Zaletel}%
  \BibitemOpen
  \bibfield  {author} {\bibinfo {author} {\bibfnamefont {M.~P.}\ \bibnamefont
  {Zaletel}}, \bibinfo {author} {\bibfnamefont {R.~S.~K.}\ \bibnamefont
  {Mong}}, \bibinfo {author} {\bibfnamefont {C.}~\bibnamefont {Karrasch}},
  \bibinfo {author} {\bibfnamefont {J.~E.}\ \bibnamefont {Moore}},\ and\
  \bibinfo {author} {\bibfnamefont {F.}~\bibnamefont {Pollmann}},\ }\bibfield
  {title} {\bibinfo {title} {Time-evolving a matrix product state with
  long-ranged interactions},\ }\href
  {https://doi.org/10.1103/PhysRevB.91.165112} {\bibfield  {journal} {\bibinfo
  {journal} {Phys. Rev. B}\ }\textbf {\bibinfo {volume} {91}},\ \bibinfo
  {pages} {165112} (\bibinfo {year} {2015})}\BibitemShut {NoStop}%
\bibitem [{\citenamefont {Aaronson}\ and\ \citenamefont
  {Gottesman}(2004)}]{Aaronson:2004}%
  \BibitemOpen
  \bibfield  {author} {\bibinfo {author} {\bibfnamefont {S.}~\bibnamefont
  {Aaronson}}\ and\ \bibinfo {author} {\bibfnamefont {D.}~\bibnamefont
  {Gottesman}},\ }\bibfield  {title} {\bibinfo {title} {Improved simulation of
  stabilizer circuits},\ }\href {https://doi.org/10.1103/PhysRevA.70.052328}
  {\bibfield  {journal} {\bibinfo  {journal} {Phys. Rev. A}\ }\textbf {\bibinfo
  {volume} {70}},\ \bibinfo {pages} {052328} (\bibinfo {year}
  {2004})}\BibitemShut {NoStop}%
\bibitem [{\citenamefont {García}\ and\ \citenamefont
  {Markov}(2015)}]{Garcia_2015}%
  \BibitemOpen
  \bibfield  {author} {\bibinfo {author} {\bibfnamefont {H.~J.}\ \bibnamefont
  {García}}\ and\ \bibinfo {author} {\bibfnamefont {I.~L.}\ \bibnamefont
  {Markov}},\ }\bibfield  {title} {\bibinfo {title} {Simulation of quantum
  circuits via stabilizer frames},\ }\href
  {https://doi.org/10.1109/TC.2014.2360532} {\bibfield  {journal} {\bibinfo
  {journal} {arXiv:1712.03554, IEEE Transactions on Computers}\ }\textbf
  {\bibinfo {volume} {64}},\ \bibinfo {pages} {2323} (\bibinfo {year}
  {2015})}\BibitemShut {NoStop}%
\bibitem [{\citenamefont {Bravyi}\ \emph {et~al.}(2016)\citenamefont {Bravyi},
  \citenamefont {Smith},\ and\ \citenamefont {Smolin}}]{Bravyi_2016b}%
  \BibitemOpen
  \bibfield  {author} {\bibinfo {author} {\bibfnamefont {S.}~\bibnamefont
  {Bravyi}}, \bibinfo {author} {\bibfnamefont {G.}~\bibnamefont {Smith}},\ and\
  \bibinfo {author} {\bibfnamefont {J.~A.}\ \bibnamefont {Smolin}},\ }\bibfield
   {title} {\bibinfo {title} {Trading classical and quantum computational
  resources},\ }\href {https://doi.org/10.1103/PhysRevX.6.021043} {\bibfield
  {journal} {\bibinfo  {journal} {Phys. Rev. X}\ }\textbf {\bibinfo {volume}
  {6}},\ \bibinfo {pages} {021043} (\bibinfo {year} {2016})}\BibitemShut
  {NoStop}%
\bibitem [{\citenamefont {Bravyi}\ and\ \citenamefont
  {Gosset}(2016)}]{Bravyi_2016}%
  \BibitemOpen
  \bibfield  {author} {\bibinfo {author} {\bibfnamefont {S.}~\bibnamefont
  {Bravyi}}\ and\ \bibinfo {author} {\bibfnamefont {D.}~\bibnamefont
  {Gosset}},\ }\bibfield  {title} {\bibinfo {title} {Improved classical
  simulation of quantum circuits dominated by \mbox{Clifford} gates},\ }\href
  {https://doi.org/10.1103/PhysRevLett.116.250501} {\bibfield  {journal}
  {\bibinfo  {journal} {Phys. Rev. Lett.}\ }\textbf {\bibinfo {volume} {116}},\
  \bibinfo {pages} {250501} (\bibinfo {year} {2016})}\BibitemShut {NoStop}%
\bibitem [{\citenamefont {Oseledets}\ and\ \citenamefont
  {Tyrtyshnikov}(2010)}]{Oseledets:2010}%
  \BibitemOpen
  \bibfield  {author} {\bibinfo {author} {\bibfnamefont {I.}~\bibnamefont
  {Oseledets}}\ and\ \bibinfo {author} {\bibfnamefont {E.}~\bibnamefont
  {Tyrtyshnikov}},\ }\bibfield  {title} {\bibinfo {title} {\mbox{TT-cross}
  approximation for multidimensional arrays},\ }\href@noop {} {\bibfield
  {journal} {\bibinfo  {journal} {Linear Algebra and its Applications}\
  }\textbf {\bibinfo {volume} {432}},\ \bibinfo {pages} {70} (\bibinfo {year}
  {2010})}\BibitemShut {NoStop}%
\bibitem [{\citenamefont {Savostyanov}\ and\ \citenamefont
  {Oseledets}(2011)}]{Savostyanov:2011}%
  \BibitemOpen
  \bibfield  {author} {\bibinfo {author} {\bibfnamefont {D.}~\bibnamefont
  {Savostyanov}}\ and\ \bibinfo {author} {\bibfnamefont {I.}~\bibnamefont
  {Oseledets}},\ }\bibfield  {title} {\bibinfo {title} {Fast adaptive
  interpolation of multi-dimensional arrays in tensor train format},\ }in\
  \href@noop {} {\emph {\bibinfo {booktitle} {The 2011 International Workshop
  on Multidimensional (nD) Systems}}}\ (\bibinfo {organization} {IEEE},\
  \bibinfo {year} {2011})\ pp.\ \bibinfo {pages} {1--8}\BibitemShut {NoStop}%
\bibitem [{\citenamefont {Savostyanov}(2014)}]{Savostyanov:2014}%
  \BibitemOpen
  \bibfield  {author} {\bibinfo {author} {\bibfnamefont {D.~V.}\ \bibnamefont
  {Savostyanov}},\ }\bibfield  {title} {\bibinfo {title} {Quasioptimality of
  maximum-volume cross interpolation of tensors},\ }\href@noop {} {\bibfield
  {journal} {\bibinfo  {journal} {Linear Algebra and its Applications}\
  }\textbf {\bibinfo {volume} {458}},\ \bibinfo {pages} {217} (\bibinfo {year}
  {2014})}\BibitemShut {NoStop}%
\bibitem [{\citenamefont {Dolgov}\ and\ \citenamefont
  {Savostyanov}(2020)}]{Dolgov:2020}%
  \BibitemOpen
  \bibfield  {author} {\bibinfo {author} {\bibfnamefont {S.}~\bibnamefont
  {Dolgov}}\ and\ \bibinfo {author} {\bibfnamefont {D.}~\bibnamefont
  {Savostyanov}},\ }\bibfield  {title} {\bibinfo {title} {Parallel cross
  interpolation for high-precision calculation of high-dimensional integrals},\
  }\href@noop {} {\bibfield  {journal} {\bibinfo  {journal} {Computer Physics
  Communications}\ }\textbf {\bibinfo {volume} {246}},\ \bibinfo {pages}
  {106869} (\bibinfo {year} {2020})}\BibitemShut {NoStop}%
\bibitem [{\citenamefont {N\'u\~nez Fern\'andez}\ \emph
  {et~al.}(2022)\citenamefont {N\'u\~nez Fern\'andez}, \citenamefont {Jeannin},
  \citenamefont {Dumitrescu}, \citenamefont {Kloss}, \citenamefont {Kaye},
  \citenamefont {Parcollet},\ and\ \citenamefont {Waintal}}]{Nunez-Fernandez}%
  \BibitemOpen
  \bibfield  {author} {\bibinfo {author} {\bibfnamefont {Y.}~\bibnamefont
  {N\'u\~nez Fern\'andez}}, \bibinfo {author} {\bibfnamefont {M.}~\bibnamefont
  {Jeannin}}, \bibinfo {author} {\bibfnamefont {P.~T.}\ \bibnamefont
  {Dumitrescu}}, \bibinfo {author} {\bibfnamefont {T.}~\bibnamefont {Kloss}},
  \bibinfo {author} {\bibfnamefont {J.}~\bibnamefont {Kaye}}, \bibinfo {author}
  {\bibfnamefont {O.}~\bibnamefont {Parcollet}},\ and\ \bibinfo {author}
  {\bibfnamefont {X.}~\bibnamefont {Waintal}},\ }\bibfield  {title} {\bibinfo
  {title} {Learning feynman diagrams with tensor trains},\ }\href
  {https://doi.org/10.1103/PhysRevX.12.041018} {\bibfield  {journal} {\bibinfo
  {journal} {Phys. Rev. X}\ }\textbf {\bibinfo {volume} {12}},\ \bibinfo
  {pages} {041018} (\bibinfo {year} {2022})}\BibitemShut {NoStop}%
\bibitem [{\citenamefont {Babbush}\ \emph {et~al.}(2021)\citenamefont
  {Babbush}, \citenamefont {McClean}, \citenamefont {Newman}, \citenamefont
  {Gidney}, \citenamefont {Boixo},\ and\ \citenamefont {Neven}}]{Babbush:2021}%
  \BibitemOpen
  \bibfield  {author} {\bibinfo {author} {\bibfnamefont {R.}~\bibnamefont
  {Babbush}}, \bibinfo {author} {\bibfnamefont {J.~R.}\ \bibnamefont
  {McClean}}, \bibinfo {author} {\bibfnamefont {M.}~\bibnamefont {Newman}},
  \bibinfo {author} {\bibfnamefont {C.}~\bibnamefont {Gidney}}, \bibinfo
  {author} {\bibfnamefont {S.}~\bibnamefont {Boixo}},\ and\ \bibinfo {author}
  {\bibfnamefont {H.}~\bibnamefont {Neven}},\ }\bibfield  {title} {\bibinfo
  {title} {Focus beyond quadratic speedups for error-corrected quantum
  advantage},\ }\href {https://doi.org/10.1103/PRXQuantum.2.010103} {\bibfield
  {journal} {\bibinfo  {journal} {PRX Quantum}\ }\textbf {\bibinfo {volume}
  {2}},\ \bibinfo {pages} {010103} (\bibinfo {year} {2021})}\BibitemShut
  {NoStop}%
\bibitem [{\citenamefont {Hoefler}\ \emph {et~al.}(2023)\citenamefont
  {Hoefler}, \citenamefont {Haener},\ and\ \citenamefont
  {Troyer}}]{TroyerDisentangling}%
  \BibitemOpen
  \bibfield  {author} {\bibinfo {author} {\bibfnamefont {T.}~\bibnamefont
  {Hoefler}}, \bibinfo {author} {\bibfnamefont {T.}~\bibnamefont {Haener}},\
  and\ \bibinfo {author} {\bibfnamefont {M.}~\bibnamefont {Troyer}},\
  }\href@noop {} {\bibinfo {title} {Disentangling hype from practicality: On
  realistically achieving quantum advantage}} (\bibinfo {year} {2023}),\
  \Eprint {https://arxiv.org/abs/2307.00523} {arXiv:2307.00523 [quant-ph]}
  \BibitemShut {NoStop}%
\bibitem [{\citenamefont {Koch}\ \emph {et~al.}(2019)\citenamefont {Koch},
  \citenamefont {Torrance}, \citenamefont {Kinghorn}, \citenamefont {Patel},
  \citenamefont {Wessing},\ and\ \citenamefont {Alsing}}]{Koch}%
  \BibitemOpen
  \bibfield  {author} {\bibinfo {author} {\bibfnamefont {D.}~\bibnamefont
  {Koch}}, \bibinfo {author} {\bibfnamefont {A.}~\bibnamefont {Torrance}},
  \bibinfo {author} {\bibfnamefont {D.}~\bibnamefont {Kinghorn}}, \bibinfo
  {author} {\bibfnamefont {S.}~\bibnamefont {Patel}}, \bibinfo {author}
  {\bibfnamefont {L.}~\bibnamefont {Wessing}},\ and\ \bibinfo {author}
  {\bibfnamefont {P.~M.}\ \bibnamefont {Alsing}},\ }\href
  {https://doi.org/10.48550/ARXIV.1908.04229} {\bibinfo {title} {Simulating
  quantum algorithms using fidelity and coherence time as principle models for
  error}} (\bibinfo {year} {2019}),\ \Eprint
  {https://arxiv.org/abs/arxiv:1098.04229} {arxiv:1098.04229} \BibitemShut
  {NoStop}%
\bibitem [{\citenamefont {Regev}\ and\ \citenamefont {Schiff}(2008)}]{Regev}%
  \BibitemOpen
  \bibfield  {author} {\bibinfo {author} {\bibfnamefont {O.}~\bibnamefont
  {Regev}}\ and\ \bibinfo {author} {\bibfnamefont {L.}~\bibnamefont {Schiff}},\
  }\bibfield  {title} {\bibinfo {title} {Impossibility of a quantum speed-up
  with a faulty oracle},\ }in\ \href@noop {} {\emph {\bibinfo {booktitle}
  {International Colloquium on Automata, Languages, and Programming}}}\
  (\bibinfo {organization} {Springer},\ \bibinfo {year} {2008})\ pp.\ \bibinfo
  {pages} {773--781}\BibitemShut {NoStop}%
\bibitem [{\citenamefont {Long}\ \emph {et~al.}(2000)\citenamefont {Long},
  \citenamefont {Li}, \citenamefont {Zhang},\ and\ \citenamefont {Tu}}]{Long}%
  \BibitemOpen
  \bibfield  {author} {\bibinfo {author} {\bibfnamefont {G.~L.}\ \bibnamefont
  {Long}}, \bibinfo {author} {\bibfnamefont {Y.~S.}\ \bibnamefont {Li}},
  \bibinfo {author} {\bibfnamefont {W.~L.}\ \bibnamefont {Zhang}},\ and\
  \bibinfo {author} {\bibfnamefont {C.~C.}\ \bibnamefont {Tu}},\ }\bibfield
  {title} {\bibinfo {title} {Dominant gate imperfection in \mbox{Grover's}
  quantum search algorithm},\ }\href
  {https://doi.org/10.1103/PhysRevA.61.042305} {\bibfield  {journal} {\bibinfo
  {journal} {Phys. Rev. A}\ }\textbf {\bibinfo {volume} {61}},\ \bibinfo
  {pages} {042305} (\bibinfo {year} {2000})}\BibitemShut {NoStop}%
\bibitem [{\citenamefont {Arute}\ \emph {et~al.}(2019)\citenamefont {Arute},
  \citenamefont {Arya}, \citenamefont {Babbush}, \citenamefont {Bacon},
  \citenamefont {Bardin}, \citenamefont {Barends}, \citenamefont {Biswas},
  \citenamefont {Boixo}, \citenamefont {Brandao}, \citenamefont {Buell} \emph
  {et~al.}}]{Arute2019}%
  \BibitemOpen
  \bibfield  {author} {\bibinfo {author} {\bibfnamefont {F.}~\bibnamefont
  {Arute}}, \bibinfo {author} {\bibfnamefont {K.}~\bibnamefont {Arya}},
  \bibinfo {author} {\bibfnamefont {R.}~\bibnamefont {Babbush}}, \bibinfo
  {author} {\bibfnamefont {D.}~\bibnamefont {Bacon}}, \bibinfo {author}
  {\bibfnamefont {J.~C.}\ \bibnamefont {Bardin}}, \bibinfo {author}
  {\bibfnamefont {R.}~\bibnamefont {Barends}}, \bibinfo {author} {\bibfnamefont
  {R.}~\bibnamefont {Biswas}}, \bibinfo {author} {\bibfnamefont
  {S.}~\bibnamefont {Boixo}}, \bibinfo {author} {\bibfnamefont {F.~G.}\
  \bibnamefont {Brandao}}, \bibinfo {author} {\bibfnamefont {D.~A.}\
  \bibnamefont {Buell}}, \emph {et~al.},\ }\bibfield  {title} {\bibinfo {title}
  {Quantum supremacy using a programmable superconducting processor},\
  }\href@noop {} {\bibfield  {journal} {\bibinfo  {journal} {Nature}\ }\textbf
  {\bibinfo {volume} {574}},\ \bibinfo {pages} {505} (\bibinfo {year}
  {2019})}\BibitemShut {NoStop}%
\bibitem [{\citenamefont {Zhang}\ and\ \citenamefont {Korepin}(2020)}]{Zhang}%
  \BibitemOpen
  \bibfield  {author} {\bibinfo {author} {\bibfnamefont {K.}~\bibnamefont
  {Zhang}}\ and\ \bibinfo {author} {\bibfnamefont {V.~E.}\ \bibnamefont
  {Korepin}},\ }\bibfield  {title} {\bibinfo {title} {Depth optimization of
  quantum search algorithms beyond \mbox{Grover's} algorithm},\ }\href
  {https://doi.org/10.1103/PhysRevA.101.032346} {\bibfield  {journal} {\bibinfo
   {journal} {Phys. Rev. A}\ }\textbf {\bibinfo {volume} {101}},\ \bibinfo
  {pages} {032346} (\bibinfo {year} {2020})}\BibitemShut {NoStop}%
\bibitem [{\citenamefont {SaiToh}(2013)}]{SaiToh}%
  \BibitemOpen
  \bibfield  {author} {\bibinfo {author} {\bibfnamefont {A.}~\bibnamefont
  {SaiToh}},\ }\bibfield  {title} {\bibinfo {title} {{A multiprecision C++
  library for matrix-product-state simulation of quantum computing: Evaluation
  of numerical errors}},\ }in\ \href@noop {} {\emph {\bibinfo {booktitle}
  {Journal of Physics: Conference Series}}},\ Vol.\ \bibinfo {volume} {454}\
  (\bibinfo {organization} {IOP Publishing},\ \bibinfo {year} {2013})\ p.\
  \bibinfo {pages} {012064}\BibitemShut {NoStop}%
\bibitem [{\citenamefont {Montanaro}(2016)}]{Montanaro:2016}%
  \BibitemOpen
  \bibfield  {author} {\bibinfo {author} {\bibfnamefont {A.}~\bibnamefont
  {Montanaro}},\ }\bibfield  {title} {\bibinfo {title} {Quantum algorithms: an
  overview},\ }\href {https://doi.org/10.1038/npjqi.2015.23} {\bibfield
  {journal} {\bibinfo  {journal} {npj Quantum Information}\ }\textbf {\bibinfo
  {volume} {2}},\ \bibinfo {pages} {1} (\bibinfo {year} {2016})}\BibitemShut
  {NoStop}%
\bibitem [{\citenamefont {Napp}\ \emph {et~al.}(2022)\citenamefont {Napp},
  \citenamefont {Placa}, \citenamefont {Dalzell}, \citenamefont {Brandao},\
  and\ \citenamefont {Harrow}}]{Napp:2022}%
  \BibitemOpen
  \bibfield  {author} {\bibinfo {author} {\bibfnamefont {J.~C.}\ \bibnamefont
  {Napp}}, \bibinfo {author} {\bibfnamefont {R.~L.~L.}\ \bibnamefont {Placa}},
  \bibinfo {author} {\bibfnamefont {A.~M.}\ \bibnamefont {Dalzell}}, \bibinfo
  {author} {\bibfnamefont {F.~G. S.~L.}\ \bibnamefont {Brandao}},\ and\
  \bibinfo {author} {\bibfnamefont {A.~W.}\ \bibnamefont {Harrow}},\ }\bibfield
   {title} {\bibinfo {title} {Efficient classical simulation of random shallow
  2d quantum circuits},\ }\href {https://doi.org/10.1103/PhysRevX.12.021021}
  {\bibfield  {journal} {\bibinfo  {journal} {Phys. Rev. X}\ }\textbf {\bibinfo
  {volume} {12}},\ \bibinfo {pages} {021021} (\bibinfo {year}
  {2022})}\BibitemShut {NoStop}%
\bibitem [{\citenamefont {Tindall}\ \emph {et~al.}(2024)\citenamefont
  {Tindall}, \citenamefont {Fishman}, \citenamefont {Stoudenmire},\ and\
  \citenamefont {Sels}}]{Tindall:2024}%
  \BibitemOpen
  \bibfield  {author} {\bibinfo {author} {\bibfnamefont {J.}~\bibnamefont
  {Tindall}}, \bibinfo {author} {\bibfnamefont {M.}~\bibnamefont {Fishman}},
  \bibinfo {author} {\bibfnamefont {E.~M.}\ \bibnamefont {Stoudenmire}},\ and\
  \bibinfo {author} {\bibfnamefont {D.}~\bibnamefont {Sels}},\ }\bibfield
  {title} {\bibinfo {title} {Efficient tensor network simulation of ibm's eagle
  kicked ising experiment},\ }\href
  {https://doi.org/10.1103/PRXQuantum.5.010308} {\bibfield  {journal} {\bibinfo
   {journal} {PRX Quantum}\ }\textbf {\bibinfo {volume} {5}},\ \bibinfo {pages}
  {010308} (\bibinfo {year} {2024})}\BibitemShut {NoStop}%
\bibitem [{\citenamefont {Schoning}(1999)}]{Schoning}%
  \BibitemOpen
  \bibfield  {author} {\bibinfo {author} {\bibfnamefont {T.}~\bibnamefont
  {Schoning}},\ }\bibfield  {title} {\bibinfo {title} {A probabilistic
  algorithm for \mbox{k-SAT} and constraint satisfaction problems},\ }in\
  \href@noop {} {\emph {\bibinfo {booktitle} {40th Annual Symposium on
  Foundations of Computer Science (Cat. No. 99CB37039)}}}\ (\bibinfo
  {organization} {IEEE},\ \bibinfo {year} {1999})\ pp.\ \bibinfo {pages}
  {410--414}\BibitemShut {NoStop}%
\bibitem [{\citenamefont {Gomes}\ \emph {et~al.}(2008)\citenamefont {Gomes},
  \citenamefont {Kautz}, \citenamefont {Sabharwal},\ and\ \citenamefont
  {Selman}}]{GomesSatisfiability}%
  \BibitemOpen
  \bibfield  {author} {\bibinfo {author} {\bibfnamefont {C.~P.}\ \bibnamefont
  {Gomes}}, \bibinfo {author} {\bibfnamefont {H.}~\bibnamefont {Kautz}},
  \bibinfo {author} {\bibfnamefont {A.}~\bibnamefont {Sabharwal}},\ and\
  \bibinfo {author} {\bibfnamefont {B.}~\bibnamefont {Selman}},\ }\bibfield
  {title} {\bibinfo {title} {Satisfiability solvers},\ }\href@noop {}
  {\bibfield  {journal} {\bibinfo  {journal} {Foundations of Artificial
  Intelligence}\ }\textbf {\bibinfo {volume} {3}},\ \bibinfo {pages} {89}
  (\bibinfo {year} {2008})}\BibitemShut {NoStop}%
\bibitem [{\citenamefont {Fowler}\ \emph {et~al.}(2012)\citenamefont {Fowler},
  \citenamefont {Mariantoni}, \citenamefont {Martinis},\ and\ \citenamefont
  {Cleland}}]{Fowler}%
  \BibitemOpen
  \bibfield  {author} {\bibinfo {author} {\bibfnamefont {A.~G.}\ \bibnamefont
  {Fowler}}, \bibinfo {author} {\bibfnamefont {M.}~\bibnamefont {Mariantoni}},
  \bibinfo {author} {\bibfnamefont {J.~M.}\ \bibnamefont {Martinis}},\ and\
  \bibinfo {author} {\bibfnamefont {A.~N.}\ \bibnamefont {Cleland}},\
  }\bibfield  {title} {\bibinfo {title} {Surface codes: Towards practical
  large-scale quantum computation},\ }\href
  {https://doi.org/10.1103/PhysRevA.86.032324} {\bibfield  {journal} {\bibinfo
  {journal} {Phys. Rev. A}\ }\textbf {\bibinfo {volume} {86}},\ \bibinfo
  {pages} {032324} (\bibinfo {year} {2012})}\BibitemShut {NoStop}%
\bibitem [{\citenamefont {Waintal}(2019)}]{Waintal2019}%
  \BibitemOpen
  \bibfield  {author} {\bibinfo {author} {\bibfnamefont {X.}~\bibnamefont
  {Waintal}},\ }\bibfield  {title} {\bibinfo {title} {What determines the
  ultimate precision of a quantum computer},\ }\href
  {https://doi.org/10.1103/PhysRevA.99.042318} {\bibfield  {journal} {\bibinfo
  {journal} {Phys. Rev. A}\ }\textbf {\bibinfo {volume} {99}},\ \bibinfo
  {pages} {042318} (\bibinfo {year} {2019})}\BibitemShut {NoStop}%
\bibitem [{\citenamefont {Reiher}\ \emph {et~al.}(2017)\citenamefont {Reiher},
  \citenamefont {Wiebe}, \citenamefont {Svore}, \citenamefont {Wecker},\ and\
  \citenamefont {Troyer}}]{Reiher}%
  \BibitemOpen
  \bibfield  {author} {\bibinfo {author} {\bibfnamefont {M.}~\bibnamefont
  {Reiher}}, \bibinfo {author} {\bibfnamefont {N.}~\bibnamefont {Wiebe}},
  \bibinfo {author} {\bibfnamefont {K.~M.}\ \bibnamefont {Svore}}, \bibinfo
  {author} {\bibfnamefont {D.}~\bibnamefont {Wecker}},\ and\ \bibinfo {author}
  {\bibfnamefont {M.}~\bibnamefont {Troyer}},\ }\bibfield  {title} {\bibinfo
  {title} {Elucidating reaction mechanisms on quantum computers},\ }\href
  {https://doi.org/10.1073/pnas.1619152114} {\bibfield  {journal} {\bibinfo
  {journal} {Proceedings of the National Academy of Sciences}\ }\textbf
  {\bibinfo {volume} {114}},\ \bibinfo {pages} {7555} (\bibinfo {year}
  {2017})},\ \Eprint
  {https://arxiv.org/abs/https://www.pnas.org/doi/pdf/10.1073/pnas.1619152114}
  {https://www.pnas.org/doi/pdf/10.1073/pnas.1619152114} \BibitemShut {NoStop}%
\bibitem [{\citenamefont {Fowler}(2013)}]{Fowler_2013}%
  \BibitemOpen
  \bibfield  {author} {\bibinfo {author} {\bibfnamefont {A.~G.}\ \bibnamefont
  {Fowler}},\ }\href {https://doi.org/10.48550/arXiv.1210.4626} {\bibinfo
  {title} {Time-optimal quantum computation}} (\bibinfo {year} {2013}),\
  \Eprint {https://arxiv.org/abs/1210.4626} {arXiv:1210.4626 [quant-ph]}
  \BibitemShut {NoStop}%
\bibitem [{\citenamefont {Gidney}\ and\ \citenamefont
  {Fowler}(2019)}]{Gidney_2019}%
  \BibitemOpen
  \bibfield  {author} {\bibinfo {author} {\bibfnamefont {C.}~\bibnamefont
  {Gidney}}\ and\ \bibinfo {author} {\bibfnamefont {A.~G.}\ \bibnamefont
  {Fowler}},\ }\bibfield  {title} {\bibinfo {title} {Efficient magic state
  factories with a catalyzed {$|CCZ\rangle$} to {$2|T\rangle$}
  transformation},\ }\href {https://doi.org/10.22331/q-2019-04-30-135}
  {\bibfield  {journal} {\bibinfo  {journal} {{Quantum}}\ }\textbf {\bibinfo
  {volume} {3}},\ \bibinfo {pages} {135} (\bibinfo {year} {2019})}\BibitemShut
  {NoStop}%
\bibitem [{\citenamefont {Chen}\ \emph {et~al.}(2022)\citenamefont {Chen},
  \citenamefont {Cotler}, \citenamefont {Huang},\ and\ \citenamefont
  {Li}}]{Chen:2022}%
  \BibitemOpen
  \bibfield  {author} {\bibinfo {author} {\bibfnamefont {S.}~\bibnamefont
  {Chen}}, \bibinfo {author} {\bibfnamefont {J.}~\bibnamefont {Cotler}},
  \bibinfo {author} {\bibfnamefont {H.-Y.}\ \bibnamefont {Huang}},\ and\
  \bibinfo {author} {\bibfnamefont {J.}~\bibnamefont {Li}},\ }\href
  {https://doi.org/10.48550/ARXIV.2210.07234} {\bibinfo {title} {The complexity
  of nisq}} (\bibinfo {year} {2022}),\ \Eprint
  {https://arxiv.org/abs/arxiv:2210.07234} {arxiv:2210.07234} \BibitemShut
  {NoStop}%
\bibitem [{\citenamefont {Gottesman}(1998)}]{Gottesman:1998}%
  \BibitemOpen
  \bibfield  {author} {\bibinfo {author} {\bibfnamefont {D.}~\bibnamefont
  {Gottesman}},\ }\href {https://doi.org/10.48550/ARXIV.QUANT-PH/9807006}
  {\bibinfo {title} {The \mbox{Heisenberg} representation of quantum
  computers}} (\bibinfo {year} {1998}),\ \Eprint
  {https://arxiv.org/abs/quant-ph/9807006} {quant-ph/9807006} \BibitemShut
  {NoStop}%
\bibitem [{\citenamefont {Jaques}\ \emph {et~al.}(2020)\citenamefont {Jaques},
  \citenamefont {Naehrig}, \citenamefont {Roetteler},\ and\ \citenamefont
  {Virdia}}]{Jaques}%
  \BibitemOpen
  \bibfield  {author} {\bibinfo {author} {\bibfnamefont {S.}~\bibnamefont
  {Jaques}}, \bibinfo {author} {\bibfnamefont {M.}~\bibnamefont {Naehrig}},
  \bibinfo {author} {\bibfnamefont {M.}~\bibnamefont {Roetteler}},\ and\
  \bibinfo {author} {\bibfnamefont {F.}~\bibnamefont {Virdia}},\ }\bibfield
  {title} {\bibinfo {title} {Implementing \mbox{Grover} oracles for quantum key
  search on \mbox{AES and LowMC}},\ }in\ \href@noop {} {\emph {\bibinfo
  {booktitle} {Advances in Cryptology -- EUROCRYPT 2020}}},\ \bibinfo {editor}
  {edited by\ \bibinfo {editor} {\bibfnamefont {A.}~\bibnamefont {Canteaut}}\
  and\ \bibinfo {editor} {\bibfnamefont {Y.}~\bibnamefont {Ishai}}}\ (\bibinfo
  {publisher} {Springer International Publishing},\ \bibinfo {year} {2020})\
  pp.\ \bibinfo {pages} {280--310}\BibitemShut {NoStop}%
\bibitem [{\citenamefont {Garcia-Saez}\ and\ \citenamefont
  {Latorre}(2012)}]{Latorre:2011}%
  \BibitemOpen
  \bibfield  {author} {\bibinfo {author} {\bibfnamefont {A.}~\bibnamefont
  {Garcia-Saez}}\ and\ \bibinfo {author} {\bibfnamefont {J.~I.}\ \bibnamefont
  {Latorre}},\ }\bibfield  {title} {\bibinfo {title} {An exact tensor network
  for the \mbox{3SAT} problem},\ }\href {https://doi.org/10.26421/QIC12.3-4-8}
  {\bibfield  {journal} {\bibinfo  {journal} {Quantum Information and
  Computation}\ }\textbf {\bibinfo {volume} {12}},\ \bibinfo {pages} {0283}
  (\bibinfo {year} {2012})}\BibitemShut {NoStop}%
\bibitem [{\citenamefont {Yang}\ \emph {et~al.}(2021)\citenamefont {Yang},
  \citenamefont {Zi}, \citenamefont {Wu}, \citenamefont {Guo}, \citenamefont
  {Zhang},\ and\ \citenamefont {Sun}}]{Yang}%
  \BibitemOpen
  \bibfield  {author} {\bibinfo {author} {\bibfnamefont {S.}~\bibnamefont
  {Yang}}, \bibinfo {author} {\bibfnamefont {W.}~\bibnamefont {Zi}}, \bibinfo
  {author} {\bibfnamefont {B.}~\bibnamefont {Wu}}, \bibinfo {author}
  {\bibfnamefont {C.}~\bibnamefont {Guo}}, \bibinfo {author} {\bibfnamefont
  {J.}~\bibnamefont {Zhang}},\ and\ \bibinfo {author} {\bibfnamefont
  {X.}~\bibnamefont {Sun}},\ }\href {https://doi.org/10.48550/ARXIV.2101.05430}
  {\bibinfo {title} {Efficient quantum circuit synthesis for \mbox{SAT-oracle}
  with limited ancillary qubit}} (\bibinfo {year} {2021}),\ \Eprint
  {https://arxiv.org/abs/arXiv:2101.05430} {arXiv:2101.05430} \BibitemShut
  {NoStop}%
\end{thebibliography}%

\appendix

\section{Quantum Circuits for Oracle and Diffusion Operators \label{appendix:circuit}}
We discuss here some details of the quantum circuit used in the \emph{simulations} of GA for Fig.~\ref{fig:entropy} only,
not to be confused with the \QIGA calculations.

For obtaining the entanglement entropy plot Fig.~\ref{fig:entropy}, which shows the entanglement not only between Grover iterations but also 
inside of the  substeps of the oracle and diffusion circuits, we used the circuits shown in the figure below. These circuits are for the case where
the target bitstring $w$ is known. (The target for the diffusion operator is always known, since it can be implemented as the oracle which targets
$|000...0\rangle$ pre- and post-processed by a Hadamard gate on each qubit.) The circuit pattern below uses at most three-qubit gates. 
This is in contrast to the implementation sometimes seen where the oracle is implemented
by a ``multi-controlled'' gate, which is equivalent to our observation in the main text that the oracle can always in principle be implemented
by a rank $\chi=2$ MPO (for the case of a single targeted bitstring $w$).

For the specific case of four qubits and a target bitstring $w=1011$, the oracle circuit pattern used for the Fig.~\ref{fig:entropy} simulation was:
\begin{center}
\includegraphics[width=0.7\columnwidth]{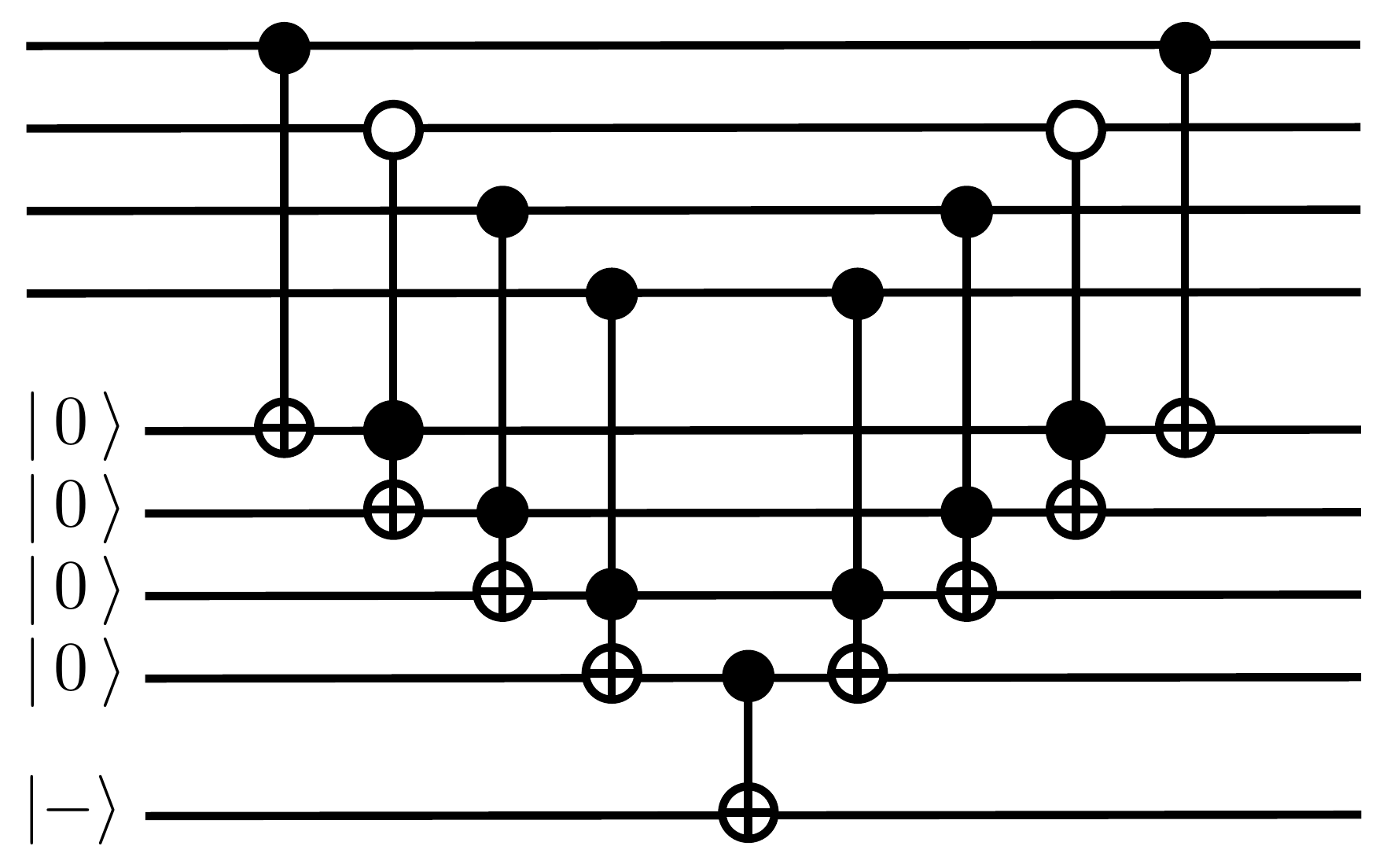}
\end{center}

For Grover's algorithm on $n$ qubits, the operator above is initialized by preparing $n+1$ additional ancilla qubits, $n$ in the $\ket{0}$ state and the $n+1$ qubit in the $\ket{-} = H\ket{1}$ state. By using Toffoli gates acting on the upper and lower registers, the ancillas are flipped to indicate that each qubit of the target bitstring has been found (upper control) and that all previous bits have been found (lower control). If so, the next ancilla qubit is flipped to 1. 

The notation of the white circle for the upper control of the second Toffoli gate stands for an ``anti-control'' meaning the gate acts only if that qubit is $\ket{0}$. This kind of control can be viewed as shorthand for:
\begin{center}
\includegraphics[width=0.5\columnwidth]{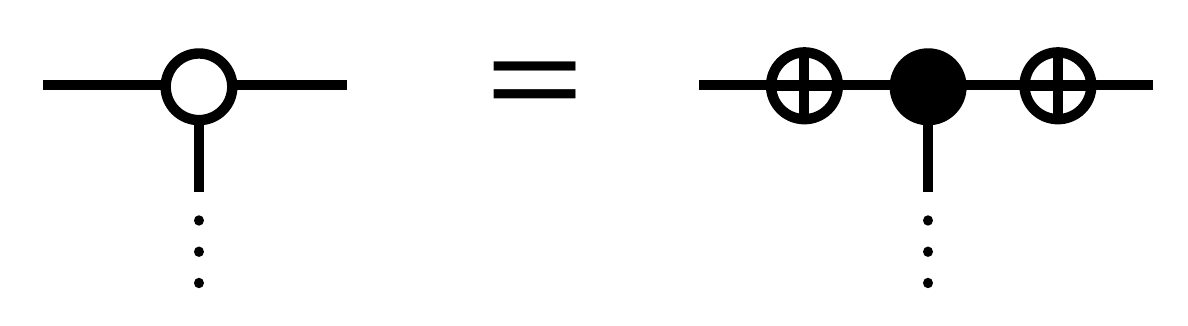}
\end{center}
that is, NOT gates on either side of a regular control.

At the center of the circuit, if the $n^\text{th}$ ancilla qubit is 1, then the $n+1$ ancilla is acted on by a NOT gate which results in a minus sign (``phase kickback mechanism'')
for the amplitude of the state with the target bitstring in the upper register. Lastly, the Toffoli gates are acted in reverse order to ``uncompute'' the ancilla register, restoring all of the ancillas to their initial product state. It is easy to check by inspection that applying the above circuit
to $\ket{b}\ket{0000-}_A$ leads to $\pm \ket{b}\ket{0000-}_A$ depending on wither $b=w$ or not.

\section{\label{sec:grover_mpo} A low rank Matrix Product Operator (MPO) for Grover's oracle}

An interesting formal observation about Grover's algorithm oracle operators 
$U_w$ is that they can be cast into the form of  a rank-$2$ MPO (rank $1+S$ in the case of multiple solutions).
We emphasize that this low-rank MPO form of the oracle was \emph{not} used in any of our numerical experiments, and 
computing it is just as hard as simulating one oracle step of Grover's algorithm (unless one is given the target bitstrings $w^\alpha$ in advance).
 
The explicit form of the MPO for a Grover oracle is
\begin{align}
U_w = 
\begin{bmatrix} 1 & 1 \end{bmatrix}
 \left( \prod_{i=1}^n M_i \right)
\begin{bmatrix} 1 \\ -2 \end{bmatrix}
\label{eq:Uw_is_mpo1}
\end{align} 
with
\begin{align}
M_i = 
\begin{bmatrix} I_i & 0                        & 0 &...\\ 
                0         & \ket{w^1_i}\bra{w^1_i} & 0 & ... \\
                0         & 0                        & \ket{w^2_i}\bra{w^2_i} & ... \\
                ...       & ...     & ... & ... \\
                ...      & 0 & 0 & \ket{w^S_i}\bra{w^S_i}
\end{bmatrix} 
\label{eq:Uw_is_mpo2}
\end{align} 
where ${\rm I}_{i}$ is the $2\times 2$ identity matrix acting on qubit $i$ and
$\ket{w^\alpha_i}\bra{w^\alpha_i}$ the projector on the bitstring $i$ of solution $\alpha$.
We emphasize that this MPO {\it exists} but its construction is not necessarily easy since, by definition, one does not have access to the solutions $w^\alpha$.
A similar MPO can be written for the diffusion operator $U_s$ with the replacement of $M_i$ by
\begin{align}
    M'_i = 
    \begin{bmatrix} I_i & 0                   \\ 
                0         & \ket{+}\bra{+} 
\end{bmatrix} 
\end{align}
 in Eq.\eqref{eq:Uw_is_mpo1}.

\end{document}